\documentclass[a4paper,11pt]{article}
\usepackage{jheppub}
\makeatletter
\gdef\@fpheader{}
\usepackage{jheppub}
\usepackage[T1]{fontenc}
\usepackage{graphicx}
\usepackage{amsmath,amssymb,amsfonts}
\usepackage{xspace}
\usepackage{subfig}
\usepackage{float}
\usepackage{color}
\usepackage{xcolor}
\usepackage{bbding}
\usepackage[utf8]{inputenc}
\usepackage{commath}
\usepackage{slashed}
\usepackage{fancyvrb}
\usepackage[utf8]{inputenc}
\usepackage[english]{babel}
\usepackage{orcidlink}
\usepackage{cancel}
\usepackage{comment}
\usepackage{hyperref}
\usepackage{tabularray}
\UseTblrLibrary{booktabs}
\usepackage{gensymb}
\usepackage{scalerel}
\usepackage[titletoc]{appendix}

\title{Anatomy of RHN DM relic in the vanilla scotogenic neutrino mass model}

\author[a]{Sujit Kumar Sahoo$^{\orcidlink{https://orcid.org/0000-0002-9014-933X}}$,}
\emailAdd{ph21resch11008@iith.ac.in}

\author[a]{Narendra Sahu$^{\orcidlink{https://orcid.org/0000-0002-9675-0484}}$,}
\emailAdd{nsahu@phy.iith.ac.in}

\author[a]{Vicky Singh Thounaojam$^{\orcidlink{https://orcid.org/0009-0001-6257-5171}}$}
\emailAdd{ph22resch01004@iith.ac.in}

\affiliation[a]{Department of Physics, Indian Institute of Technology Hyderabad, Kandi, Telangana-502285, India.}

\abstract{
The scotogeneic neutrino mass models are very popular choices to generate light neutrino masses via radiative mechanism. In these models, the particles running in the loop are distinguished from the standard model due to an imposed $\mathcal{Z}_2$ symmetry under which the loop particles are odd. Therefore, the lightest particle running in the loop can be a viable dark matter candidate. In this paper, we revisit the minimal scotogenic neutrino mass model and study the anatomy of right handed neutrino (RHN) DM relic, taking into account contributions from self-annihilation, co-annihilation, conversion-driven processes, as well as production via the freeze-in mechanism. We impose the constraints from direct detection and collider searches of DM including anomalous magnetic moment of muon, charged lepton flavor violation and low-energy neutrino oscillation data to show that the lightest RHN can be a viable DM in the mass range: $M_{h}/2\lesssim M_{\rm DM}\lesssim2000 {\rm GeV}$ (thermal DM) and $0.1 ~{\rm GeV}\lesssim M_{\rm DM}\lesssim 1000 {\rm GeV}$ (non-thermal DM), where $M_h$ denotes the Standard Model Higgs mass and $M_{\rm DM}$ is the RHN dark matter mass. We also find the displaced vertex signatures of long lived particles which can be probed at future colliders.
}

\begin{document}
	
\maketitle
\flushbottom

%%%%%%%%%%%%%%%%%%%%%%%%%%%%%%%%%%%%%%%%%%%%%%%%%%%%%%%%%%%%%%%%%%%%%%%%%%%%%%%%%%%%%%%%%%
\section{Introduction}\label{sec:Introduction}
The Scotogenic model \cite{Ma:2006km} represents a simple yet highly appealing extension of the Standard Model (SM), in which neutrino masses are generated radiatively at the one-loop level by incorporating three right-handed neutrinos ($N_{1,2,3}$) and a scalar doublet $\eta$ which belong to a $\mathcal{Z}_2$-odd dark sector. In contrast to the conventional seesaw mechanism—where the smallness of neutrino masses (Majorana) originates from the presence of an ultra-heavy mass scale—the radiative nature of neutrino mass generation in the Scotogenic model avoids such high scales and instead yields a compact and transparent structure of the form
\begin{equation*}
    m_\nu \sim \frac{M_{\rm DM}}{16\pi^2}\times (\text{Yukawa})^2 \times \lambda_5,%(\text{mass splitting}) ,.
\end{equation*}
where $\lambda_5$ represents the strength of lepton number violation such that in the limit, $\lambda_5\to 0$, $m_{\nu}\to 0$.
Furthermore, the lightest dark-sector particle running in the loop can naturally serve as a viable dark matter (DM) candidate. In the literature, both scalar \cite{Barbieri:2006dq}
and fermionic \cite{Kubo:2006yx,Vicente:2014wga,Liu:2022byu,Avila:2025qsc,Molinaro:2014lfa,Hessler:2016kwm,Baumholzer:2019twf,Heeck:2022rep,Bonilla:2019ipe,DeRomeri:2022cem,Chun:2023vbh,Borah:2020wut,Borah:2018smz,Roy:2025moo,Borah:2021khc,Borah:2021pet,Borah:2021qmi,Borah:2023hqw,Ganguly:2023jml,Borah:2024gql}particles have been extensively studied as viable DM candidates within the vanilla scotogenic framework, with several works demonstrating mechanisms capable of reproducing the observed relic density while remaining consistent with existing phenomenological constraints.

In the fermionic sector, the lightest right-handed neutrinos (RHN) behaves as a viable DM. However, it faces different challenges, predominantly in achieving the correct relic density ($\Omega_{\rm DM}\simeq0.12$ \cite{Planck:2018vyg}) across a wide region of parameter space, since its annihilation channels are strongly suppressed by the small Yukawa couplings required to satisfy charged lepton flavor violation (cLFV) constraints \cite{MEGII:2023ltw}, neutrino oscillation data \cite{Super-Kamiokande:1998kpq,SNO:2001kpb,DoubleChooz:2011ymz,DayaBay:2012fng,RENO:2012mkc}, the muon anomalous magnetic moment \cite{Aliberti:2025beg,Muong-2:2025xyk}, and electroweak precision tests \cite{Frank:2021pkc}. Often new particles are introduced to bring down the RHN DM relic abundance to correct ball park \cite{Bonilla:2019ipe,DeRomeri:2022cem,Chun:2023vbh}.

In this work, we study the anatomy of RHN DM relic within the minimal scotogenic setup and demonstrate that the observed relic density can still be successfully reproduced in regions of parameter space where it is conventionally assumed to be unachievable, by considering self-annihilation (SA), co-annihilation (CA) \cite{Griest:1990kh} and conversion-driven \cite{DAgnolo:2017dbv,Heeck:2022rep,Paul:2024prs,Paul:2025spm,Garny:2017rxs,Maity:2019hre,Liu:2025swd} processes. We identify viable regions that simultaneously satisfy all relevant phenomenological constraints, including low-energy neutrino oscillation data, bounds from cLFV, electroweak precision measurements via the oblique $S$ and $T$ parameters, and the anomalous magnetic moment of the muon. The parameter space is also compatible with the RHN DM searches. From the indirect detection perspective, RHN DM is largely unconstrained due to the p-wave suppression of its annihilation cross section. Moreover, direct detection rates are loop suppressed, rendering them well below current experimental sensitivity. However, the parameter space compatible with the RHN DM relic density is associated with an interesting collider signature. In particular, the scalar doublet can be produced copiously through its Higgs and gauge portal interactions. The subsequent decay of its charged component into charged leptons and the singlet fermion can give rise to displaced vertex signatures, thereby providing a complementary probe of the viable parameter space.

The remainder of the paper is organized as follows. In Section~\ref{sec:model}, we present the model and discuss existing constraints arising from vacuum stability, corrections to electroweak precision parameters, the anomalous magnetic moment, charged lepton flavor violation, and neutrino oscillation data. In Section~\ref{sec:Relic_Density}, we analyze the dark matter phenomenology, including the relic abundance and dark matter searches. The computation of the spin-independent and spin-dependent direct detection cross sections is presented in Section~\ref{sec:DD}. In Section~\ref{sec:DV}, we study the displaced-vertex signatures arising from the decay of the charged scalar component into RHN DM and a charged lepton. Finally, we summarize our findings and conclude in Section~\ref{sec:conclusion}.

\section{The Model}\label{sec:model}
In a vanilla scotogenic model \cite{Ma:2006km}, to generate light neutrino mass, the SM is augmented with three generations of Majorana singlet fermions, $N_i(i=1,2,3)$ and a scalar doublet, $\eta=(\eta^+~\eta^0)^T$ (with hypercharge, $Y=+1$, where the electromagnetic charge is defined as $Q=I_3+Y/2$). With this particle content, a discrete $\mathcal{Z}_2$ symmetry is imposed under which SM particles are even and all other particles are odd. The lightest $\mathcal{Z}_2$ odd particle serves as the DM candidate. In our analysis, we consider the lightest RHN, $N_1$ to be the DM candidate. The charge assignments are provided in Table \ref{tab:particles}.

\begin{table}[H]
		\small
		\begin{center}
			\begin{tabular}{||@{\hspace{0cm}}c@{\hspace{0cm}}|@{\hspace{0cm}}c@{\hspace{0cm}}|@{\hspace{0cm}}c@{\hspace{0cm}}||@{\hspace{0cm}}c@{\hspace{0cm}}||}
				\hline
				\hline
				\begin{tabular}{c}
                {\bf ~~~~Symmetry~~~~}\\
				%{\bf ~~~~ Gauge~~~~}\\
					{\bf ~~~~Group~~~~}\\ 
					\hline
					
					$SU(2)_{L}$\\ 
					\hline
					$U(1)_{Y}$\\ 
					\hline
					$Z_2$\\ 
				\end{tabular}
				&
				\begin{tabular}{c|c}
					\multicolumn{2}{c}{\bf Fermion Fields}\\
					\hline
					~~~$L$~~~& ~$N$~  \\
					\hline
					$2$&$1$\\
					\hline
					$-1$&$0$\\
					\hline
					$+$&$-$\\
				\end{tabular}
				&
				\begin{tabular}{c|c}
					\multicolumn{2}{c}{\bf Scalar Field}\\
					\hline
					~~~$H$~~~& ~~~$\eta$~~~\\
					\hline
					$2$&$2$\\
					\hline
					$1$&$1$\\
                    \hline
					$+$&$-$\\
				\end{tabular}\\
				\hline
				\hline
			\end{tabular}
			\caption{Particle content of the scotogenic neutrino mass model.}
			\label{tab:particles}
		\end{center}    
	\end{table}

The relevant Lagrangian of the model is given by,
\begin{equation}
    \mathcal{L}\supseteq M_N\overline{N^C}N -y_N \overline{L}\tilde{\eta}N + h.c.- V_{\rm scalar},
\end{equation}
where $\tilde{\eta}=i\sigma_2\eta^*$. We have suppressed the lepton flavor indices and the generation indices of $N$. The scalar potential is given by,
\begin{eqnarray}
    V_{\rm scalar}&=& -\mu^2_h H^\dagger H + \lambda_h (H^\dagger H)^2 + \mu^2_\eta \eta^\dagger \eta + \lambda_\eta (\eta^\dagger \eta)^2\nonumber\\
    &{}&+\lambda_3 (\eta^\dagger\eta)(H^\dagger H)+\lambda_4(H^\dagger\eta)(\eta^\dagger H)\nonumber\\
    &{}&+\frac{\lambda_5}{2}((H^\dagger\eta)(H^\dagger\eta)+h.c.)
\end{eqnarray}
The electroweak symmetry breaks when the SM Higgs acquires a vacuum expectation value (VEV): $v$. The masses of physical scalar states are then given by,
\begin{eqnarray}
    M^2_h&=& \lambda_hv^2\\
    M^2_{\eta^{\pm}}&=& \mu^2_\eta + \frac{\lambda_3}{2}v^2\\
    M^2_{\eta_R}&=& \mu^2_\eta + \frac{\lambda_3+\lambda_4+\lambda_5}{2}v^2\\
    M^2_{\eta_I}&=& \mu^2_\eta + \frac{\lambda_3+\lambda_4-\lambda_5}{2}v^2.
\end{eqnarray}

\subsection{Vacuum Stability and Perturbative bound}
The scalar potential must be bounded from below to avoid instability. This leads to the following conditions on quartic couplings \cite{Kannike:2012pe,Belyaev:2016lok}:
\begin{eqnarray}
    2\sqrt{\lambda_h \lambda_\eta}+\lambda_3\geq&&0,\\
    2\sqrt{\lambda_h \lambda_\eta}+\lambda_3+\lambda_4-|\lambda_5|\geq&&0.
\end{eqnarray}
We also use a conservative perturbative bound on the scalar quartic couplings to a maximum value of $4\pi$.

\subsection{Electroweak precision tests}\label{sec:ew_precision}

The presence of the additional scalar doublet $\eta$ introduces new gauge interactions, which in turn modify the SM electroweak gauge-boson propagators at the one-loop level. These loop-induced effects are parametrized as corrections to the electroweak precision test parameters or conventionally known as oblique parameters: $S,T$ and $U$ \cite{Peskin:1990zt,Peskin:1991sw,Grimus:2007if}. Following the analysis reported in \cite{Frank:2021pkc}, in the limit $U=0$, the $S$ and $T$ parameters are constrained to $0.00\pm0.07$ and $0.05\pm0.06$ at $95\%$ C.L., respectively. The expressions of $S$ and $T$ in this model are given in Appendix \ref{app:oblique}.

We choose our parameters such that $M_{\eta_R}$ and $\lambda_{4,5}$ are treated as free inputs, while the masses of $\eta_I$ and $\eta^+$ are determined through
\begin{equation}\label{eq:param_choice}
    M_{\eta_I} = \sqrt{M_{\eta_R}^2 - \lambda_5 v^2},\,\,
    M_{\eta^+} = \sqrt{M_{\eta_R}^2 - \frac{(\lambda_4 + \lambda_5)}{2} v^2}.
\end{equation}

\begin{figure}[h]
    \centering
    \includegraphics[width=0.6\linewidth]{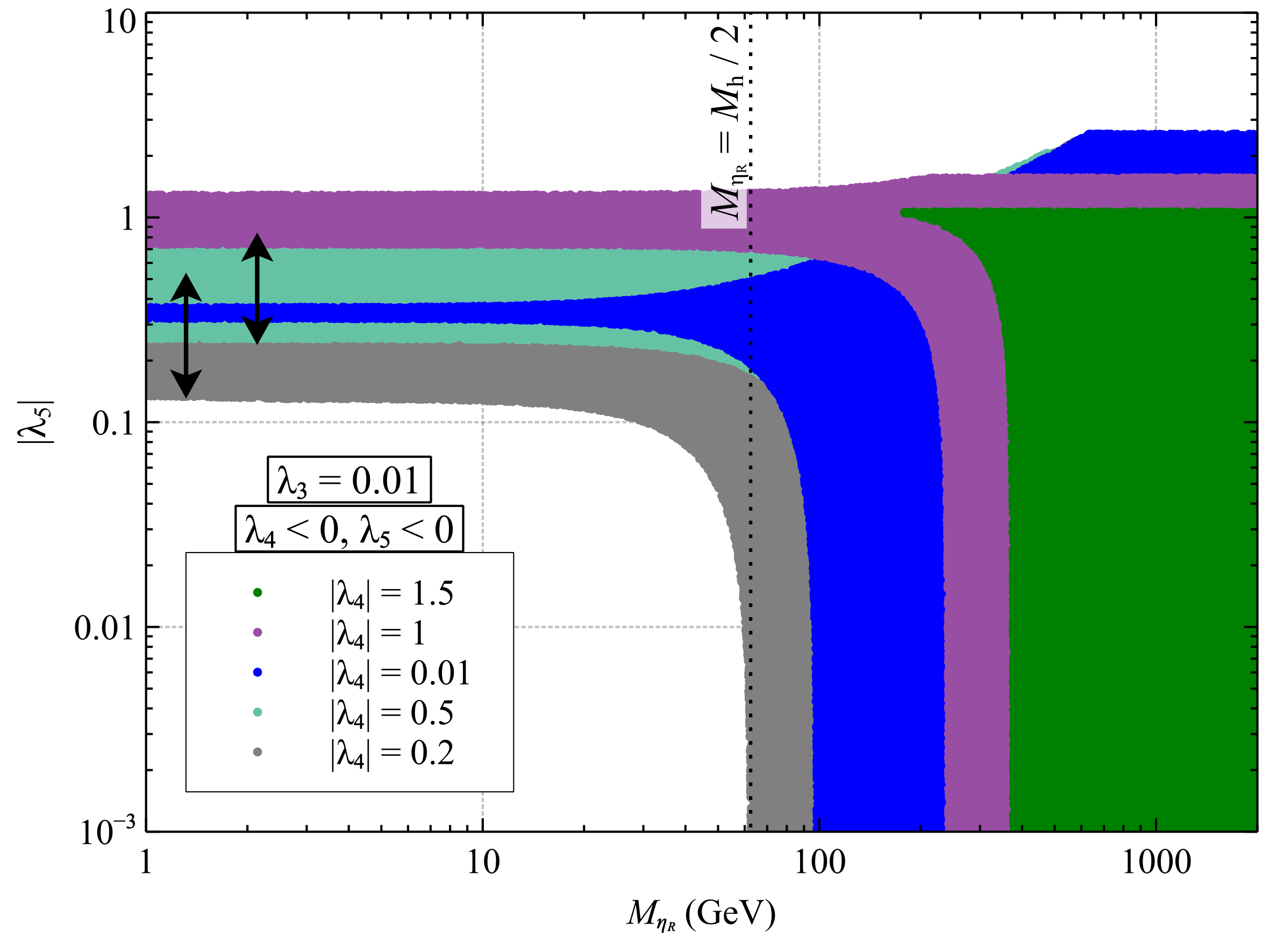}
    \caption{Behaviour of $|\lambda_{5}|$ vs $M_{\eta_R}$ for various $\lambda_{4}$ values. Left to the line of $M_{\eta_{\rm R}}=M_{\rm h}/2$ corresponds to the disallowed region from Higgs Invisible decay (see Appendix \ref{app:HID}). Here we choose $\lambda_{4,5}<0$ and $\lambda_3=10^{-2}$.}
    \label{fig:l5_vs_MetaR}
\end{figure}

With this choice, any value of $\lambda_3$ can be absorbed into the mass parameter $\mu_{\eta}$ without altering $M_{\eta^+}$. In Fig.~\ref{fig:l5_vs_MetaR}, we show the parameter space consistent with electroweak precision constraints in the plane of $|\lambda_5|$ and $M_{\eta_R}$ for different variations of $\lambda_4$, where we have used the constraints from LEP (i.e. $M_{\eta^+}>100$ GeV) \cite{Babu:2019mfe} and $Z$-invisible decay (i.e. $M_{\eta_R}+M_{\eta_I}>M_Z$) \cite{ParticleDataGroup:2024cfk}. Here, we choose $\lambda_3=10^{-2}$. For this choice of parameter, we get the maximum value of $|\lambda_5|$ allowed is $\sim2.5$. Further increase in $\lambda_3$ allows relatively larger $|\lambda_5|$ values.

\begin{figure}[h]
    \centering
    \includegraphics[width=0.4\linewidth]{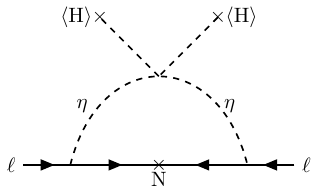}
    \caption{One loop realization of Majorana neutrino mass.}
    \label{fig:neutrnomass}
\end{figure}

\subsection{Neutrino mass generation}\label{sec:nu_mass}
In the effective theory, the neutrino mass is generated at one-loop level \cite{Ma:2006km} as shown in Fig. \ref{fig:neutrnomass}.
In this model, the neutrino mass matrix is given as:

\begin{eqnarray}\label{eq:numass}
\left(m_{\nu}\right)_{\alpha\beta}&=&\sum_{i=1}^3\frac{{y}_{\alpha i}{y}_{i\beta}^*}{16\pi^2} M_{N_i}\left[\frac{M_{\eta_R}^2}{M^2_{\eta_R}-M_{N_i}^2}\log\left(\frac{M^2_{\eta_R}}{M_{N_i}^2}\right)-\frac{M_{\eta_I}^2}{M^2_{\eta_I}-M_{N_i}^2}\log\left(\frac{M^2_{\eta_I}}{M_{N_i}^2}\right)\right].
\end{eqnarray}

The above equation takes the structure of
\begin{equation}
\left(m_\nu\right)_{\alpha\beta}=\left(y^T\Lambda y\right)_{\alpha\beta},
\end{equation}
where the $3\times3$ diagonal matrix, $\Lambda$ is given by,
\begin{equation}
\Lambda=
    \begin{pmatrix}
        \Lambda_1 & 0 & 0\\
        0 & \Lambda_2 & 0\\
        0 & 0 & \Lambda_3
    \end{pmatrix}.
\end{equation}
with
\begin{eqnarray}
    \Lambda_i&=&\frac{1}{16\pi^2} M_{N_i}\left[\frac{M_{\eta_R}^2}{M^2_{\eta_R}-M_{N_i}^2}\log\left(\frac{M^2_{\eta_R}}{M_{N_i}^2}\right)  -\frac{M_{\eta_I}^2}{M^2_{\eta_I}-M_{N_i}^2}\log\left(\frac{M^2_{\eta_I}}{M_{N_i}^2}\right)\right],
\end{eqnarray}

In our numerical analysis, the lightest RHN is considered to be the DM with mass $M_{N_1}(=M_{\rm DM})$. We fix the heavier RHN masses $M_{N_{2,3}}$ at $3~\rm TeV$. For numerical computation purpose, we define $\Delta M_R = M_{\eta_R} - M_{\rm DM}$,
\begin{equation*}
    \begin{split}
        \Delta M_{I}&= M_{\eta_I}-M_{\rm DM}\\
    &=\sqrt{\left(M_{\rm DM}+\Delta M_{R}\right)^2-\lambda_5v^2}-M_{\rm DM},
    \end{split}
\end{equation*} 
and
\begin{equation*}
    \begin{split}
        \Delta M^+&= M_{\eta^+}-M_{\rm DM}\\
    &= \sqrt{\frac{1}{2}\left[\left(M_{\rm DM}+\Delta M_{R}\right)^2-\left(\lambda_5+\lambda_4\right)v^2\right]}-M_{\rm DM}.
    \end{split}
\end{equation*}

Using the Casas-Ibarra parametrization \cite{Casas:2001sr}, the Yukawa coupling can be expressed as,
\begin{eqnarray}\label{eq:yukawa}
y&=&\sqrt{\Lambda^{-1}}R\sqrt{\hat{m}_\nu}U_{\rm PMNS}\\
&=&\begin{pmatrix}
    y_{e1} & y_{e2} & y_{e3}\\
    y_{\mu1} & y_{\mu2} & y_{\mu3}\\
    y_{\tau1} & y_{\tau2} & y_{\tau3}\\
\end{pmatrix}\nonumber
\end{eqnarray}
where $\hat{m}_\nu$ is the diagonal neutrino mass matrix and $U_{\rm PMNS}$ is the Pontecorvo-Maki-Nakagawa-Sakata matrix \cite{ParticleDataGroup:2024cfk}. The matrix $R$ is in general a complex orthogonal matrix. A typical value of $R$ is given by\footnote{For most of our analysis, we have chosen the matrix $R$ to vary randomly.}:

{
\begin{equation}\label{eq:Rmatrix}
    R=\begin{pmatrix}
        1.371+i0.452 & -1.057+i0.925& 0.349+i1.029\\
        0.533-i0.691 &1.262+i0.321 & 0.067-i0.549\\
        0.327-i0.767 & 0.556+i1.030& 1.510-i0.213
    \end{pmatrix}
\end{equation}
}
\iffalse
can be parametrized as,
{\scriptsize
\begin{equation}
    R=\begin{pmatrix}
        1 & 0& 0\\
        0 &\cos{\alpha} & \sin{\alpha}\\
        0 & -\sin{\alpha}& \cos{\alpha}
    \end{pmatrix}
    \begin{pmatrix}
        \cos{\beta}& 0 & \sin{\beta}\\
        0 & 1 & 0\\
        -\sin{\beta} & 0 & \cos{\beta}
    \end{pmatrix}
    \begin{pmatrix}
        \cos{\gamma}&\sin{\gamma}&0\\
        -\sin{\gamma}&\cos{\gamma}&0\\
        0 & 0 & 1
    \end{pmatrix},
\end{equation}
}
where $\alpha,\beta$ and $\gamma$ represent complex angles. The real and imaginary parts of $\alpha,\beta$ and $\gamma$ are randomly varied within $[-\pi,\pi]$.
\fi
We have used the best-fit values of neutrino oscillation parameters from \cite{deSalas:2020pgw} for the rest of our analysis as given in Table \ref{tab:tab2}. With this choice of parameter space, the Yukawa coupling given in Eq. (\ref{eq:yukawa}) can be written as a function of $(M_{\eta_R},M_{\eta_I}, M_{N_i})$.

\begin{table}[h]
\begin{center}
	\begin{tabular}{|@{\hspace{0cm}}c@{\hspace{0cm}}|@{\hspace{0cm}}c@{\hspace{0cm}}|}
		\hline 
        Parameters & Best-fit Values\\
		\hline
        \hline
		$\Delta m_{21}^2[10^{-5}\rm eV^2]$&7.5\\
        \hline
		$\Delta m_{31}^2[10^{-3}\rm eV^2]$&2.55\\
        \hline
		$\sin^2\theta_{12}$&34.3\\
        \hline
		$\sin^2\theta_{23}$&49.26\\
        \hline
		$\sin^2\theta_{13}$&8.53\\
        \hline
		$\delta$&$194^{o}$\\
		\hline
    \end{tabular}
	\caption{The best-fit values of the neutrino oscillation parameters \cite{deSalas:2020pgw} for normal ordering of the neutrino mass spectrum.}
	\label{tab:tab2}
\end{center}
\end{table}

\subsection{Anomalous Magnetic Moment of Muon and Charged Lepton flavor violating interaction}

\subsubsection{$\mathbf{(g-2)_\mu}$}
\begin{figure}[h]
    \centering
    \includegraphics[width=0.5\linewidth]{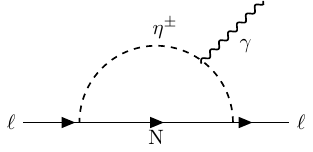}
    \caption{Feynman diagram for charged lepton flavor violation.}
    \label{fig:lfv_fynnmann}
\end{figure}
In this setup, the additional singlet fermions and the charged scalar contribute to $(g-2)_\mu$ at one-loop as illustrated in Fig. \ref{fig:lfv_fynnmann}. The contribution to the anomalous magnetic moment $\Delta a_{\mu}$ via the loop-integral for $\ell=\mu$ is given as \cite{Lindner:2016bgg}:
\begin{equation}
    \Delta a_\mu=\sum_{i}\frac{-2 |y_{\mu i}|^2}{8\pi^2}\frac{M_\mu^2}{M_{\eta^+}^2}I_{\mu},
\end{equation}
where $I_{\mu}$ is defined for $\epsilon_{i}=M_{N_{i}}/M_\mu$, $\delta=M_\mu/M_{\eta^+}$ as 
\begin{equation}
    I_{\mu}=\int^1_0dx\frac{x^2(1-x)}{(\epsilon_{i}\delta)^2(1-x)(1-\epsilon_{i}^{-2}x)+x}
\end{equation}

The Fermilab collaboration has recently updated the combined average of Muon anomalous magnetic moment with improved precision, which now reads $a_\mu (\rm exp)=116\times10^{-12}$. Additionally, the lattice QCD community has significantly minimized the uncertainty in the leading order hadronic vacuum polarization contribution to $\Delta a_\mu$. Recent calculations indicate that the SM prediction is now consistent with the updated experimental value within 1$\sigma$ uncertainty, $\Delta a_\mu=a_\mu^{\rm exp}-a_\mu^{\rm SM}=(39\pm64)\times10^{-11}$ \cite{Aliberti:2025beg,Muong-2:2025xyk}.

\subsubsection{\textbf{cLFV}}

\begin{figure}[h]
    \centering
    \includegraphics[width=0.45\linewidth]{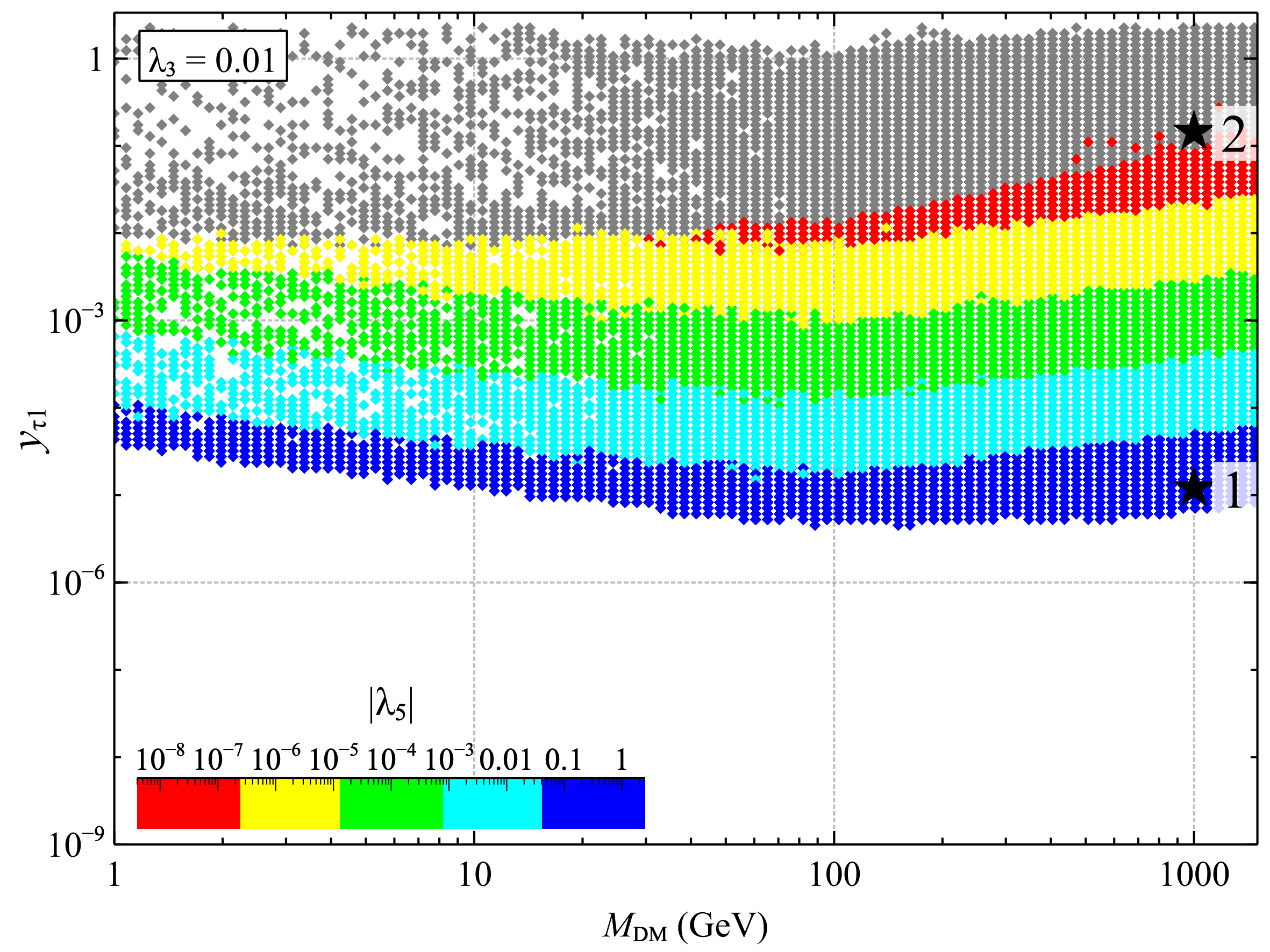}
    \includegraphics[width=0.45\linewidth]{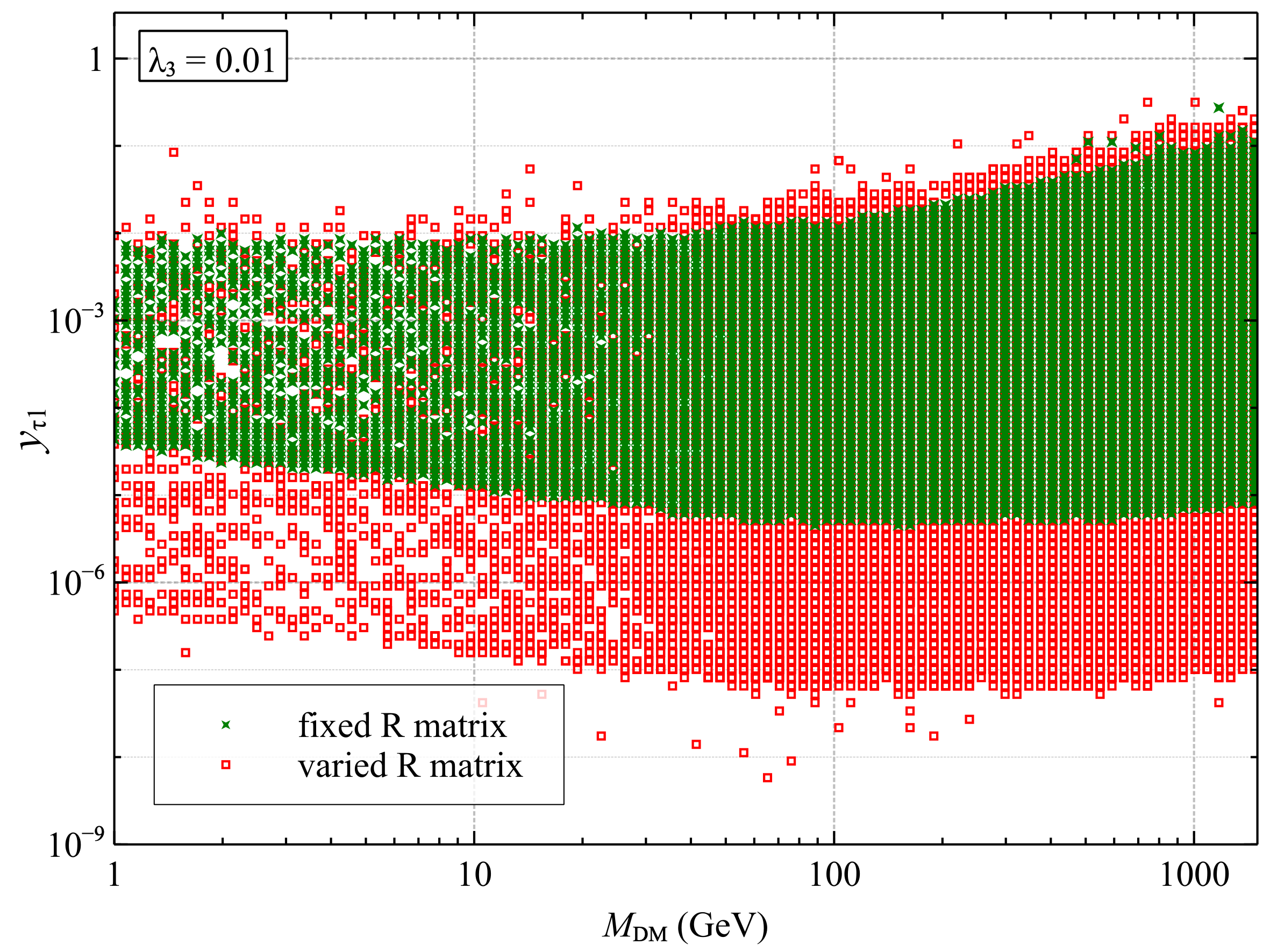}
    \caption{$y_{\tau 1}$ is shown as a function of $M_{DM}$. In the \textit{left} panel, the gray points satisfy the neutrino oscillation data, perturbativity bound, $(g-2)_\mu$ and  electro-weak precision tests except the cLFV bounds, while the colored points satisfy all the constraints including cLFV constraints. The color band indicates the $|\lambda_5|$ values. The \textit{right} panel shows the allowed region of $y_{\tau 1}$ for a fixed (green) and varied (red) $R$ matrix. We adopt $\lambda_{4,5} < 0$ throughout.}
    \label{fig:NMmassparams}
\end{figure}

In the SM, charged lepton flavor violating (cLFV) decays arise at the one-loop level and are suppressed by the extremely small neutrino masses \cite{Petcov:1976ff}. Consequently, their predicted rates lie many orders of magnitude below the reach of current experimental sensitivities, especially MEG II \cite{MEGII:2023ltw} reports an upper bound on the branching ratio of $\mu \rightarrow e\gamma$ to $3.1 \times 10^{-13}$ at 90\% C.L. Thus, the observation of any cLFV process, such as the radiative decay $\mu \rightarrow e\gamma$, would constitute a definitive signature for the BSM search. In the present model, such cLFV transitions are generated at the one-loop level, as illustrated in Fig.~\ref{fig:lfv_fynnmann}. The expression for the branching ratio of $\mu \rightarrow e\gamma$ is provided in Eq.~(\ref{eq:lfv}) \cite{Ma:2001mr,Vicente:2014wga}.
\begin{equation}\label{eq:lfv}
   {\rm Br}(\mu\rightarrow e\gamma)=\frac{3 (4 \pi )^3 \alpha}{4 G_F^2}\left[\frac{-1}{2{(4 \pi)}^2} \frac{y_{\mu i} y_{e i}^{*}}{M^2_{\eta^+}}f\left(\frac{M^2_{N_i}}{M^2_{\eta^+}}\right)\right]^2, 
\end{equation}
where $\alpha=1/137$ is the fine-structure constant, $G_F= 1.166\times 10^{-5}~\text{GeV}^{-2}$ is the Fermi's constant and the loop function $f(x)$ is given by
\begin{equation}
   f(x)=\frac{1-6x+3 x^2+2 x^3 - 6x^2\log(x)}{6 (1-x)^4}.
\end{equation}
By imposing constraints from electroweak precision tests, $(g-2)_\mu$ and cLFV, and using the CI parametrization (Eq.~\ref{eq:yukawa}), we determine the Yukawa couplings which satisfy the light neutrino mass spectrum $\hat{m}_{\nu_1,\nu_2,\nu_3}=\{0.01,8.6,50\}$ meV. We vary $|\lambda_{5,4}|$ over the ranges $\{10^{-12}-4\pi\},\{10^{-3}-4\pi\}$ and $\Delta M_R$ over $\{10^{-4}-10^2\}~\text{GeV}$. Fig.~\ref{fig:NMmassparams} presents the allowed Yukawa coupling, particularly for the $\tau$ flavor, as a function of $N_1$ mass ($M_{\rm DM}$). The \textit{left} panel corresponds to the typical $R$ matrix given in Eq.~(\ref{eq:Rmatrix}). The gray colored points indicate the parameter space excluded by current cLFV bounds, while the colored points represent the variation of $\lambda_5$, illustrating the inverse correlation between the Yukawa couplings and $\lambda_5$ required to generate correct neutrino masses. As can be seen from Eq.~(\ref{eq:yukawa}), the choice of the $R$ matrix has a significant impact on the Yukawa couplings. To perform a more general analysis, we vary the matrix $R$ randomly as given in Eq.~(\ref{eq:Rparameterization}), where the mixing angles are varied in the range $\{10^{-6},\pi\}$. The resulting allowed region for the Yukawa coupling $y_{\tau 1}$ is shown by the red points in the \textit{right} panel of Fig.~\ref{fig:NMmassparams}, while the green points correspond to the typical choice of $R$ matrix as given in Eq.~(\ref{eq:Rmatrix}), consistent with the \textit{left} panel. We find that the fixed $R$ matrix permits a minimum value of $|\lambda_5|$ of $4.1\times10^{-9}$, whereas allowing the $R$ matrix to vary randomly extends this minimum value to $6\times10^{-12}$.

For illustration purpose, we give two benchmark points ($[\star1]$ and $[\star2]$), where the BP$[\star1]$ correspond to the parameter choice $:\{\lambda_3,\lambda_4,\lambda_5,M_{\rm DM},M_{\eta_R}\}=\{0.01,-0.1,-1.91,1000~{\rm GeV},1005~{\rm GeV}\}$ which gives the Yukawa coupling given by
\begin{equation*}
    y=10^{-6}\begin{pmatrix}
        -0.69-i~2.90 & -1.61+i~3.03 & -2.33-i~1.05\\
        -12.07-i~0.01 & 3.82+i~4.12 & -1.08+i~14.45\\
        -5.43+i~5.89 & 5.01-i~2.84 & 4.43+i~9.63
    \end{pmatrix}.
\end{equation*}
We calculated the corresponding $(g-2)_\mu$ and cLFV contribution are found to be 
\begin{equation*}
    \{ (g-2)_\mu,~{\rm Br}(\mu \rightarrow e\gamma) \}=\{6.11 \times10^{-28},~1.32 \times 10^{-29} \}.
\end{equation*}
These values are exceedingly small, indicating that the contributions from these processes are negligible when the Yukawa couplings are of the order $\mathcal{O}(10^{-6})$. Such small Yukawa couplings are due to the choice of large $\lambda_5$. On the other hand, with a choice of small $|\lambda_5|$, the Yukawa couplings can be enhanced to $\mathcal{O}(0.1)$. We provide another BP$[\star2]$ corresponding to the parameter choice $:~\{\lambda_3,\lambda_4,\lambda_5,M_{\rm DM},M_{\eta_R}\}=\{0.01,-0.1,-1.29\times10^{-8},1000~{\rm GeV},1005~{\rm GeV}\}$. This results in the Yukawa coupling given by
\begin{equation*}
    y=10^{-2}\begin{pmatrix}
        -0.83-i~3.46 & -1.94+i~3.65 & -2.80-i~1.26\\
        -14.41-i~0.01 & 4.59+i~4.95 & -1.30+i~17.39\\
        -6.49+i~7.03 & 6.03-i~3.41 & 5.32+i~11.58
    \end{pmatrix}
\end{equation*}
The corresponding $(g-2)_\mu$ and cLFV contribution are found to be
\begin{equation*}
    \{ (g-2)_\mu,~{\rm Br}(\mu \rightarrow e\gamma) \}=\{8.78 \times10^{-20},~2.87 \times 10^{-13} \}.
\end{equation*}

Although the above BPs are shown in the left panel of Fig.~\ref{fig:NMmassparams} for illustration purpose, we consider all the points given in the right panel of Fig.~\ref{fig:NMmassparams} to calculate the relic density in the next section.

\section{Relic Density of DM}\label{sec:Relic_Density}
In the minimal scotogenic scenario, we take the singlet fermion $N_1$ to be the lightest particle in the dark sector, which is naturally stable due to the imposed $\mathcal{Z}_2$ symmetry and is therefore treated as the DM candidate. For simplicity, the heavier generations of RHN $N_2$ and $N_3$ are assumed to be significantly heavier than $N_1$, ensuring that their thermal histories do not affect the evolution of $N_1$. In our study, we fix the lightest neutrino mass to $0.01~{\rm meV}$ and investigate how the quartic couplings $\lambda_{3,4,5}$, together with the mass splitting between the DM and the next-to-lightest stable particle (NLSP), influence the resulting DM relic density. With this setup, to investigate the DM phenomenology, we consider the hierarchy:
\begin{equation}\label{eq:massorder}
    M_{\rm DM} < M_{\eta_R} < M_{\eta_I/\eta^+} \ll M_{N_{2,3}}.
\end{equation}
In this configuration—where the singlet fermion acts as DM—the channels contributing to DM depletion include self-annihilation (SA) into leptons, co-annihilation (CA) with the scalar doublet components ($\eta_R, \eta_I$ and $\eta^\pm$), co-scattering (CS) with SM particles, and decay/inverse-decay processes as shown in Appendix~\ref{app:FD}.

In order to consistently account for all relevant contributions, we divide the particle content into three sectors. Sector~1 contains the singlet DM candidate $N_1$, while sector~2 consists of the remaining dark-sector particles $\eta_R$, $\eta_I$, $\eta^+$, $N_2$ and $N_3$. As discussed earlier, the contributions from $N_{2,3}$ are suppressed due to the chosen mass hierarchy. Sector~0 comprises all SM particles. The comoving number densities of sector~1 and sector~2 are defined as $Y_1 = n_{N_1}/s$ and $Y_2 = (n_{\eta_R} + n_{\eta_I} + n_{\eta^+})/s$, respectively, where $n_i$ denotes the number density of the $i$th species and $s = 2\pi^2/45~g_{*s}(T)~T^3$ is the entropy density. The evolution of $Y_1$ and $Y_2$ is governed by the coupled Boltzmann equations (BEs):
\begin{eqnarray}
	\frac{dY_1}{dT} &=&   \frac{1}{3\mathcal{H}}\frac{ds}{dT} \left[    \langle \sigma_{1100} v \rangle ( Y_1^2 - {Y_1^{\rm eq}}^2) +    \langle \sigma_{1122} v \rangle \left( Y_1^2 - Y_2^2  \frac{{Y_1^{\rm eq}}^2}{{Y_2^{\rm eq}}^2}\right)  + \langle \sigma_{1200} v \rangle ( Y_1 Y_2 - Y_1^{\rm eq}Y_2^{\rm eq})\right. \nonumber\\
	&&+\left.  \langle \sigma_{1222} v \rangle \left( Y_1 Y_2 - Y_2^2   \frac{Y_1^{\rm eq}}{Y_2^{\rm eq}} \right) -\langle \sigma_{1211} v \rangle \left( Y_1 Y_2 - Y_1^2   \frac{Y_2^{\rm eq}}{Y_1^{\rm eq}} \right)
	-\frac{ \Gamma_{2\rightarrow 1}}{s}\left( Y_2 -Y_1 \frac{Y_2^{\rm eq}}{Y_1^{\rm eq}}  \right)        \right] ,\nonumber\\
		\label{eq:Y1}
\end{eqnarray}
\begin{eqnarray}
	\frac{dY_2}{dT} &=&   \frac{1}{3\mathcal{H}}\frac{ds}{dT}\left[    \langle \sigma_{2200} v \rangle ( Y_2^2 - {Y_2^{\rm eq}}^2) -    \langle \sigma_{1122} v \rangle \left( Y_1^2 - Y_2^2  \frac{{Y_1^{\rm eq}}^2}{{Y_2^{\rm eq}}^2}\right) +  \langle \sigma_{1200} v \rangle ( Y_1 Y_2 - Y_1^{\rm eq}Y_2^{\rm eq}) \right. \nonumber \\
	&&- \left. \langle \sigma_{1222} v \rangle \left( Y_1 Y_2 - Y_2^2   \frac{Y_1^{\rm eq}}{Y_2^{\rm eq}} \right)
	+\langle \sigma_{1211} v \rangle \left( Y_1 Y_2 - Y_1^2   \frac{Y_2^{\rm eq}}{Y_1^{\rm eq}} \right)  + \frac{ \Gamma_{2\rightarrow 1}}{s}\left( Y_2 -Y_1 \frac{Y_2^{\rm eq}}{Y_1^{\rm eq}}  \right)        \right],\nonumber\\
	\label{eq:Y2}
\end{eqnarray}
where $\langle \sigma_{\alpha\beta\gamma\delta} v\rangle$ are the thermally averaged cross-sections for processes involving the annihilation of particles of sectors $\alpha\beta\rightarrow \gamma\delta$ ($\alpha,\beta,\gamma,\delta\in\{0,1,2\}$), which is given by \cite{Gondolo:1990dk,Alguero:2022inz}:\begin{eqnarray}\label{eq:sigmavth}
    \langle \sigma_{\alpha\beta\gamma\delta} v\rangle=\frac{T}{8m_{\alpha}^2m_{\beta}^2K_{2}(\frac{m_\alpha}{T})K_{2}(\frac{m_\beta}{T})}\int_{(m_\alpha+m_\beta)^2}^\infty\sigma_{\alpha\beta\rightarrow\gamma\delta}(s)\big(s-(m_\alpha+m_\beta)^2\big)\sqrt{s}K_1\bigg(\frac{\sqrt{s}}{T}\bigg)ds.\nonumber\\ 
    \end{eqnarray}
In Eqs (\ref{eq:Y1}) and (\ref{eq:Y2}), $Y_i^{\rm eq}\left(=n_i^{\rm eq}/s\right)$ is the equilibrium abundance for $i$-th species, $\mathcal{H}=1.66\sqrt{g_*}T^2/M_{\rm Pl}$ is the  Hubble parameter with $M_{\rm Pl}=1.22\times10^{19}$ GeV being the Planck mass. The term $\Gamma_{2\rightarrow 1}$ in BEs is the conversion term, which includes both the interaction rate of the co-scattering process as well as the decay/inverse-decay and is given  by

\begin{eqnarray}\label{eq:gamma21}
    \Gamma_{2\rightarrow1}=\sum_i\Gamma_{\eta_i\rightarrow N_1,\rm SM}\frac{K_1(M_{\eta_{i}}/T)}{K_2(M_{\eta_i}/T)}+\langle \sigma_{2010} v \rangle n^{\rm eq}_{\rm SM},
\end{eqnarray}
where $\eta_{i}\in\{\eta_{R},\eta_{I},\eta^{+}\}$, $\Gamma_{2\rightarrow N_1,\rm SM}$ includes the decay rate of sector 2 particles to sector 1 particle, $\langle\sigma_{2010}v\rangle$ denotes the thermally averaged cross-sections of the co-scattering processes. The total DM relic is $Y_{\rm DM}=Y_1+Y_2$. In most of the parameter space, $Y_2$ remains under-abundant due to the large interactions (large quartic coupling as well as gauge interactions) of the sector-2 particles, unless they are very heavy. However, we note that these sector-2 particles play non-trivial role in bringing the DM relic to the correct ball-park via the conversion-driven term.

Before solving the BEs given in Eqs (\ref{eq:Y1}) and (\ref{eq:Y2}) to evaluate the relic density, we specify the independent parameters relevant for DM relic abundance:
\begin{equation*}
    \{ M_{\rm DM},~\lambda_{4},~\lambda_{5},~\Delta M_{R} \}.
\end{equation*}

\subsection{Thermal Relic density via pure self-annihilation}\label{sec:SA}
\begin{figure}[H]
    \centering
    \includegraphics[width=0.47\linewidth]{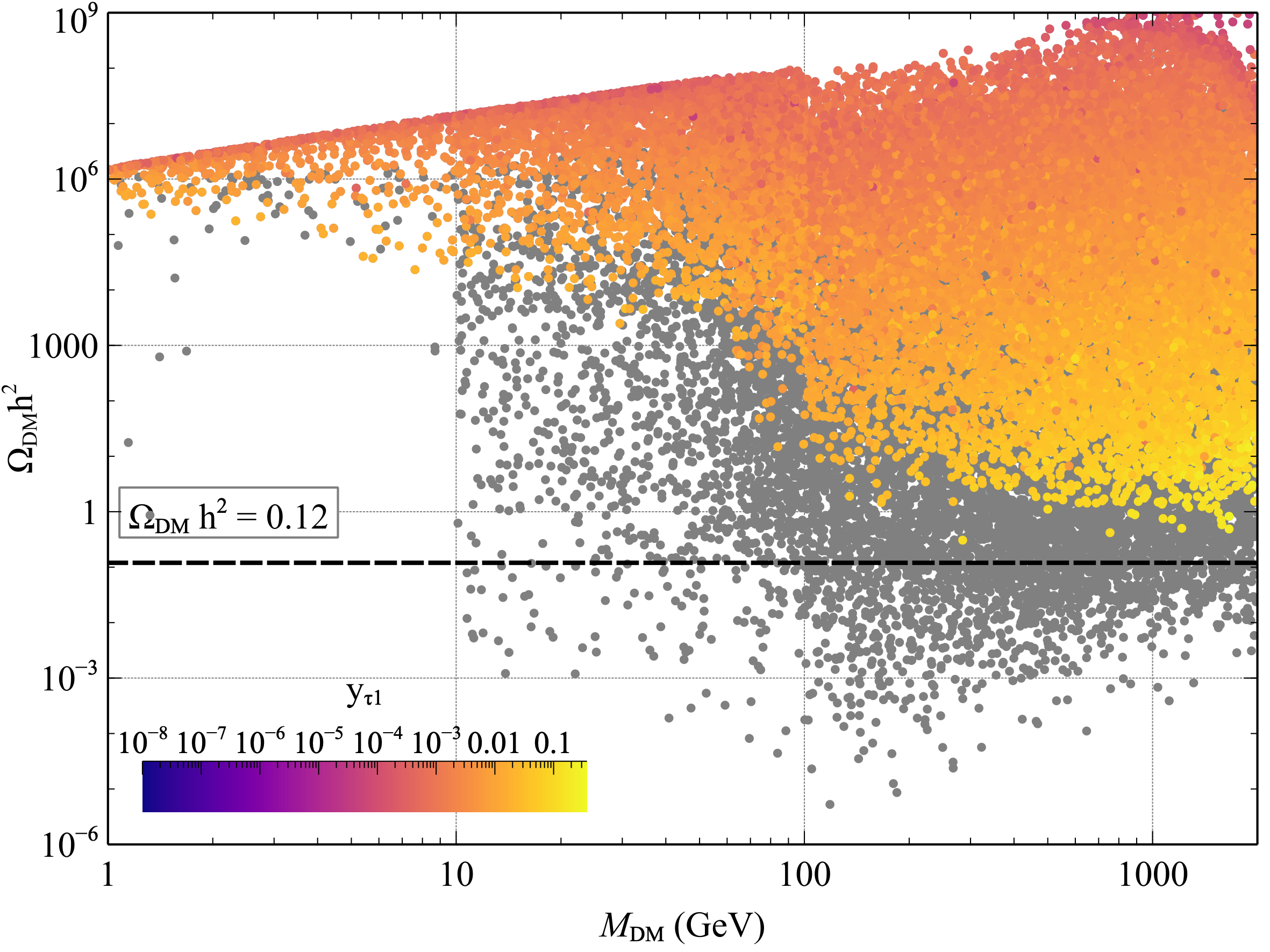}
    \caption{The DM relic density is shown as a function of $M_\text{DM}$. The 
    gray colored points satisfy all the experimental constraints (the neutrino oscillation data, perturbativity bound, $(g-2)_\mu$ and  electro-weak precision tests) except the cLFV bound, while the colored points satisfy all the above constraints including cLFV.}
    \label{fig:pureannihilation}
\end{figure}
In this section, we investigate the DM relic abundance arising solely from the self-annihilation processes of $N_1$ as illustrated in Appendix~\ref{app:FD} (Fig.~\ref{fig:feynSA1}).

To explore the complete parameter space consistent with the observed DM relic density through pure self-annihilation\footnote{Here, we switched off the co-annihilation and conversion driven processes such that the relic density will be generated via its self-annihilation only.}, we utilize Eq.~\ref{eq:yukawa}, which allows the generation of Yukawa couplings compatible with the neutrino oscillation data, while satisfying all other experimental constraints (electro-weak precision test, $(g-2)_\mu$, cLFV and perturbativity). Accordingly, we consider the viable value of $|\lambda_5|$ in the range $[10^{-12},4\pi]$, and randomly vary the orthogonal matrix $R$, then perform a relic-density scan using \texttt{micrOMEGAs}~\cite{Alguero:2023zol}.

The DM relic density in this framework is predominantly dictated by the annihilation channel $N_1 N_1 \rightarrow \ell_\alpha \ell_\beta$, mediated by the scalar $\eta$. The corresponding annihilation cross section exhibits a pronounced dependence on both the Yukawa couplings $y_{\alpha 1}$ and the mediator masses ($M_{\eta_{R,I}}~\text{and}~ M_{\eta^+}$). In particular, larger Yukawa couplings substantially amplify the annihilation rate, thereby leading to an efficient depletion of the DM abundance, whereas a heavier mediator suppresses the annihilation cross section and consequently enhances the relic density. Therefore, reproducing the observed DM abundance necessitates a delicate interplay between the strength of the Yukawa interactions and the mediator mass scale. 
As shown in Fig.~\ref{fig:pureannihilation}, the cLFV constraints severely restrict this parameter space. The gray colored points which satisfy all the experimental constraints (electro-weak precision test, $(g-2)_\mu$ and perturbativity) except cLFV can potentially give rise to the correct relic density in a certain parameter space. However, when we impose the cLFV constraint (colored points), we found that none of the points satisfy the correct relic density. In fact, these points correspond to over-abundant relic density. The reason is that, in our analysis, the Yukawa couplings ($y_{e1},y_{\mu 1}, y_{\tau 1}$), which are generated by the CI parametrization are of the similar order. As a result, when the cLFV demands the couplings to be small, the corresponding DM relic density becomes over-abundant. This problem can be overcome if we demand a hierarchy among the Yukawa couplings: $y_{e 1} \ll y_{\mu 1} \lesssim y_{\tau 1}$. In that case, the product of couplings $|y_{e1}.y^*_{\mu 1}|$ which enter into the cLFV calculation can be small while keeping $y_{\tau 1}$ larger, thus leading to correct relic density points via the annihilation through $y_{\tau 1}$ coupling. The details can be found in \cite{Vicente:2014wga,Liu:2022byu,Avila:2025qsc}.

As discussed above, we assume the Yukawa couplings ($y_{e1},y_{\mu 1}, y_{\tau 1}$) to be of the similar order. As a result, we could not satisfy the cLFV and the DM relic density simultaneously. However, by adding co-annihilation processes (see Sec.~\ref{sec:SA+CA}), we can obtain parameter space satisfying both the cLFV and DM relic density constraints without maintaining any hierarchy among the Yukawa couplings. In fact, throughout the rest of our analysis, we adhere to the similar order of Yukawa couplings.

\subsection{Thermal Relic density via self-annihilation and co-annihilation}\label{sec:SA+CA}
\begin{figure}[H]
    \centering
    \includegraphics[width=0.49\linewidth]
    {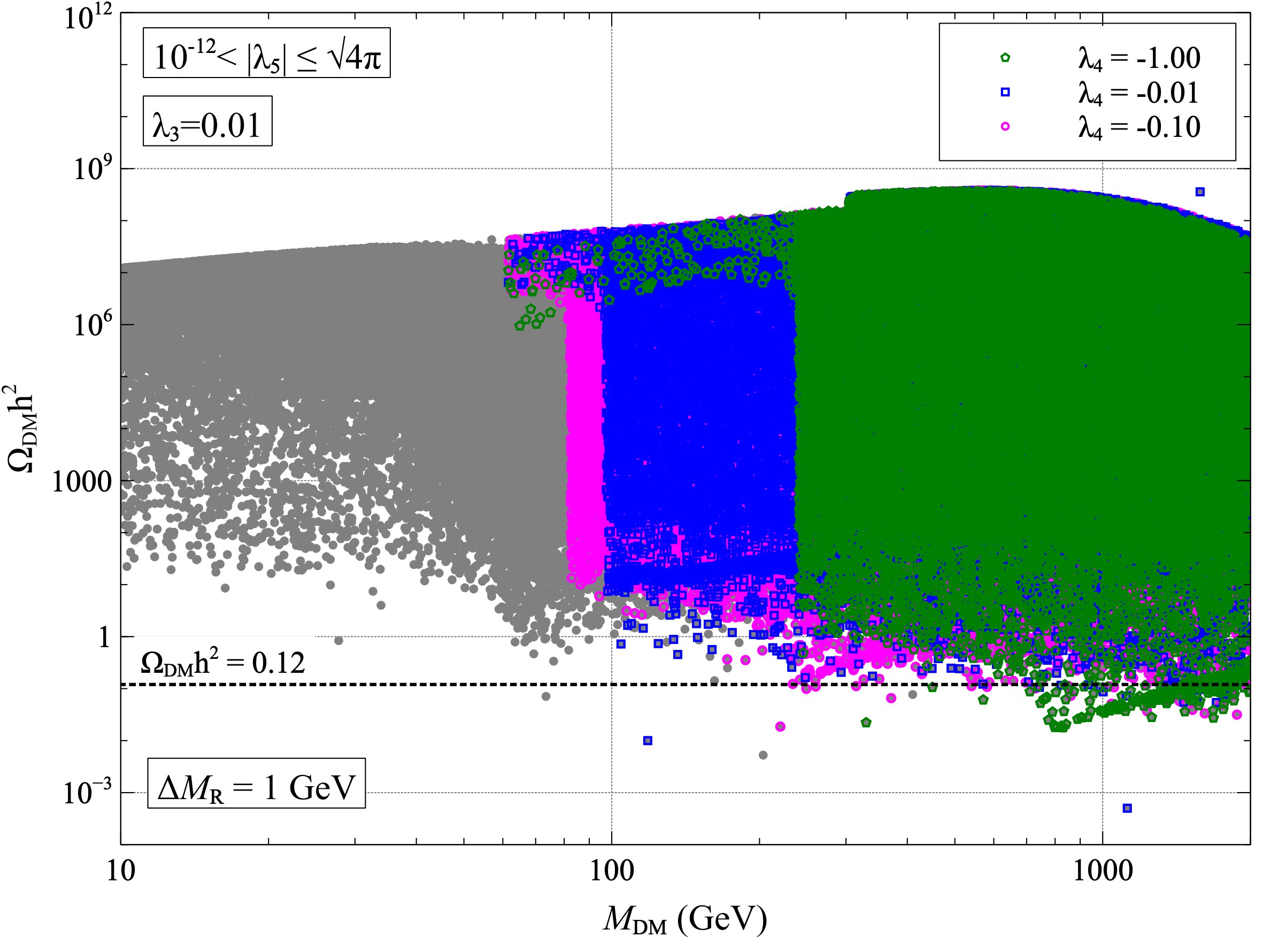}
    \includegraphics[width=0.49\linewidth]
    {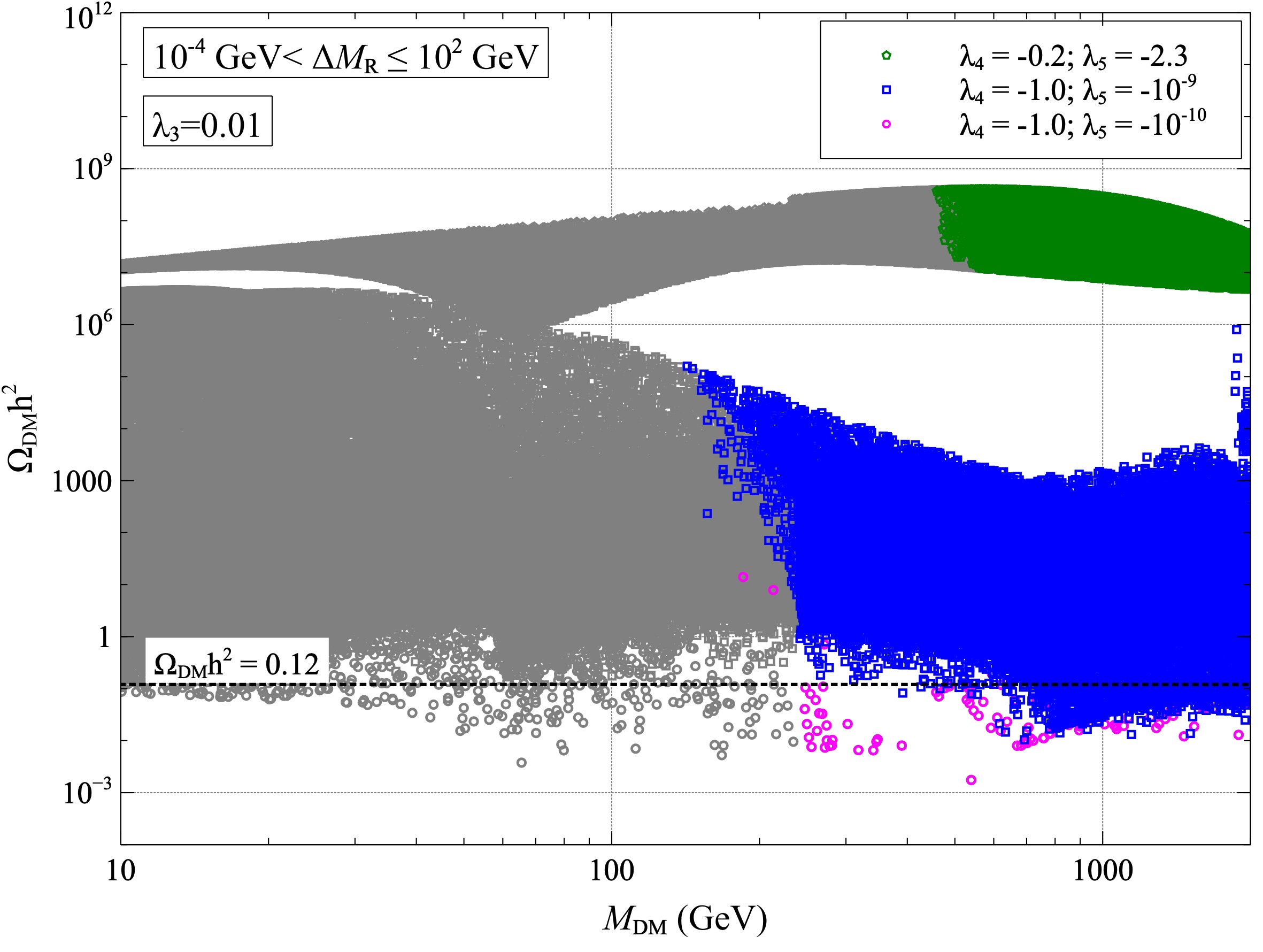}
    \caption{Relic density is shown as a function of the DM mass. The \textit{left} panel illustrates the variation with $|\lambda_5|$ for a fixed mass splitting of $\Delta M_{\rm R}=1~\rm GeV$, with three representative choices of $\lambda_4$ indicated by different colors. The \textit{right} panel shows the variation with $\Delta M_{\rm R}$ for three representative choices of $|\lambda_{4,5}|$, also distinguished by colors. All points satisfy the constraints from neutrino oscillations data, perturbativity, $(g-2)_\mu$ and cLFV bounds, while the gray colored points in both panels denote the parameter space excluded by Higgs and $Z$ invisible decay limits, the $S$ and $T$ electroweak precision parameters, and LEP constraints.}
    \label{fig:relic_CA01}
\end{figure}
In this section, we investigate the DM relic abundance arising from both the annihilation and co-annihilation processes. The corresponding Feynman diagrams are provided in Figs.~\ref{fig:feynSA1} and \ref{fig:feynCA1} in Appendix \ref{app:FD}. While solving the BEs in Eqs.~(\ref{eq:Y1}) and (\ref{eq:Y2}), we set the conversion-driven term $\Gamma_{2\to1}$ to zero. We utilize Eq.~(\ref{eq:yukawa}) to generate the Yukawa couplings by randomly varying the R matrix. We ensure that the generated couplings satisfy the experimental constraints from the neutrino oscillation data, $(g-2)_\mu$, cLFV and perturbativity bounds. For simplicity, we fixed $\lambda_3=10^{-2}$ and then perform a relic-density scan using \texttt{micrOMEGAs}. The resulting relic density as a function of the DM mass is shown in Fig.~\ref{fig:relic_CA01}.
In the left panel of Fig.~\ref{fig:relic_CA01}, we study the variation of relic density as a function of $|\lambda_5|$ ($\lambda_5<0$) in the range $[10^{-12},4\pi]$. To achieve this, we took three typical values of $\lambda_4$ such as $\lambda_4=-0.01 {(\rm blue)},-0.1 {(\rm magenta)},-1 {(\rm green)}$ for a fixed $\Delta M_{R}=1~\rm GeV$. 

We note that the interaction rate for DM annihilation processes scales as $\propto y_{\alpha 1}^4$, whereas the rates for co-annihilation processes involving the scalar doublet scale as $\propto y_{\alpha 1}^2 \lambda_x^2$ ($x\in\{R,I,3\}$). Here, $\lambda_R = \lambda_3 + \lambda_4 + \lambda_5$ governs co-annihilation with $\eta_R$, $\lambda_I = \lambda_3 + \lambda_4 - \lambda_5$ governs co-annihilation with $\eta_I$, and $\lambda_3$ controls co-annihilation with $\eta^+$.

As illustrated in Fig.~\ref{fig:relic_CA01} (\textit{left} panel), achieving the observed relic density is challenging in the low DM mass region ($M_{\rm DM}<250~\rm GeV$) for a fixed mass splitting of $\Delta M_R = 1~\mathrm{GeV}$. However in the large DM mass region ($M_{\rm DM} \gtrsim 250~{\rm GeV}$), it is possible to achieve the correct relic density via the efficient co-annihilation. This can be understood from the right panel of Fig.~\ref{fig:NMmassparams}, where the larger Yukawa couplings are allowed with increase in DM mass, resulting in larger co-annihilation rate for heavier DM masses. The efficient co-annihilation additionally requires the co-annihilating partner $\eta$ to remain in thermal equilibrium with the SM bath. As discussed earlier, this condition is governed by the couplings $\lambda_{R,I,3}$. It is evident from the \textit{left} panel of Fig.~\ref{fig:relic_CA01} that the green points, corresponding to $\lambda_4=-1$, yield the largest $|\lambda_R|$. Consequently, the observed relic density can be realized for comparatively larger DM masses. In contrast, the blue and magenta points, corresponding to smaller values of $|\lambda_4|$ and hence slightly smaller $|\lambda_R|$, extend the viable relic-density region towards lower DM masses. While sufficiently large $\lambda_{R,I,3}$ values are essential for co-annihilation to be effective, the mass splitting $\Delta M_R$ is equally important.

The \textit{right} panel of Fig.~\ref{fig:relic_CA01} shows the relic density as a function of the DM mass, with $\Delta M_R$ varied over the range $\{10^{-4}\text{ GeV},10^2\text{ GeV}\}$. The green points, corresponding to the largest values of $|\lambda_R|$, remain overabundant across the entire parameter space. This behavior is expected since the associated large $|\lambda_5|$ values suppress the allowed Yukawa couplings. In contrast, the blue points correspond to smaller $|\lambda_5|$, allowing larger Yukawa couplings and consequently more efficient annihilation. As a result, the viable parameter space yielding the observed relic density is significantly larger. Even for a further smaller $|\lambda_5|$ case (magenta points), the parameter space is further larger.

In both the left and right panel of Fig.~\ref{fig:relic_CA01}, the gray colored points are ruled out from Higgs and $Z$ invisible decay limits, the $S$ and $T$ electroweak precision parameters, and LEP constraints.

\subsection{Thermal Relic density via conversion-driven processes}\label{sec:CS+DID}

\begin{figure}[h]
    \centering
    \includegraphics[width=0.6\linewidth]{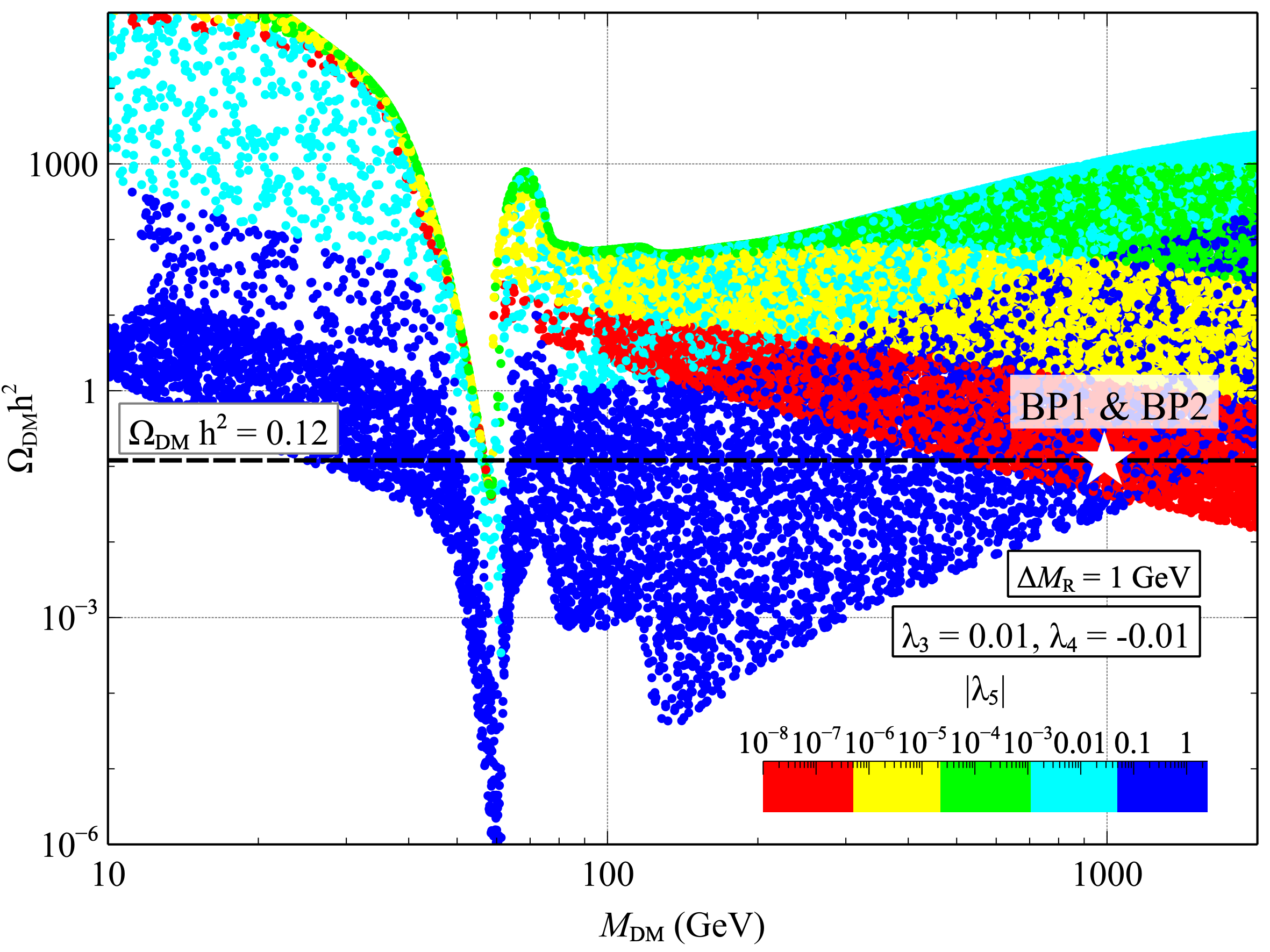}
    \caption{Relic density as a function of the DM mass for $\Delta M_{\rm R} = 1~\text{GeV}$, $\lambda_3 = 0.01$ and $\lambda_4=-0.01$. The colored scattered points represent different values of $|\lambda_5|$($\lambda_5$ chosen to be $<0$) as shown in the figure inset. These points are generated using a fixed $R$ matrix given in Eq.~(\ref{eq:Rmatrix}), while satisfying neutrino oscillation data, $(g-2)_\mu$, cLFV and LEP bounds. We have provided the evolution plot in Fig. \ref{fig:BP01} for BP1 and BP2, which are marked with a white colored $\star$.}
    \label{fig:relic_lambda}
\end{figure}

\begin{figure}
    \centering
    \includegraphics[width=0.6\linewidth]
    {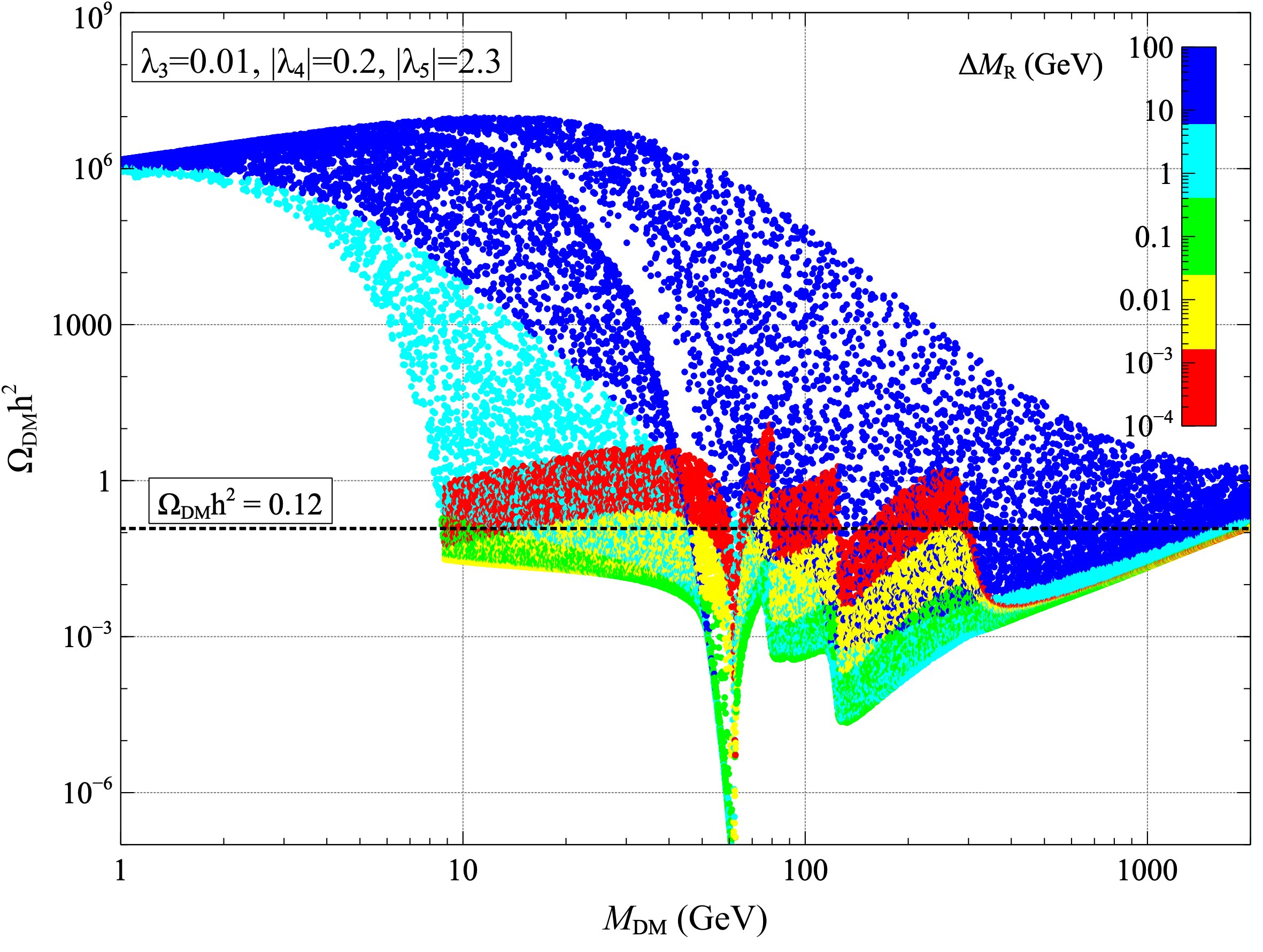}
    \caption{Variation of the relic density with $\Delta{M_R}$ in the ($\Omega_{\rm DM}h^2,M_{\rm DM}$) plane for the choice $\lambda_{3,4,5} = \{ 0.01, -0.2, -2.3\} $(see Fig.~\ref{fig:relic_CA01}), this time including conversion driven processes. As before, these points are generated using a fixed $R$ matrix given in Eq.~(\ref{eq:Rmatrix}) and all relevant constraints discussed in Fig.~\ref{fig:relic_lambda} are imposed.}
    \label{fig:relic_DeltaM}
\end{figure}

\begin{figure}[h]
    \centering
    \includegraphics[width=0.49\linewidth]{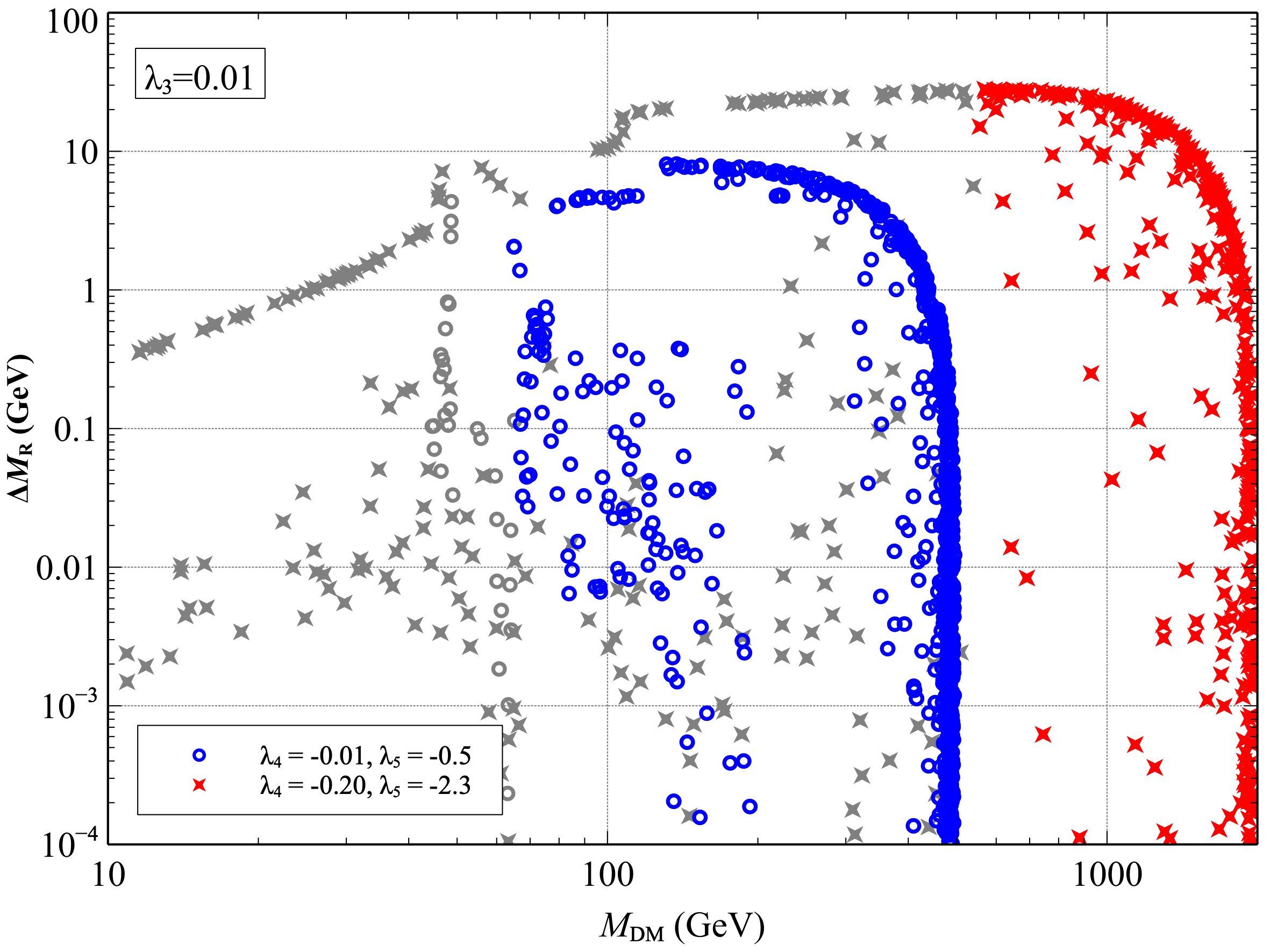}
    \includegraphics[width=0.49\linewidth]{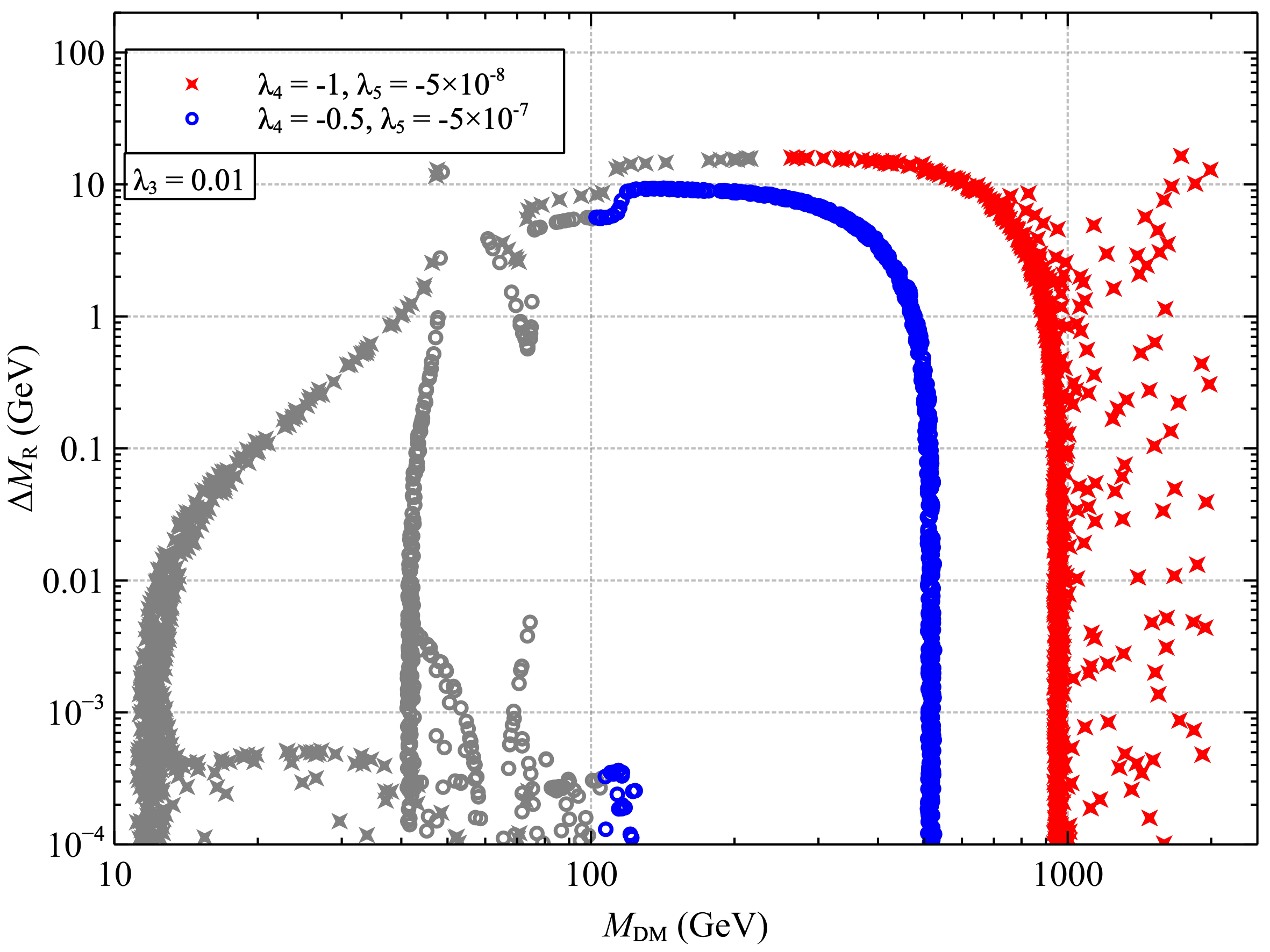}
    \caption{The correct relic density points are shown in the plane of $M_{\rm DM}$ and $\Delta M_{R}$, for three sets of $\lambda_{4,5}$ combinations given in the figure inset of each plot. The gray color points (in both \textit{left} and \textit{right} panel) are disallowed from the LEP, Higgs and Z-invisible decay, and electroweak corrections constraints. The \textit{left} panel shows the correct relic density parameter space corresponding to large $\lambda_5$ values, while the \textit{right} plot corresponds to small $|\lambda_5|$ values.}
    \label{fig:summary01}
\end{figure}

In this section, we study the effect of conversion-driven processes on RHN DM relic density, which is given by the $\Gamma_{2\to1}$ term in Eq.~(\ref{eq:gamma21}). We solve the BEs (\ref{eq:Y1}) and (\ref{eq:Y2}), without switching off any term and perform a relic-density scan using \texttt{micrOMEGAs}. We generate the Yukawa couplings using Eq.~(\ref{eq:yukawa}) to reproduce the neutrino oscillation data while satisfying other experimental bounds including $(g-2)_\mu$, LEP, cLFV, and perturbativity by fixing the R matrix as given in Eq.~(\ref{eq:Rmatrix}). We note that, for demonstration purpose, we used the fixed R matrix in Figs.~\ref{fig:relic_lambda} and \ref{fig:relic_DeltaM}. However, to delineate the final parameter space, we vary the R matrix randomly in Fig.~\ref{fig:summary01} so that a larger parameter space can be obtained while satisfying additional constraints from Higgs and $Z-$ invisible decay and electroweak precision test. We keep $\lambda_3=0.01$ throughout this analysis.

From Fig.~\ref{fig:relic_lambda}, we observe that the correct DM relic abundance can be achieved over a significantly wider range of DM masses when conversion-driven processes are included (with $\Gamma_{2\to1}$ kept non-zero), in contrast to the results shown in Fig.~\ref{fig:relic_CA01}. In particular, for the choice $\lambda_3=0.01$, $\lambda_4=-0.01$ and $\Delta M_{R}=1$ GeV, represented by the blue points in the left panel of Fig.~\ref{fig:relic_CA01}, the relic density is satisfied in a limited parameter space when conversion-driven effects are excluded.
As evident from Eq.~(\ref{eq:gamma21}), the quantity $\Gamma_{2\to1}$ receives contributions from both co-scattering processes, denoted as \texttt{"2010"} (see Fig.~\ref{fig:feyncoscattering} in Appendix~\ref{app:FD}), and decay/inverse decay processes of the form $\eta_i \leftrightarrow N_1 + {\rm SM}$ (see Fig.~\ref{fig:feyndecay} in Appendix~\ref{app:FD}), where $\eta_i \in {\eta_R, \eta_I, \eta^+}$. The interaction rate for co-scattering processes scales as $n_{\rm SM}^{\rm eq} \times y_{\alpha 1}^2$, while that for decay and inverse decay processes scales as $y_{\alpha 1}^2$.
As shown previously in Fig.~\ref{fig:NMmassparams}, the Yukawa couplings become larger for smaller values of $|\lambda_5|$. Consequently, small $|\lambda_5|$ enhances the interaction rates of both co-scattering and decay/inverse decay processes, whereas for large $|\lambda_5|$ the corresponding rates are significantly reduced. It is important to emphasize that the conversion-driven processes do not directly deplete the total dark sector abundance, but rather convert the relic abundance between the two dark sectors (i.e., sector-1 and sector-2) with the help of SM particles. We further note that, in addition to the respective annihilation channels \texttt{"1100"} and \texttt{"2200"}, co-annihilation processes denoted by \texttt{"1200"} play a crucial role in maintaining thermal equilibrium between sector-1, sector-2, and the SM thermal bath. This is because the \texttt{"1200"} processes scale as $\propto y_{\alpha 1}^2$ and are therefore especially effective in regions of parameter space with sizable Yukawa couplings. The \texttt{"2200"} processes involve gauge interactions and quartic scalar interactions, in addition to Yukawa interactions, enabling sector-2 particles to remain in equilibrium with the SM bath for a longer period. As long as the sector-2 is in thermal equilibrium, the conversion-driven processes can efficiently deplete the DM relic density via \texttt{"1020"} followed by \texttt{"2200"} processes, leading to the correct relic density over a large region of parameter space, as evidenced by Fig.~\ref{fig:relic_lambda}.

\begin{figure}[h]
    \centering
    \includegraphics[width=0.49\linewidth]{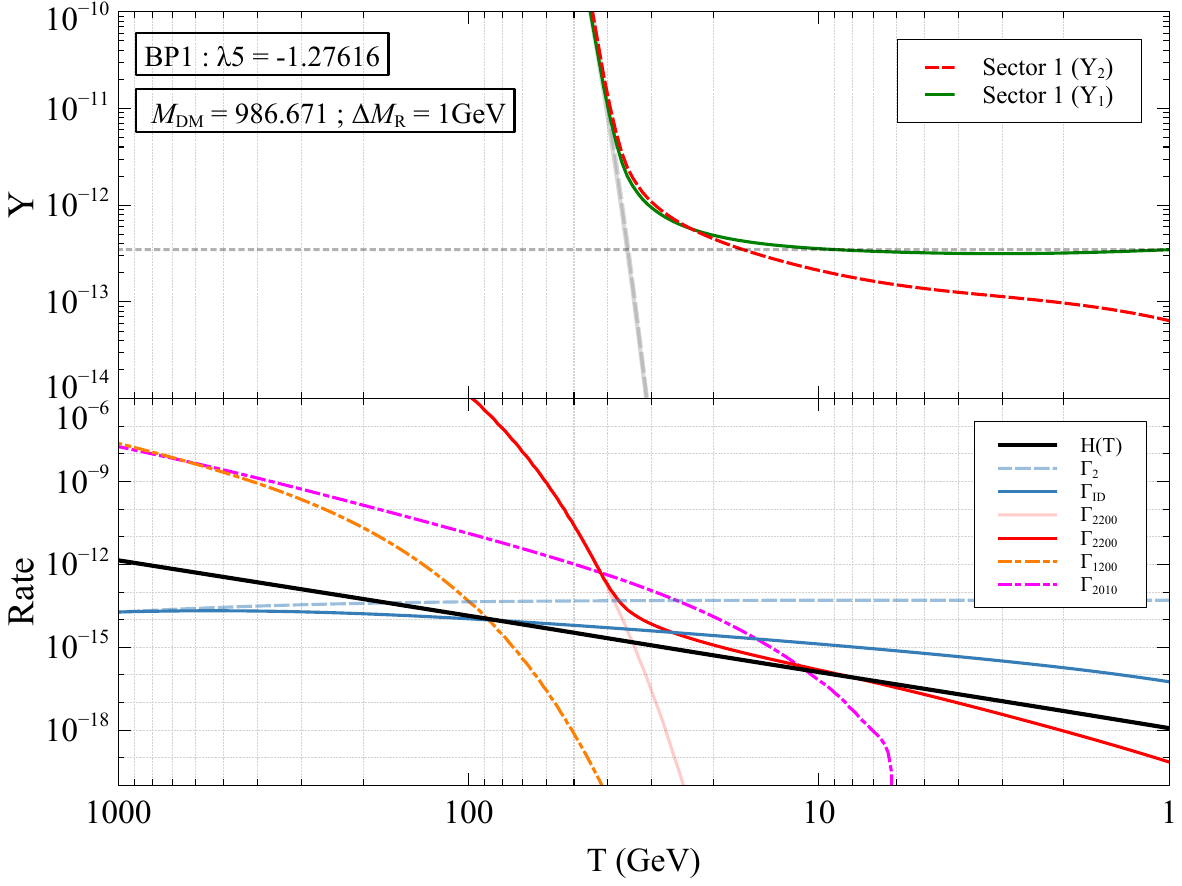}
    \includegraphics[width=0.49\linewidth]{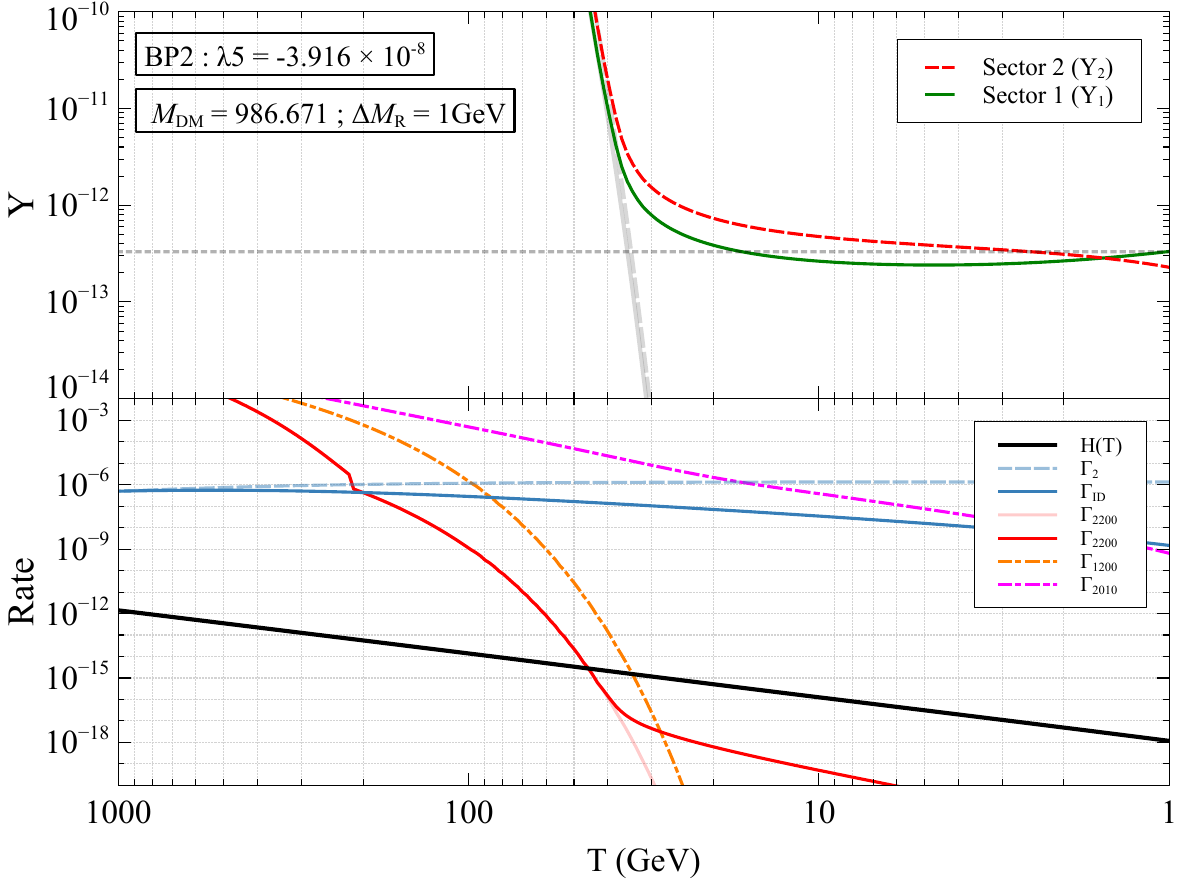}
    \caption{Evolution of sector-1 and sector-2 particles are shown in the upper panel of the plot and their corresponding interaction rates are given in the lower panel of the plot for BP1 (\textit{left}) and BP2 (\textit{right}).}
    \label{fig:BP01}
\end{figure}

For further clarification, we present the evolution plots in Fig.~\ref{fig:BP01} for two benchmark points: BP1 ($M_{\rm DM}=986.671~\mathrm{GeV}$, $\Delta M_{R}=1~\mathrm{GeV}$, $\lambda_5=-1.27$) and BP2 ($M_{\rm DM}=986.671~\mathrm{GeV}$, $\Delta M_{R}=1~\mathrm{GeV}$, $\lambda_5=-3.91\times10^{-8}$), as listed in Table~\ref{tab:BP}. Both benchmark points reproduce the observed DM relic density, although the processes involved are different. The \textit{left} panel of Fig.~\ref{fig:BP01} shows the evolution of the dark sector abundances ($Y_1$ and $Y_2$), along with the corresponding interaction rates for BP1, while the \textit{right} panel displays the same quantities for BP2. For BP1, the relatively large value of $|\lambda_5|$ ensures that the \texttt{"2200"} processes remain active for a longer epoch, thereby maintaining thermal equilibrium between the sector-2 particles and the SM thermal bath. With an appropriate choice of $\lambda_5$ (and hence the Yukawa couplings) and the mass splitting $\Delta M_R$, the rate of conversion-driven processes can be tuned to be just sufficient to deplete the DM relic abundance to bring it to the correct ball park, until the \texttt{"2200"} processes eventually decouple. On the other hand, for BP2 we choose a small value of $|\lambda_5|$, which corresponds to comparatively large Yukawa couplings. In this case, the \texttt{"2200"} processes maintain thermal equilibrium primarily through gauge interactions, but decouple at an earlier temperature compared to BP1. In contrast, the enhanced Yukawa couplings significantly increase the rates of the \texttt{"1200"} processes as well as the conversion-driven interactions. The \texttt{"1200"} processes enable both sector-1 and sector-2 particles to remain in thermal equilibrium for a relatively longer duration than that sustained by the \texttt{"2200"} processes alone. During this epoch, the conversion-driven processes efficiently deplete the DM relic abundance, thereby bringing it into the correct ball park.

As discussed at the beginning of Section~\ref{sec:Relic_Density}, and owing to the chosen signs of $\lambda_4$ and $\lambda_5$, $\eta_R$ emerges as the next-to-lightest stable particle. Consequently, the conversion-driven processes are also governed by the parameter $\Delta M_{\rm R}$ as well. In particular, the co-scattering rate for the process \texttt{"2010"} and the inverse decay rate (see Eq.~(\ref{eq:inverse_rate})) are exponentially suppressed by the factor $e^{-\frac{\Delta M_{\rm R}}{T}}$. Accordingly, in Fig.~\ref{fig:relic_DeltaM}, we present the variation of the relic density with $\Delta M_{\rm R}$ for all DM masses in the range [1–2000]~GeV. Interestingly, two distinct behaviors emerge depending on the value of $\Delta M_{\rm R}$. This can be understood as follows. For the given choice of $\lambda_4=-0.2$ and $\lambda_5=-2.3$, the Yukawa couplings are extremely small. In this case, we can neglect the rate of co-scattering process in comparison to that of decay/inverse decay process (see Appendix~\ref{app:coscattering}). As presented in Eq.~(\ref{eq:inverse_rate}), the inverse decay rate is proportional to $\Delta M_{\rm R}^2 \times e^{-\Delta M_{\rm R}/T}$. As a result, when $\Delta M_{\rm R}$ decreases from 100~GeV to  1~GeV, the relic-density pattern shifts downward, as expected. This behavior arises because a smaller $\Delta M_{\rm R}$ enhances the inverse decay rate, leading to a more efficient depletion of the DM relic abundance, until sector-2 particles decouple. However, upon further reducing $\Delta M_{\rm R}$ from 1~GeV to $10^{-4}$ GeV, the decay/inverse decay rate is suppressed due to a smaller $\Delta M_{\rm R}$. As a result, the conversion from sector-1 to sector-2 particles are inefficient, thus leading to a relatively larger DM relic density. This behavior can be easily read from Fig.~\ref{fig:relic_DeltaM}

\begin{figure*}
    \centering
    \includegraphics[width=0.9\linewidth]
    {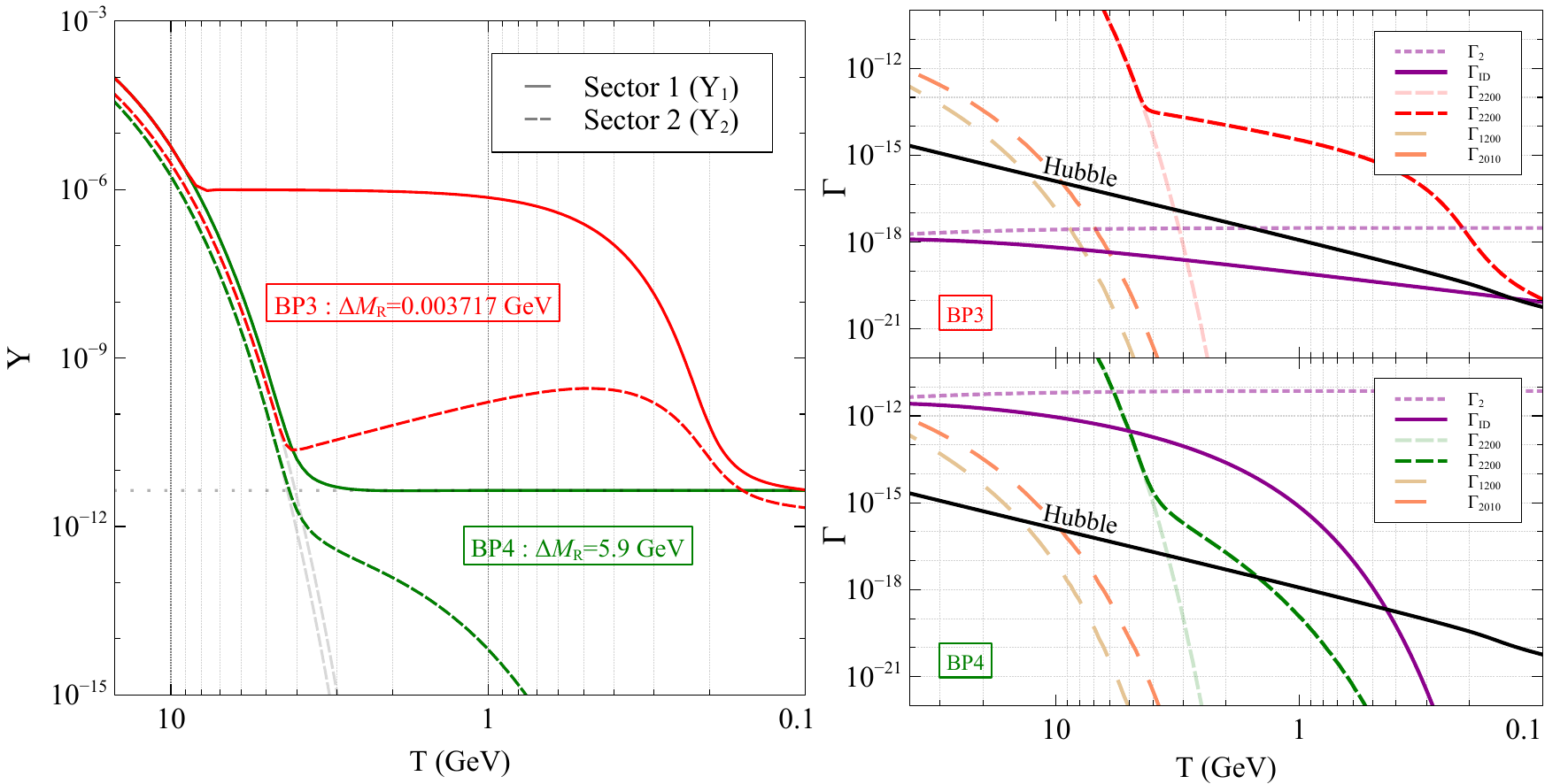}
    \caption{\textit{Left} panel: Evolution of the sector~1 and sector~2 particle abundances for two representative values of $\Delta M_R$.
    \textit{Right} panel: Comparison of the relevant interaction rates with the Hubble expansion rate, including the co-annihilation and co-scattering rate between sector 1 and 2, the self-annihilation rate of the sector~2 particle incorporating decay and inverse decay effects, and the corresponding hypothetical scenario in which decay is switched off.}
    \label{fig:rate_evolution}
\end{figure*}
To gain further insight into the relic density evolution in these two scenarios, we present the abundance evolution and the corresponding interaction rates for two benchmark points, BP3 and BP4 (see Table~\ref{tab:BP}), as shown in Fig.~\ref{fig:rate_evolution}. These two benchmark points, which differ only in $\Delta M_{\rm R}$ for the same DM mass ($M_{\rm DM}=100~{\rm GeV}$), both yield the correct DM relic density. For relatively large $\Delta M_R=5.9~{\rm GeV}$ (represented by green color contours in the \textit{left} panel of Fig.~\ref{fig:rate_evolution}), the sector-1 particle decouples early with a relatively larger abundance. Since, the decay width is sizable, ensuring that conversion-driven processes remain efficient thereby depleting the DM relic abundance. Conversely, as $\Delta M_R$ decreases to $\Delta M_R=0.003717~{\rm GeV}$ (represented by red color contours in the \textit{left} panel of Fig.~\ref{fig:rate_evolution}), the decay width becomes highly suppressed. Although conversion processes remain active, the inverse decay rate—being proportional to $\Delta M_R$—is no longer fast enough to maintain equilibration of $N_1$, especially when the $\eta$ particle is still in thermal equilibrium. However, the inverse-decay processes populate the $\eta$ abundance which also increases the rate of \texttt{"2200"} processes and further reduce the DM abundance. This situation persists until the inverse decay rate becomes comparable to the Hubble expansion rate while $\eta$ is in equilibrium, thereby bringing the final DM relic abundance to the correct ballpark. The above discussion can be conferred by comparing the various rates with the Hubble expansion rate as shown in the \textit{right} panel of the Fig.~\ref{fig:rate_evolution}.

\begin{table}[h]
\begin{center}
	\begin{tabular}{|@{\hspace{0.3cm}}c@{\hspace{0.3cm}}|@{\hspace{0cm}}c@{\hspace{0cm}}|@{\hspace{0cm}}c@{\hspace{0cm}}|@{\hspace{0.2cm}}c@{\hspace{0.2cm}}|@{\hspace{0.2cm}}c@{\hspace{0.2cm}}|}
		\hline 
        \hline
        {} & $M_{\rm DM}$ (GeV) & $\Delta M_{\rm R}$  (GeV) & $\lambda_4$ & $\lambda_5$\\
		\hline
        \hline
		\textbf{BP1} & $986.671$ & 1 & -0.01 & -1.27\\
        \hline
        \textbf{BP2} & $986.671$ & $1$ & -0.01 & -$3.8\times10^{-8}$\\
        \hline
		\textbf{BP3} & 100 & 0.003717 & -0.2 & -0.5\\
        \hline
		\textbf{BP4} & 100 & 5.9 & -0.2 & -0.5\\
		\hline
        \hline
    \end{tabular}
	\caption{Benchmark points with $\lambda_3=0.01$.}
	\label{tab:BP}
\end{center}
\end{table}

Having demonstrated that conversion-driven processes can successfully reproduce the observed DM relic density over a wide region of parameter space, we emphasize that Figs.~\ref{fig:relic_lambda} and \ref{fig:relic_DeltaM} were generated using the fixed $R$ matrix given in Eq.~\ref{eq:Rmatrix}. These figures illustrate the dependence of the relic density on the viable parameters $\lambda_5$ and $\Delta M_R$, respectively, through their impact on the Yukawa couplings. As discussed in the previous subsections on self-annihilation (\ref{sec:SA}) and co-annihilation (\ref{sec:SA+CA}), allowing the $R$ matrix to vary significantly enlarges the viable parameter space, leading to a large range of parameters consistent with the observed relic density.

We allow R to vary randomly and present the final parameter space yielding the correct dark matter relic density in the $\Delta M_{\rm R}$–$M_{\rm DM}$ plane for two representative sets of $\lambda_4$ and $\lambda_5$, as shown in Fig.~\ref{fig:summary01}. Throughout this analysis, we fix $\lambda_3 = 10^{-2}$ as before. The \textit{left} panel illustrates the scenario with large $|\lambda_5|$ (equivalently, small Yukawa couplings), whereas the \textit{right} panel corresponds to small $|\lambda_5|$ (large Yukawa couplings).
Focusing on the \textit{left} panel, the two colored scattered points represent distinct choices of the $\{\lambda_4,\lambda_5\}$ parameter sets, as indicated in the figure inset.

In the \textit{left} panel, the red star-shaped points correspond to $|\lambda_R|=0.43$, represent the parameter space that simultaneously reproduces the observed relic density while satisfying all the experimental constraints discussed so far. For this choice of parameters, viable DM masses lie in the range $550$--$2000$ GeV. In contrast, reducing $|\lambda_R|$ to $0.051$ (blue circle points) shifts the allowed DM mass range to $62.5$--$500$ GeV, thereby covering a substantial portion of the parameter space not accessible to the red-star benchmark. The same analysis is repeated for smaller values of $|\lambda_5|$, as shown in the \textit{right} panel. In this case, $|\lambda_R|$ is primarily determined by $\lambda_4$, while the Yukawa couplings continue to play an important role in identifying the viable regions of parameter space. It is instructive to compare the parameter space covered by the different benchmark choices.

In both panels, larger values of $|\lambda_R|$ favor heavier DM masses, whereas smaller values of $|\lambda_R|$ shift the viable region toward lighter DM masses. Interestingly, the statistically significant points are concentrated along contour-like structures in the $\Delta M_R$--$M_{\rm DM}$ plane. Such contours can, in principle, be reproduced even by fixing the $R$ matrix to a single representative form throughout the scan. As before, the blue circle points and the red star points satisfy all the relevant experimental constraints. The gray points are excluded by the Higgs and $Z$ invisible decay limit, LEP constraint and electro-weak precision test.

We note that in Fig.~\ref{fig:summary01}, we have used a fixed value of $\lambda_{4,5}$ values to obtain different patches for the correct DM relic density in the plane of $\Delta M_{\rm R}$ vs $M_{\rm DM}$. However, if we vary the $|\lambda_4|$ and $|\lambda_5|$ within the permissible limits, then the correct thermal DM relic can be achieved for all $M_{\rm DM}\gtrsim M_h/2$ while satisfying all the phenomenological constraints discussed throughout the paper.

\subsection{Relic Density of DM via Freeze-in Mechanism}

In Eq.~(\ref{eq:numass}), we have presented the neutrino mass matrix including contributions from three RHNs, leading to three light neutrino mass eigenstates. However, neutrino oscillation data require at least two non-zero light neutrino masses, which can be satisfied with a minimum of two RHNs. Accordingly, assuming the lightest neutrino mass to be vanishingly small, we consider only the contributions from $N_2$ and $N_3$ to the neutrino mass matrix. With this choice, $N_1$ does not participate in neutrino mass generation, as illustrated in Fig.~\ref{fig:neutrnomass}. Equivalently, the Yukawa coupling $y_{\alpha 1}$ associated with $N_1$ is taken to be extremely small, typically $\lesssim\mathcal{O}(10^{-7})$, thus ensuring that $N_1$ never reaches thermal equilibrium. In this scenario, the relic density of the DM ($N_1$) can be generated via the freeze-in mechanism \cite{Molinaro:2014lfa,McDonald:2001vt,McDonald:2008ua,Hall:2009bx}.

\begin{figure}[h]
    \centering
     \includegraphics[width=0.6\linewidth]{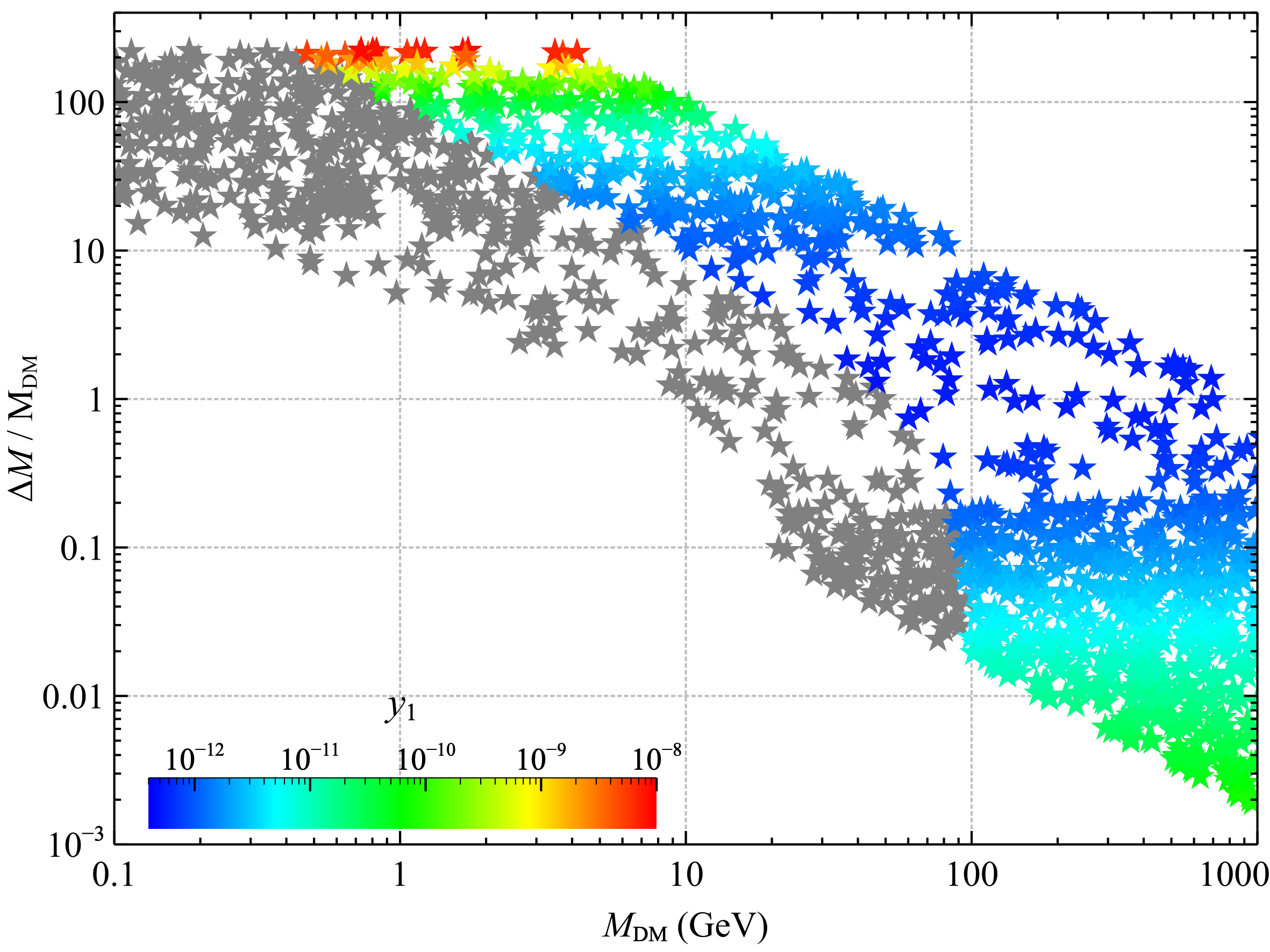}
    \caption{Allowed parameter space yielding the correct DM relic density through the freeze-in mechanism. All displayed points meet the BBN constraint, corresponding to an $\eta$ decay lifetime $<1$ sec. We also apply LEP limits (gray points), given that the $\eta$ masses are taken to be identical with $\Delta M=\Delta M_{\rm R}=\Delta M_{\rm I}=\Delta M^\pm$.}
    \label{fig:FI}
\end{figure}

In our setup, the non-thermal production of $N_1$ can be realized from the decay of $Z_2$ odd scalars ($\eta_R,\eta_I,\eta^\pm$) and $N_{2,3}$. Moreover, the annihilation of SM leptons and $Z_2$ odd scalars can also produce $N_1$ relic. However these processes are heavily suppressed as the relevant cross-section is proportional to the fourth power of the Yukawa coupling $y_{\alpha 1}$. The production of $N_1$ from $N_{2,3}$ proceeds via three body decay and hence this production channel is suppressed as well. Hence, the freeze-in production for DM is dominated by the decay of the scalars. The equation governing the evolution of the DM from the scalar decay is given by
\begin{equation}\label{eq:FI}
    \frac{dY_{N_1}}{dx}=\frac{1}{x \mathcal{H}(T)}\sum_{i=R,I,\pm}\langle\Gamma_{\eta_i}\rangle Y^{\rm eq}_{\eta_i},
\end{equation}
where, $x=M_{N_1}/T$ and $Y^{\rm eq}_{\eta_i}$ is the equilibrium abundance of $\eta_i$ and the average decay width, $\langle \Gamma_{\eta} \rangle=\Gamma_{\eta}(K_1(x)/K_2(x))$. The above equation assumes that the DM production happens while the $\eta_i$ is in thermal equilibrium.

In Fig.~\ref{fig:FI}, we present the parameter space yielding the correct relic abundance in the $\Delta M/M_{\rm DM}$ vs $M_{\rm DM}$ plane, obtained by solving Eq.~(\ref{eq:FI}). The color bar indicates the value of the Yukawa coupling $y_1$, where, for simplicity, we assume Yukawa couplings, $ y_{e1} = y_{\mu1} = y_{\tau1}=y_1 $. In addition, we take $M_{\eta_R} = M_{\eta_I} = M_{\eta^\pm}$, which is a good approximation in the limit of small quartic couplings.

For the equilibrium abundance $Y_{\eta}^{\rm eq}$, we recall that in a freeze-in scenario the characteristic production temperature is $T_{\rm FI} \simeq M_{\rm DM}$. Consequently, the equilibrium abundance of the $\eta$ particle can be estimated as $Y_{\eta}^{\rm eq} \propto e^{-\frac{\Delta M}{M_{\rm DM}}}$. From the Fig. \ref{fig:FI}, it is evident that the Boltzmann suppression is significant in the region of $\Delta M/M_{\rm DM}>1$, while it becomes milder when the same ratio is less than 1. As follows from Eq.~(\ref{eq:FI}), for a fixed $Y_{\eta}^{\rm eq}$, the DM production rate is proportional to the decay width $\Gamma_\eta$. Therefore, in the large $\Delta M/M_{\rm DM}$ regime, the DM yield is suppressed due to the smaller $Y_{\eta}^{\rm eq}$, requiring a larger Yukawa coupling $y_1$ to reproduce the observed relic abundance. Conversely, in the small $\Delta M/M_{\rm DM}$ region, the Boltzmann suppression is relaxed and hence the final DM yield is decided by the production rate, $\Gamma_\eta$. As $\Delta M$ decreases, the decay width reduces, thereby requiring a larger Yukawa coupling to efficiently produce the correct DM yield.

\section{Direct Detection}\label{sec:DD}
\begin{figure}[h]
    \centering
    \includegraphics[width=0.45\linewidth]{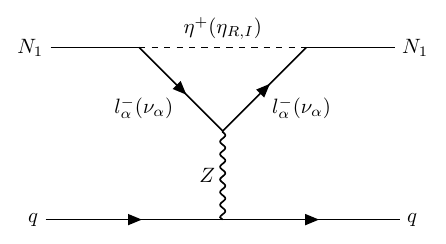}
    \includegraphics[width=0.45\linewidth]{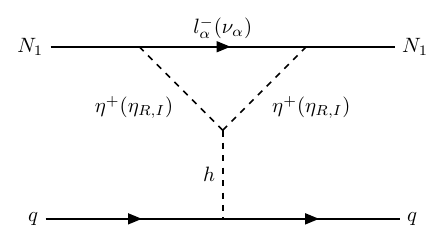}
    \caption{Feynman diagram representing the spin-dependent (left figure) and the spin-independent (right figure) process relevant for direct detection.}
    \label{fig:dd_diagram}
\end{figure}

\begin{figure}[h]
    \centering
    \includegraphics[width=0.49\linewidth]{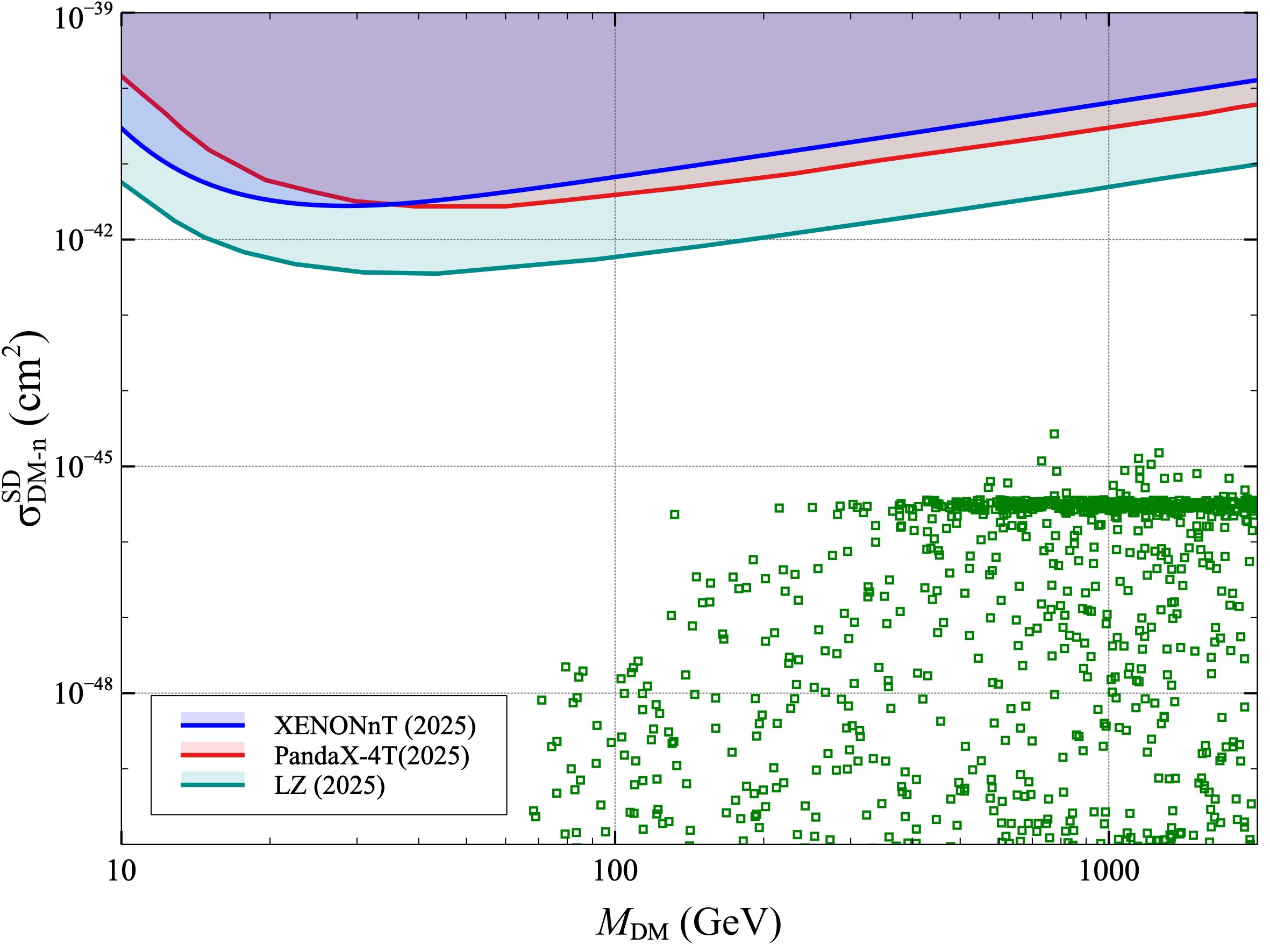}
    \includegraphics[width=0.49\linewidth]{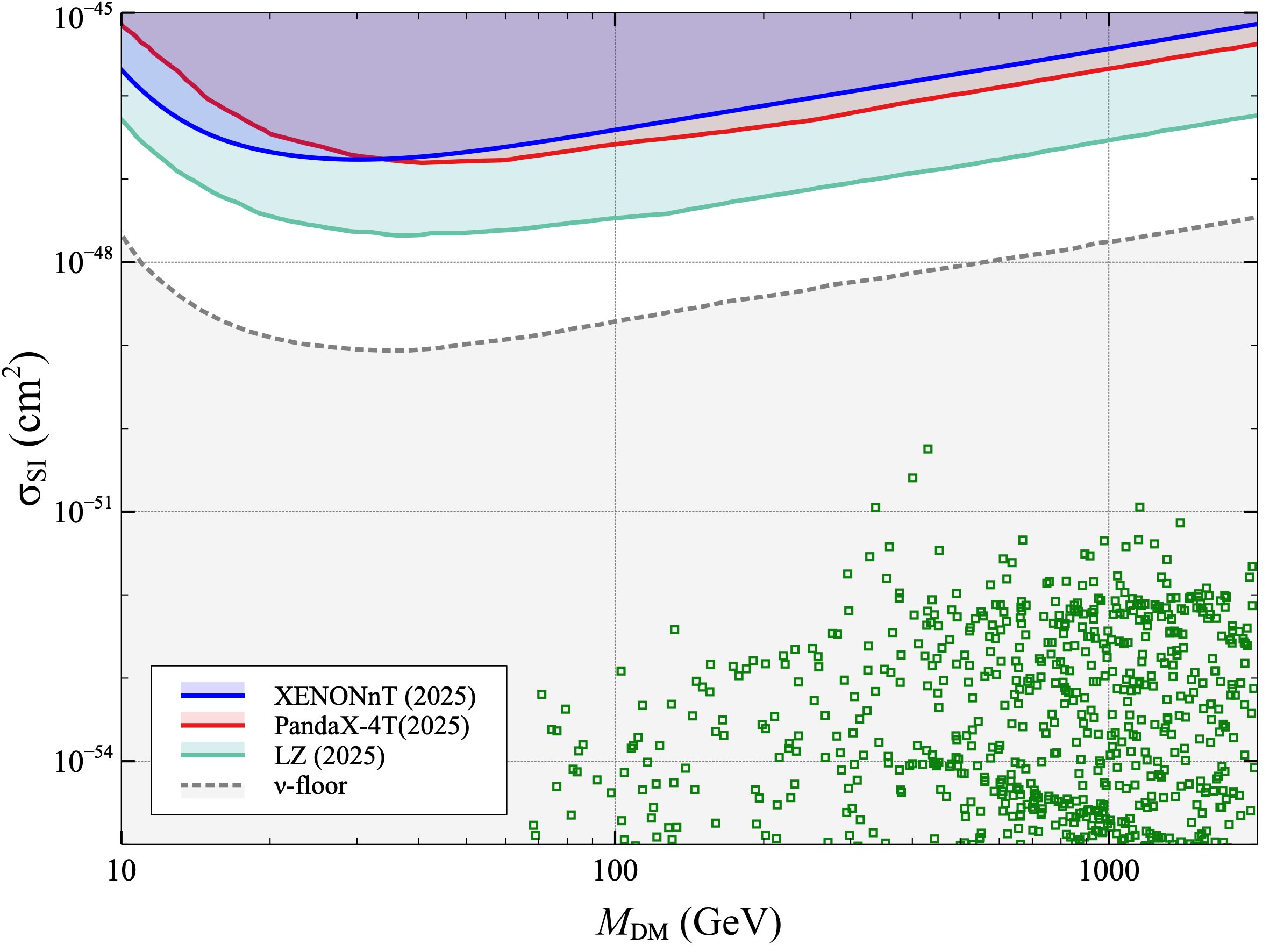}
    \caption{\textit{Left} Panel: Spin-dependent DM--neutron scattering cross section as a function of the DM mass for the benchmark parameter sets $\{\lambda_{4,5}\}$=$\{ [10^{-3},4\pi],[10^{-10},4\pi]\}$,  $\Delta M_{\rm R}$ in the range $[10^{-4} {~\rm GeV},10^{2} {~\rm GeV}]$ and $\lambda_3=0.01$. Here we computed the direct detection cross sections (green) only for the points satisfying the correct relic density and compared with the current exclusion limits from XENONnT \cite{XENON:2025vwd}, LZ \cite{LZ:2024zvo}, and PandaX-4T \cite{PandaX:2024qfu}. \textit{Right} Panel: Spin-independent DM--nucleon scattering cross section for the same benchmark parameter sets, shown together with the latest experimental limits from current direct detection experiments.}
    \label{fig:dd_bound}
\end{figure}

Since the dark matter particle $N_1$ couples to the SM only through the single Yukawa interaction $y_{\alpha 1} \overline{\ell}\tilde{\eta}N_1$, there is no tree-level process for direct detection. As a result, direct detection arises only at the one-loop level, with diagrams mediated by either gauge bosons or the Higgs boson $h$. In particular, the $Z$-boson–mediated diagram shown in Fig.~$\ref{fig:dd_diagram}$ induces an effective axial-vector interaction of the form $\xi_q \overline{N_1}\gamma^\mu \gamma^5 N_1 \overline{q}\gamma_\mu \gamma^5 q$, where $\xi_q$ is given in Ref.~\cite{Ibarra:2016dlb}. 
\begin{equation}
    \xi_q=\frac{y_{1}^2a_q}{32 \pi^2 M_Z^2} \left[ (v_l+a_l)\mathcal{G}_2\left(\frac{M_{N1}}{M_{\eta}^+}\right) +(v_\nu+a_\nu) \left ( \mathcal{G}_2\left(\frac{M_{N1}}{M_{\eta_R}}\right) +\mathcal{G}_2\left(\frac{M_{N1}}{M_{\eta_I}}\right) \right) \right]
\end{equation}
where $y_1^2=\sum_{\alpha=e,\mu,\tau}y_{\alpha 1}^2$, $a_l=\frac{-g}{2 c_W}\frac{1}{2}$, $v_l=\frac{-g}{2 c_W}\left( \frac{1}{2}-2 s_W^2 \right)$, $v_\nu=a_\nu=\frac{g}{2c_W}\frac{1}{2}$, $a_q=\frac{1}{2}-\frac{4}{3}s_{W}^2\left( -\frac{1}{2} +\frac{2}{3}s_{W}^2 \right)$ \cite{Babu:2025czb} for $q=u,c,t(d,s,b)$, and the loop function $\mathcal{G}_2(x)$ is given by
\begin{equation}
    \mathcal{G}_2(x)=-1+\frac{2(x+(1-x){\rm ln}(1-x))}{x^2}.
\end{equation}
Since the effective coupling $\xi_q$ depends solely on the Yukawa coupling $y_{\alpha 1}$ and is independent of the quartic couplings $\lambda_{3,4}$, the above process becomes relevant only for very small $\lambda_5$. The resulting spin-dependent cross section per nucleon $N$ is then given by \cite{Jungman:1995df}.
\begin{equation}
    \sigma^{SD}_{{\rm DM-n}}=\frac{16}{\pi}\frac{M_{N_1}m_N^2}{(M_{N_1}+m_N)^2}J_N(J_N+1)\xi_N^2,
\end{equation}
where $\xi_N=\sum_{q=u,d,s} \Delta^N_q \xi_q$ with $\Delta^N_u=0.842,\Delta^N_d=-0.427$ and $\Delta^N_s=0.085$ \cite{HERMES:2006jyl}. Fig.~$\ref{fig:dd_bound}$ displays the current experimental limits on the spin-dependent dark matter–neutron scattering cross section. Since we vary the orthogonal matrix R randomly, our scan covers the whole range of Yukawa couplings from $6\times10^{-9}$ to $0.31$ as shown in Fig.~\ref{fig:NMmassparams}. Even with the largest possible Yukawa couplings, the direct detection prediction of the present model lies well below the current sensitivity of the experiments shown.

It is important to note that for sufficiently large values of the quartic couplings $\lambda_{3,4,5}$, the spin-independent direct detection interaction corresponding to the diagram shown in the right panel of Fig.~$\ref{fig:dd_diagram}$ becomes relevant. The resulting effective operator is of the form $\Lambda_q \overline{N_1}N_1 \overline{q}q$, where $\Lambda_q$ is given by
\begin{equation}
    \Lambda_q=-\frac{y_1^2}{16 \pi^2 M_h^2 M_{N_1}} \left[ \lambda_3 \mathcal{G}_1 \left(\frac{M_{N_1}^2}{M_{\eta^+}^2} \right)+\frac{\lambda_R}{2} \mathcal{G}_1 \left(\frac{M_{N_R}^2}{M_{\eta_R}^2} \right)+\frac{\lambda_I}{2} \mathcal{G}_1 \left(\frac{M_{N_1}^2}{M_{\eta_I}^2} \right) \right] m_q .
\end{equation}
Here, the loop function $\mathcal{G}_1(x)$ is defined as
\begin{equation}
    \mathcal{G}_1(x)=\frac{x+(1-x)\ln(1-x)}{x}.
\end{equation}
The spin-independent scattering cross section of $N_1$ off a proton is then given by
\begin{equation}
    \sigma_{SI}=\frac{4}{\pi}\frac{M_{N_1}m_p^2}{(M_{N_1}+m_p)^2}m_p^2 \left( \frac{\Lambda_q}{m_q} \right)^2 f_p^2 ,
\end{equation}
where $m_p$ denotes the proton mass and $f_p \approx 0.3$ \cite{Hoferichter:2017olk} is the scalar form factor. This contribution is subject to constraints from current and future dark matter direct detection experiments.

However, larger values of $\lambda_5$ are disfavored since they correspond to smaller Yukawa couplings, leading to a strong suppression of the spin-independent direct detection cross section. Enhancing the direct detection signal therefore requires larger Yukawa couplings, which in turn implies that $\lambda_5$ must be very small. While the remaining quartic couplings $\lambda_{3,4}$ can in principle be increased, large values of $\lambda_4$ are disfavored due to constraints from displaced vertex signatures associated with $\eta^\pm$ decays (see Section~\ref{sec:DV}). Consequently, the only viable parameter for enhancing the signal is the quartic coupling $\lambda_3$. 
%\textcolor{blue}{needs to be updated}
Even for sizable $\lambda_3$ and small $\lambda_5$, the resulting cross section remains below the current experimental sensitivity and lies below the neutrino floor, as illustrated in the right panel of Fig.~\ref{fig:dd_bound}.

\section{Displaced Vertex Signature}\label{sec:DV}
When a sufficiently long lived particle produced via collision of SM particles at colliders travels some distance and then decays at a point away from the point of collision, may leave displaced vertex signature. In such cases, the presence of the charged leptons or jets as the decay final state particles can be detected and reconstructed by dedicated analysis \cite{CMS:2024trg,ATLAS:2022gbw,MATHUSLA:2019qpy}. This is a clean signature of long-lived particle which is different form SM particles.
\begin{figure}[h]
    \centering
     \includegraphics[width=0.48\linewidth]{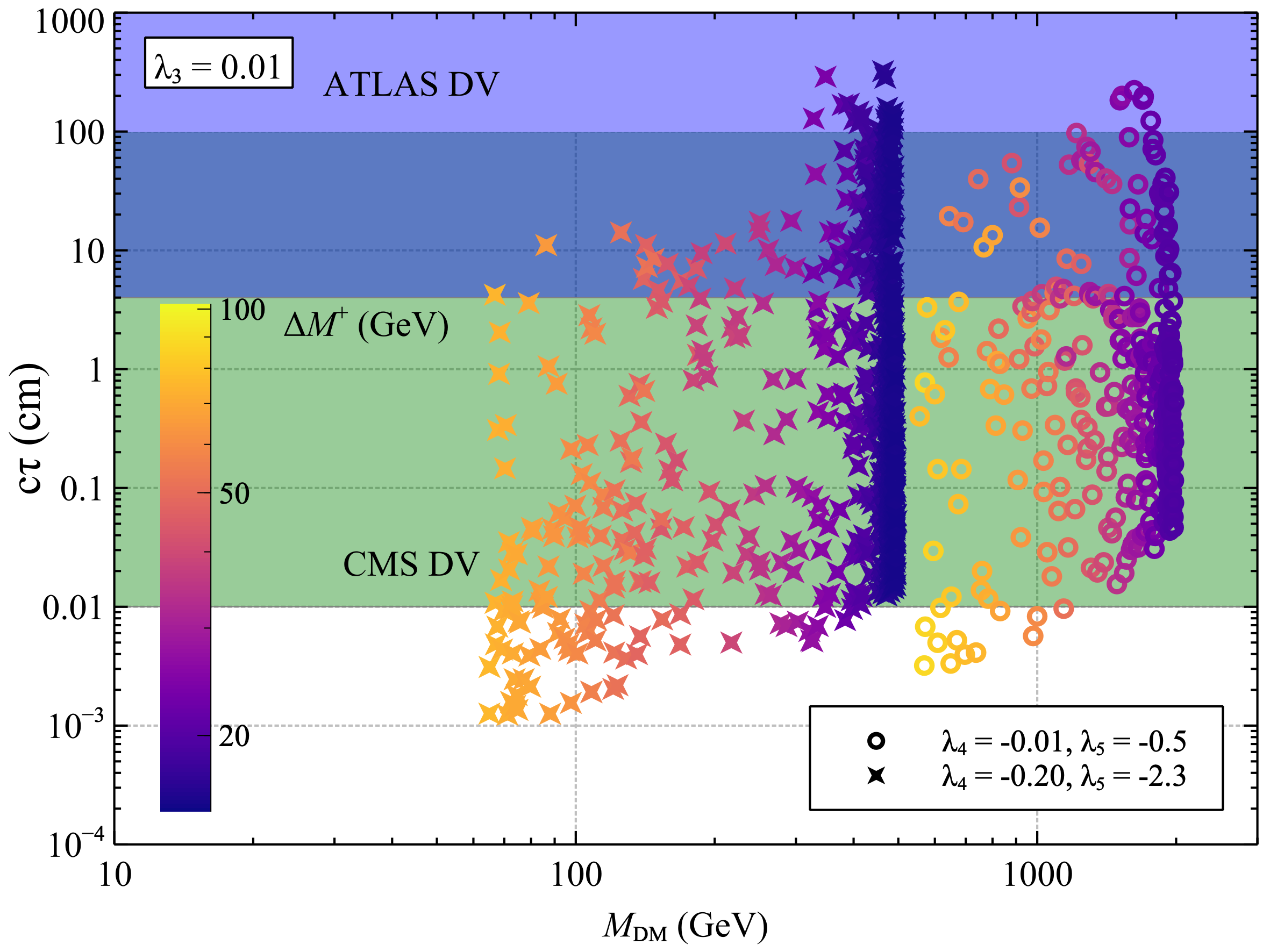}
     \includegraphics[width=0.48\linewidth]{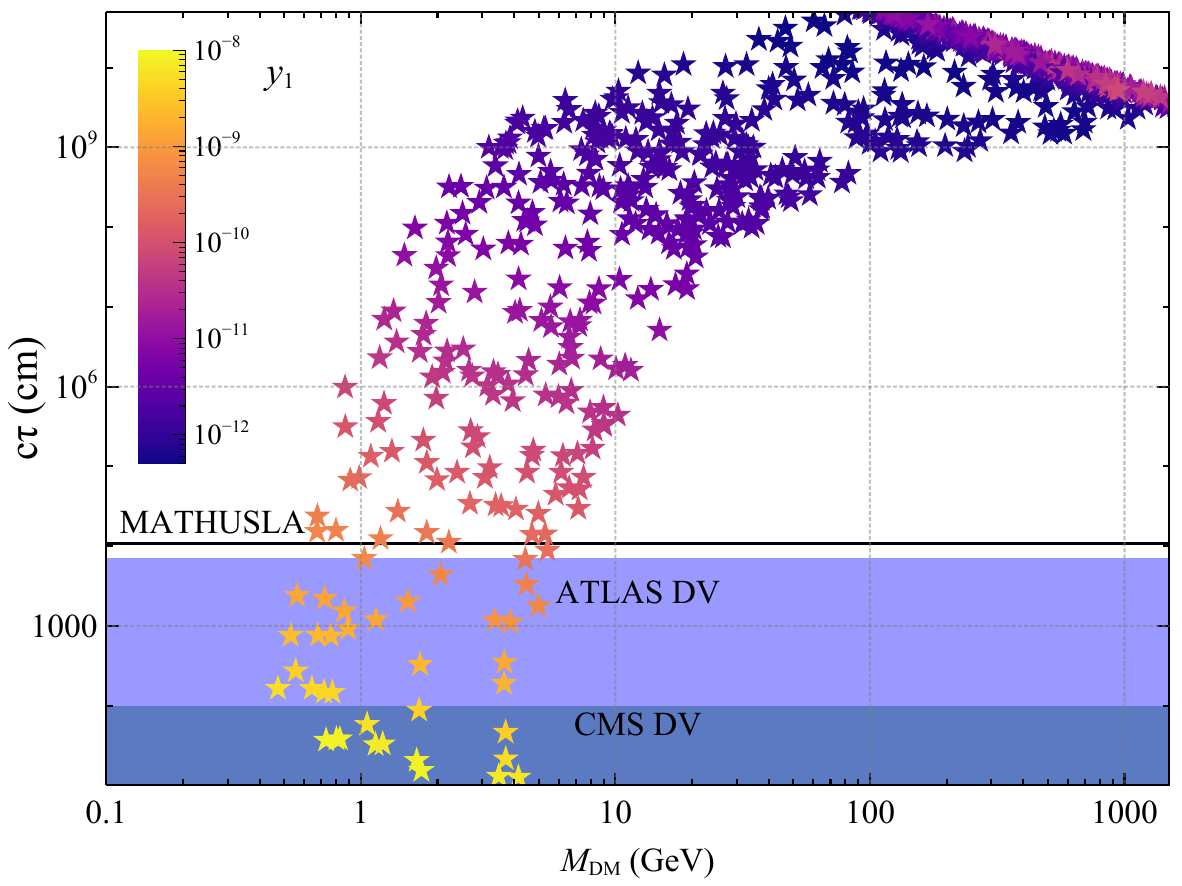}
    \caption{The decay length, $c\tau$ is shown as a function of $M_{\rm DM}$. In the \textit{left} panel, $c\tau$ is shown for the thermal DM relic satisfying points, where $\Delta {M^{+}}$ is shown in the color band. Three sets of $\lambda_{4,5}$ are provided in the figure inset, those are shown as $\circledcirc$, $\star$ and $\triangle$ shape. In the \textit{right} panel, we show the $c\tau$ for the DM relic points produced via freeze-in mechanism. The sensitivities of CMS (0.01 cm - 100 cm) and ATLAS (4 cm - 72.4 m) are shown in colored shaded region and that for MATHUSLA (107 m - 108 m) is shown in solid line.}
    \label{fig:DV}
\end{figure}
In our setup, the charged component of scalar doublet, $\eta^+$ can be produced via Higgs portal as well as through gauge interactions at colliders, which, subsequently, can decay to $N_1$ and charged lepton (with mass $M_{\alpha}$) with a decay width given in Eq. (\ref{eq:decay_width_eta}).
In the limit of $M_\alpha\ll M_{N_1}$, and replacing $M_{\eta^+}$ with $(M_{N_1}+\Delta M^{+})$, the decay width reduces to,

\begin{equation}\label{eq:DVdecay}
   \Gamma_{\eta^+\rightarrow N_{1}l_{\alpha}^{+}}\approx \frac{y_{\alpha1}^2}{8\pi}\frac{\Delta {M^{+}}^2\left(2M_{\rm DM}+\Delta M^{+}\right)^2}{\left(M_{\rm DM}+\Delta M^{+}\right)^3}.
\end{equation}

An enhanced decay length requires a suppressed decay width. From Eq.~(\ref{eq:DVdecay}), achieving such suppression necessitates small Yukawa coupling (this can be realized with a large $|\lambda_5|$), as well as a small $\Delta {M^{+}}$ and a large $M_{\rm DM}$. We show the $c\tau$ (in units of cm) for all the points satisfy correct relic density via freeze-out mechanism in \textit{left} panel of Fig.~\ref{fig:DV}, where the color scale denotes the mass splitting $\Delta {M^{+}}$. In this case, the Yukawa coupling can not be arbitrarily small, leading to a small displaced vertex length. On the other hand, if the relic of the DM is produced via freeze-in mechanism, then the Yukawa couplings can be small, which can give rise large displaced vertex length. This can be easily seen from the \textit{right} panel of the Fig. \ref{fig:DV}, where the color scale represents the Yukawa coupling, $y$.

\section{Conclusions}\label{sec:conclusion}
The scotogenic model offers a minimal and unified framework in which neutrino masses are generated radiatively through interactions with dark-sector particles, while the same discrete symmetry that forbids tree-level neutrino masses simultaneously guarantees the stability of a viable DM candidate. Its strong multi-messenger testability makes this framework particularly appealing, and it has been widely explored to identify regions of parameter space consistent with both neutrino mass generation and the observed DM relic abundance while satisfying electroweak parameters, neutrino oscillation data, cLFV, LEP and muon anomalous magnetic moment.

We have chosen the quartic couplings $\lambda_4$ and $\lambda_5$ to be negative, which ensures that $M_{\eta_R}$ remains the NLSP in the dark sector. The neutrino mass requirement implies that there is a see-saw relation between $|\lambda_5|$ and Yukawa couplings. Smaller $|\lambda_5|$ implies larger Yukawa couplings, however such large Yukawa couplings are severely constrained by the cLFV bound. Therefore, pure self-annihilation of DM is not a viable solution to get the correct DM relic unless we maintain a hierarchy among the Yukawa couplings ($y_{e 1} \ll y_{\mu 1} \lesssim y_{\tau 1}$). Nevertheless, we find that the observed relic abundance can be consistently reproduced across the full dark matter mass range once co-annihilation and conversion-driven processes are incorporated in the Boltzmann equations.

The conversion-driven mechanism is realized through co-scattering as well as decay and inverse decay processes. While co-scattering contributes to the depletion of the dark matter number density, its impact is largely subdominant compared to that arising from decay and inverse decay. Efficient depletion via conversion-driven processes is ensured as long as the $\eta_R$ particle remains in thermal equilibrium with the SM bath. We further observe that the correct relic density can be obtained for two distinct values of $\Delta M_{R}$, a feature that can be naturally understood from the behavior of the inverse decay rate, which scales with both $\Gamma_{\eta_R}$ and the Boltzmann suppression factor $e^{-\Delta M_{\rm R}/T}$. In addition, for a sufficiently large dark matter mass, the correct relic density can be achieved over a broad range of $\Delta M_{ R}$, with the viable mass being strongly correlated with the choice of $\lambda_{3,4,5}$ or, equivalently $\lambda_R$. Although a thermal relic is viable over a wide dark matter mass range, we identify regions—particularly at low dark matter masses—where thermal equilibrium with the SM is not attained. In such cases, the observed relic abundance can instead be generated via the freeze-in mechanism.

Finally, despite a substantial portion of the parameter space being excluded by constraints from the $S$ and $T$ parameters, cLFV, LEP searches, $(g-2)_\mu$, and Higgs invisible decay limits, we identify sizable regions that remain accessible at colliders, most notably through displaced vertex signatures. Such signals are enhanced for large $\lambda_5$, large $M_{\rm DM}$, and small $\lambda_4$. While these regions can be effectively probed at collider experiments, we find that current direct detection searches do not impose additional constraints, as the relevant interactions arise only at the loop level.

%%%%%%%%%%%%%%%%%%%%%%%%%%%%%%%%%%%%%%%%%%%%%%%%%%%%%%%%%%%%%%%%%%%%%%%%%%%%%%%%%%%%%%%%%%
\section*{Acknowledgments}
We are grateful to Bhupal Dev for his useful suggestion on DM direct detection. We acknowledge helpful conversations with colleagues during the International Workshop - WHEPP (2025).

%%%%%%%%%%%%%%%%%%%%%%%%%%%%%%%%%%%%%%%%%

%%%%%%%%%%%%%%%%%%%%%%%%%%%%%%%%%%%%%%%%%

%
\appendix

\section{Functions in oblique parameters}\label{app:oblique}
The oblique parameters, $S$ and $T$ are expressed as \cite{Chen:2015gaa,Batra:2022pej,Batra:2022arl},
\begin{equation}
    S=\frac{1}{24 \pi}\left[(s_w^2-c_w^2)G(z_{\eta^+},z_{\eta^+})+G(z_{\eta_R},z_{\eta_I})+\Tilde{G}(z_h)+\log\left(\frac{M_h^2 M^2_{\eta_R} M^2_{\eta_I}}{M_{\eta^+}^6}\right)-\Tilde{G}(z_{h_{\rm ref}})-\log\left(\frac{M_{h_{\rm ref}}^2}{M_{\eta^+}^2}\right)\right],
\end{equation}
\begin{eqnarray}
\alpha T&=&\frac{1}{16 \pi^2 v^2}\left[F(M_{\eta^+}^2,M_{\eta_R}^2)+F(M_{\eta^+}^2,M_{\eta_I}^2)-F(M_{\eta_R}^2,M_{\eta_I}^2)+3\left(F(M_Z^2,M_h^2)-F(M_Z^2,M_h^2)\right)\right.\nonumber\\
    &&~~~~~~~~~~~~~~~~\left.-3\left(F(M_Z^2,M^2_{h_{\rm ref}})-F(M_Z^2,M^2_{h_{\rm ref}})\right)\right],
\end{eqnarray}
where $z_a=M_a^2/M^2_Z$, $M_{h_{\rm ref}}=M_{h}$, and the loop functions are given by,
    \begin{eqnarray}
        G(x,y)=&{}&\frac{-16}{3}+5(x+y)-2(x-y)^2+3\left[\frac{x^2+y^2}{x-y}-x^2+y^2+\frac{(x-y)^3}{3}\right]\log\left(\frac{x}{y}\right)\nonumber\\
        &{}&~~~~~~~~~~~~~+\left[1-2(x+y)+(x-y)^2\right]f(x+y-1,1-2(x+y)+(x-y)^2)\\
        \Tilde{G}(x)=&{}&\frac{-79}{3}+9x-2x^2+\left(-10+18x-6x^2+x^3-9\frac{x+1}{x-1}\right)\log x+(12-4x+x^2)f(x,x^2-4x)
    \end{eqnarray}
    \begin{equation}
        f(z,w)=\begin{cases}
    \sqrt{w}\ln\left|\frac{z-\sqrt{w}}{z+\sqrt{w}}\right| &\Leftarrow  w> 0\\
    0 &\Leftarrow  w=0\\
    2\sqrt{-w}\arctan\left(\frac{\sqrt{-w}}{2}\right) &\Leftarrow w<0
\end{cases}
    \end{equation}
    \begin{equation}
        F(x,y)=\frac{x+y}{2}-\frac{xy}{x-y}\log\left(\frac{x}{y}\right)
    \end{equation}

\section{Higgs Invisible Decay}\label{app:HID}

Owing to the scalar couplings $\lambda_{3,4,5}$, the SM Higgs boson $h$ can decay into the neutral components $\eta_R$ and $\eta_I$ of the inert doublet. In contrast, the decay $h \to \eta^{+}\eta^{-}$ is kinematically forbidden because values $M_{\eta^+} < 100~\mathrm{GeV}$ are strongly excluded by LEP limits. Since only the neutral channels remain viable, we define the effective couplings governing the two decay modes as
\begin{equation*}
    \lambda_R = \lambda_3 + \lambda_4 + \lambda_5, \qquad
\lambda_I = \lambda_3 + \lambda_4 - \lambda_5 .
\end{equation*}
The corresponding partial decay widths are
\begin{equation}
    \Gamma_{h\to \eta_i \eta_i} =
\frac{\lambda_i^2}{16\pi M_h}
\sqrt{1 - \frac{4 M_{\eta_i}^2}{M_h^2}} \qquad (i = R, I).
\end{equation}

Within the Standard Model, the total Higgs width is $4.1~\mathrm{MeV}$ \cite{LHCHiggsCrossSectionWorkingGroup:2016ypw}, and any additional contribution from new physics appears in the invisible decay branching ratio. The observed upper limits on the Higgs invisible branching fraction, as reported by the CMS \cite{CMS:2022qva} and ATLAS \cite{ATLAS:2022yvh} collaborations, are $18\%$ and $14.5\%$ at the $95\%$ C.L., respectively. These bounds are expected to improve to approximately $10\%$ for both experiments. Before identifying the allowed parameter space consistent with Higgs invisible decay bounds, we clarify our sign conventions for the relevant couplings.

We take $\lambda_5 < 0$, which ensures that $M_{\eta_I} > M_{\eta_R}$ at all times. Additionally, choosing $\lambda_4 < 0$ guarantees the mass ordering $M_{\eta^+} > M_{\eta_R}$. This mass hierarchy automatically satisfies the LEP constraint on charged scalars and avoids contributions to the invisible decay width of the $Z$ boson, provided that $M_{\eta_R} + M_{\eta_I} > M_Z$. This choice also allows $M_{\eta_R}$ to remain relatively small while eliminating the decay $h \to \eta_I \eta_I$ when kinematically inaccessible.

\begin{figure}[H]
    \centering
    \includegraphics[width=0.6\linewidth]{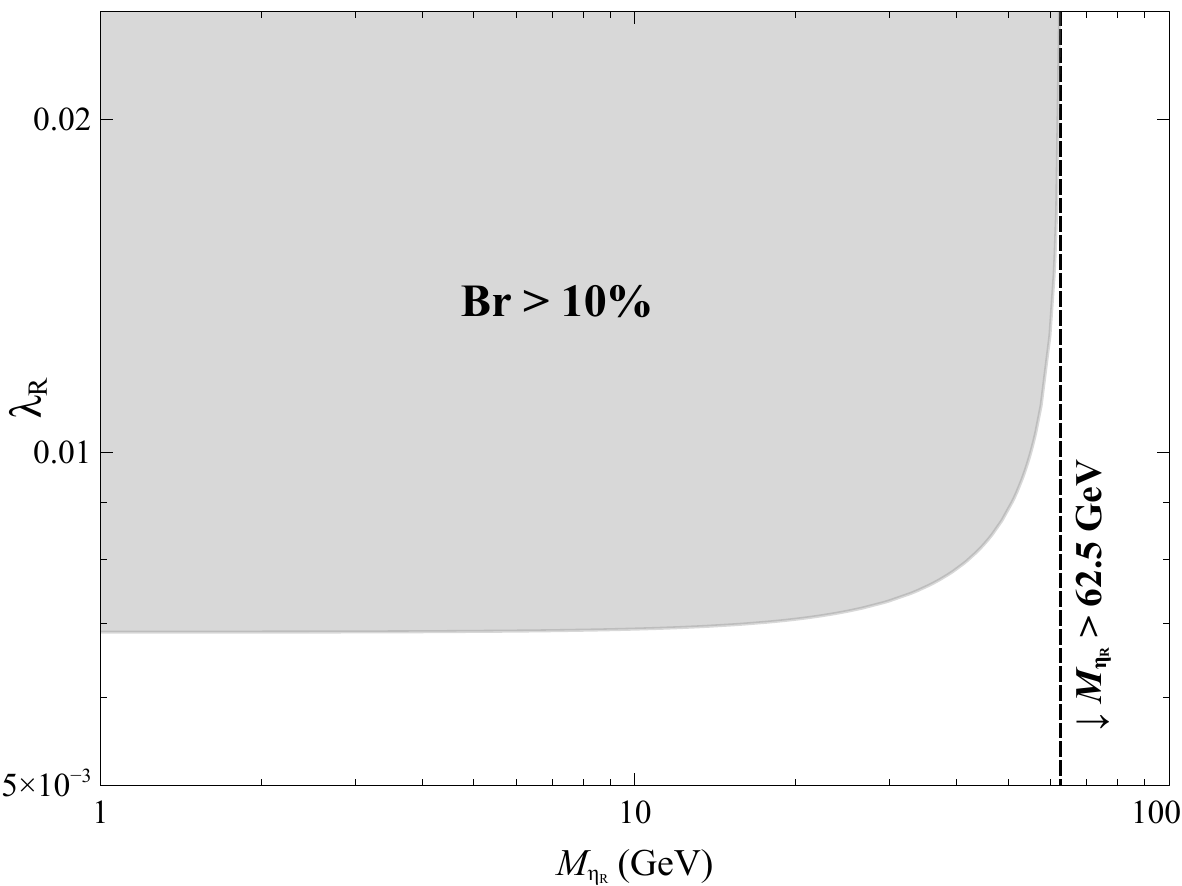}
    \caption{Higgs Invisible Decay}
    \label{fig:HID}
\end{figure}

With these considerations, Fig. \ref{fig:HID} illustrates the constraints from Higgs invisible decay in the $M_{\eta_R}-\lambda_R$ parameter plane.

\section{Parameterization of $R$ matrix}
The $R$ matrix given in Eq.~(\ref{eq:yukawa}) can be parameterized as
\begin{equation}\label{eq:Rparameterization}
    R=\begin{pmatrix}
        1 & 0& 0\\
        0 &\cos{\alpha} & \sin{\alpha}\\
        0 & -\sin{\alpha}& \cos{\alpha}
    \end{pmatrix}
    \begin{pmatrix}
        \cos{\beta}& 0 & \sin{\beta}\\
        0 & 1 & 0\\
        -\sin{\beta} & 0 & \cos{\beta}
    \end{pmatrix}
    \begin{pmatrix}
        \cos{\gamma}&\sin{\gamma}&0\\
        -\sin{\gamma}&\cos{\gamma}&0\\
        0 & 0 & 1
    \end{pmatrix},
\end{equation}
where the $\alpha,\beta$ and $\gamma$ are the mixing angles.

\section{Feynman Diagrams}\label{app:FD}
\begin{figure}[H]
    \centering
    \includegraphics[width=0.26\linewidth]{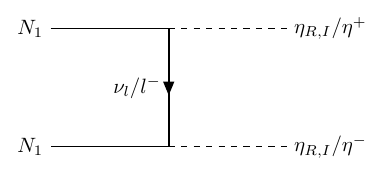}
    \includegraphics[width=0.25\linewidth]{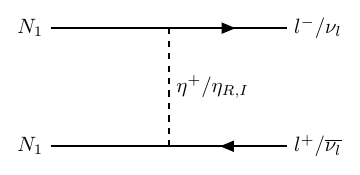}
    \includegraphics[width=0.25\linewidth]{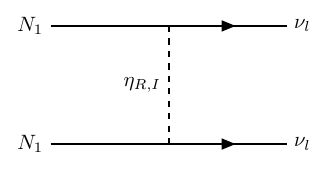}
    \caption{Feynman diagrams corresponding to \texttt{"1100"} processes.}
    \label{fig:feynSA1}
\end{figure}

\begin{figure}[H]
    \centering
    \includegraphics[width=0.25\linewidth]{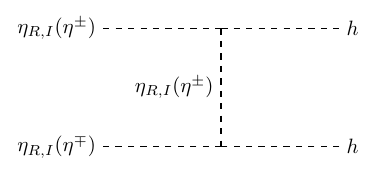}

    \includegraphics[width=0.25\linewidth]{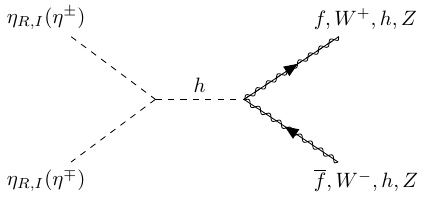}
    \includegraphics[width=0.15\linewidth]{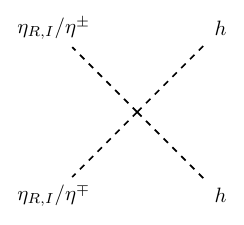}
    
    \caption{Feynman diagrams corresponding to \texttt{"2200"} processes.}
    \label{fig:feynSA2}
\end{figure}

\begin{figure}[H]
    \centering
    \includegraphics[width=0.25\linewidth]{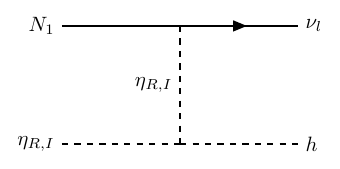}
    \includegraphics[width=0.25\linewidth]{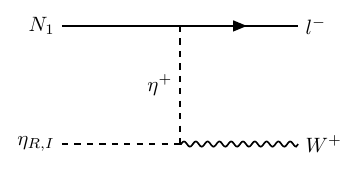}
    \includegraphics[width=0.25\linewidth]{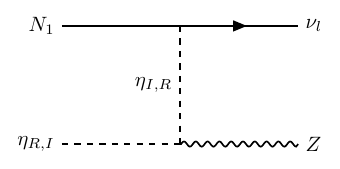}
    \includegraphics[width=0.25\linewidth]{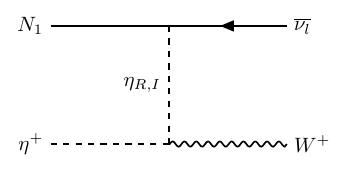}
    \includegraphics[width=0.25\linewidth]{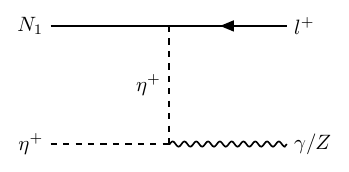}
    \includegraphics[width=0.25\linewidth]{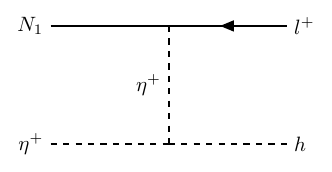}
    \includegraphics[width=0.25\linewidth]{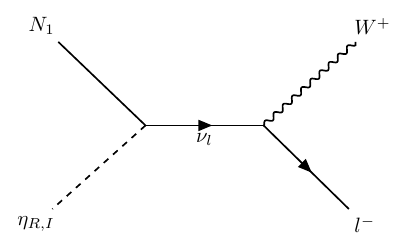}
    \includegraphics[width=0.25\linewidth]{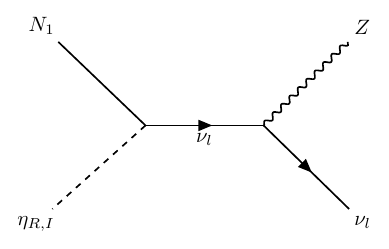}
    \includegraphics[width=0.25\linewidth]{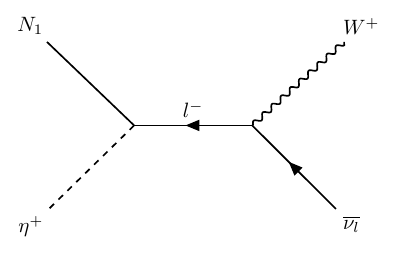}
    \includegraphics[width=0.25\linewidth]{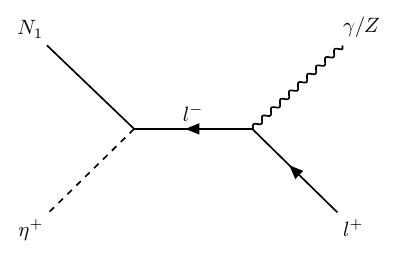}
    \includegraphics[width=0.25\linewidth]{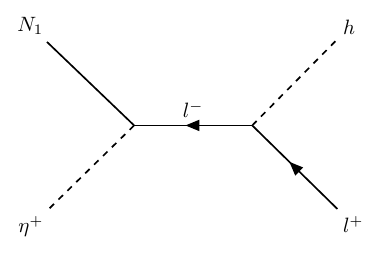}
    \caption{Feynman diagrams corresponding to \texttt{"1200"} processes.}
    \label{fig:feynCA1}
\end{figure}

\textbf{Self-Annihilation (SA) of DM:}
\\
We present the Feynman diagrams corresponding to the self-annihilation of sector-1 particle (i.e. $N_1$) in Fig. \ref{fig:feynSA1}.
\textbf{Annihilation and co-annihilation among the sector-2 particles:}
\\
We present the Feynman diagrams corresponding to the self-annihilation and co-annihilation among the sector-2 particles (i.e. $\eta_R,\eta_I$ and $\eta^\pm$) in Fig. \ref{fig:feynSA2}.
\textbf{Co-Annihilation (CA) among DM and sector-2 particles:}
\\
We present the Feynman diagrams corresponding to the co-annihilation among the sector-1 and sector-2 particles in Fig. \ref{fig:feynCA1}.

\begin{figure}[H]
    \centering
    \includegraphics[width=0.25\linewidth]{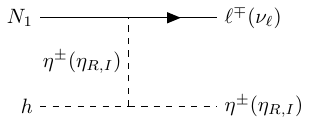}
    \includegraphics[width=0.25\linewidth]{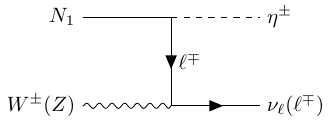}
    \includegraphics[width=0.25\linewidth]{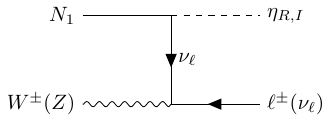}
    \includegraphics[width=0.25\linewidth]{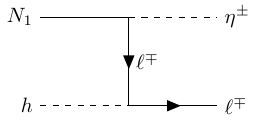}
    \caption{Feynman diagram corresponding to \texttt{"2010"} process.}
    \label{fig:feyncoscattering}
\end{figure}

\textbf{Co-Scattering (CS):}\\
We present the Feynman diagrams corresponding to the co-scattering of sector-1 particle with sector-0 particles to sector-2 and sector-0 particles in Fig. \ref{fig:feyncoscattering}.

\begin{figure}[H]
    \centering
    \includegraphics[width=0.25\linewidth]{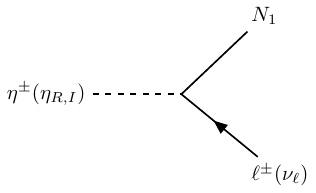}
    \caption{Feynman diagram representing decay of $\eta$ particle to DM ($N_1$) and leptons.}
    \label{fig:feyndecay}
\end{figure}

\textbf{Decay and Inverse Decay (D\& ID):}
\\
We show the Feynman diagram corresponding to the decay of sector-2 particle to sector-1 particle in Fig. \ref{fig:feyndecay}.

\section{Decay and Inverse Decay}\label{app:InvDecay}
In this section, we consider a general framework involving two dark sector particles: a singlet scalar $\eta$ and a singlet fermion $N_{1}$, with mass hierarchy $M_{\eta} > M_{N_{1}}$. We assume that $N_{1}$ does not couple directly to the SM particles, but interacts solely with $\eta$ through decay and inverse–decay processes. The scalar $\eta$, on the other hand, is assumed to interact efficiently with the SM thermal bath, thereby maintaining chemical equilibrium with the SM plasma.
The simultaneous presence of decay and inverse–decay processes, together with the thermal equilibration of $\eta$ with the SM bath, enables a depletion of the relic abundance of $N_{1}$. To elucidate this behavior, we begin by analyzing the evolution of the comoving number densities, defined as $Y_{i} = n_{i}/s$ (with $s = T^{3}$), for each species ($\eta$ and $N_{1}$). The evolution equations take the form:
\iffalse
For a process related to the decay of particle A to two final state particles B and C, the number density of B can be tracked by solving the Boltzmann equation
\begin{equation}
    \frac{dn_B}{dt}+3 \mathcal{H}n_B =\langle \Gamma_A \rangle \left(n_A - \frac{n_B}{n_B^{\rm eq}}.\frac{n_C}{n_C^{\rm eq}}.n_A^{\rm eq}\right)
\end{equation}

For the scalar $\eta$ decaying to both the DM candidate $N_1$ and SM leptons, we have the following equations in terms of $Y_{i}=n_{i}/s$ (for $s=T^3$) and $x=M_{N_1}/T$ to track the evolution of both $\eta$ and $N_1$:
\fi
\begin{eqnarray}
    \frac{dY_{\eta}}{dx}&=&-\frac{\Gamma_{\eta}}{x \mathcal{H}(x)} \left(Y_{\eta}-Y_{N_1}\frac{Y_{\eta}^{\rm eq}}{Y_{N_1}^{\rm eq}} \right)+ (\eta~\eta \leftrightarrow SM~SM),\label{eq:decay+fo1}
\end{eqnarray}
\begin{equation}
\frac{dY_{N_1}}{dx}=\frac{\langle \Gamma_{\eta} \rangle}{x \mathcal{H}(x)}  \left(Y_{\eta}-Y_{N_1}\frac{Y_{\eta}^{\rm eq}}{Y_{N_1}^{\rm eq}} \right),
    \label{eq:decay+fo2}
\end{equation}
\iffalse
\begin{equation}
        \frac{dY_{N_1}}{dx}=\frac{\langle \Gamma_{\eta} \rangle}{x \mathcal{H}(x)}  \left(Y_{\eta}-Y_{N_1}\frac{Y_{\eta}^{\rm eq}}{Y_{N_1}^{\rm eq}} \right)+ (N_1 N_1 \leftrightarrow SM~SM),
    \label{eq:decay+fo}
\end{equation}
\fi
where $x=M_{N_1}/T$, the average decay width, $\langle \Gamma_{\eta} \rangle=\Gamma_{\eta}(K_1(x)/K_2(x))$ and the decay width $\Gamma_{\eta}$ is given by
\begin{equation*}
    \begin{split}
        \Gamma_\eta & = \frac{y_{\alpha 1}^2 M_\eta}{8 \pi} \left(1-\frac{M_{N_1}^2}{M_\eta^2} \right)^2\\
        & = \frac{y_{\alpha 1}^2 \Delta M^2}{8 \pi} \times \frac{(2M_{N_1}+\Delta M)^2}{(M_{N_1}+\Delta M)^3}
    \end{split}
\end{equation*}
where $\Delta M=M_\eta-M_{N_1}$ is the mass splitting between the DM and the parent particle. We also include the $2 \to 2$ annihilation processes of $N_1$ to account for any initial relic abundance present before decay and inverse-decay effects become significant. The above equation suggests that the abundance of CDM $N_1$ can be lowered if $Y_{N1}$ attains a fixed number, and the ratio $Y_{N1}/Y_{N1}^{\rm eq}$ gets larger than 1, thereby reducing freeze-out abundance of CDM $N_1$. This scenario can be better explained if we rewrite the above equation (\ref{eq:decay+fo2}) as 
\begin{equation}
    \begin{split}
        \frac{dY_{N_1}}{dx}&=\frac{\langle \Gamma_{\eta} \rangle}{x \mathcal{H}(x)} \frac{Y_\eta^{\rm eq}}{Y_{N_1}^{\rm eq}} \left(Y_{\eta}~\frac{Y_{N_1}^{\rm eq}}{Y_{\eta}^{\rm eq}}-Y_{N_1} \right)+ (N_1 N_1 \leftrightarrow SM~SM)\\
        &= -\frac{\Gamma_{\rm ID}}{x \mathcal{H}(x)} \left(Y_{N_1}-Y_{N_1}^{\rm eq}~\frac{Y_\eta}{Y_{\eta}^{\rm eq}} \right)+ (N_1 N_1 \leftrightarrow SM~SM)
    \end{split}
    \label{eq:INVERSE_DECAY}
\end{equation}

\begin{figure}[H]
    \centering
    \includegraphics[width=0.6\linewidth]{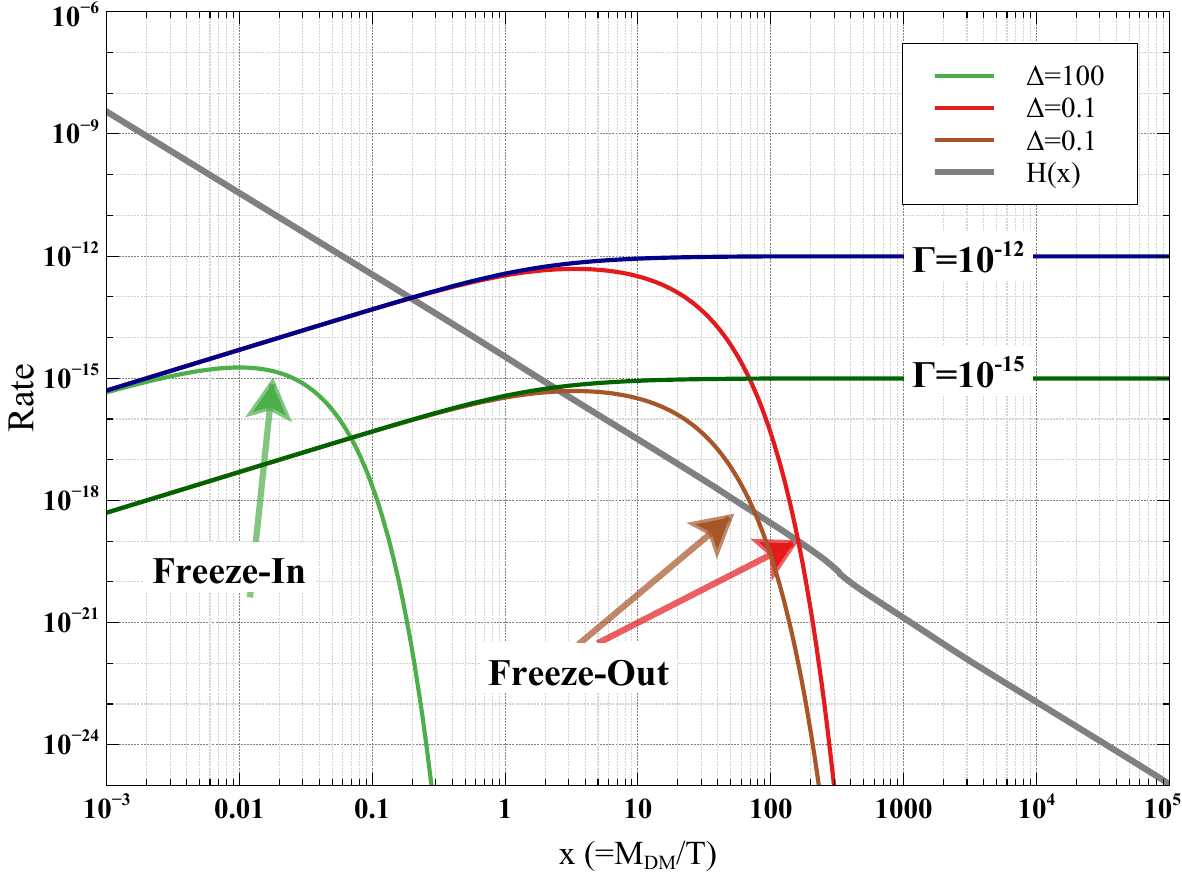}
    \caption{Rate of Inverse Decay as a function of $\Delta$, where $\Delta=\frac{\Delta M}{M_{N_1}}$.}
    \label{fig:RELIC}
\end{figure}

Eqn.~\ref{eq:INVERSE_DECAY} mimics a standard freeze-out scenario, however linear in $Y_{N_1}$, and with $\Gamma_{\rm ID}$ mimicking thermally averaged cross-section for the decay and inverse decay, with its expression given by
\begin{equation}
    \begin{split}
        \Gamma_{\rm ID}& =\langle \Gamma_\eta \rangle \frac{Y_\eta^{\rm eq}}{Y_{N_1}^{\rm eq}}\\
        & = \langle \Gamma_\eta \rangle \times \frac{g_\eta}{g_{N1}}\sqrt{1+\frac{\Delta M}{M_{N_1}}}\left(\frac{23 + x \frac{\Delta M}{M_{N_1}}}{15+8 x} \right)~e^{-x \frac{\Delta M}{M_{N_1}}}
        \label{eq:inverse_rate}
    \end{split}
\end{equation}
The figure~\ref{fig:RELIC} illustrates the freeze-in freeze-out scenario.

\section{Effect of Co-scattering on Relic Density of DM}\label{app:coscattering}
\begin{figure}[h]
    \centering
     \includegraphics[width=0.48\linewidth]{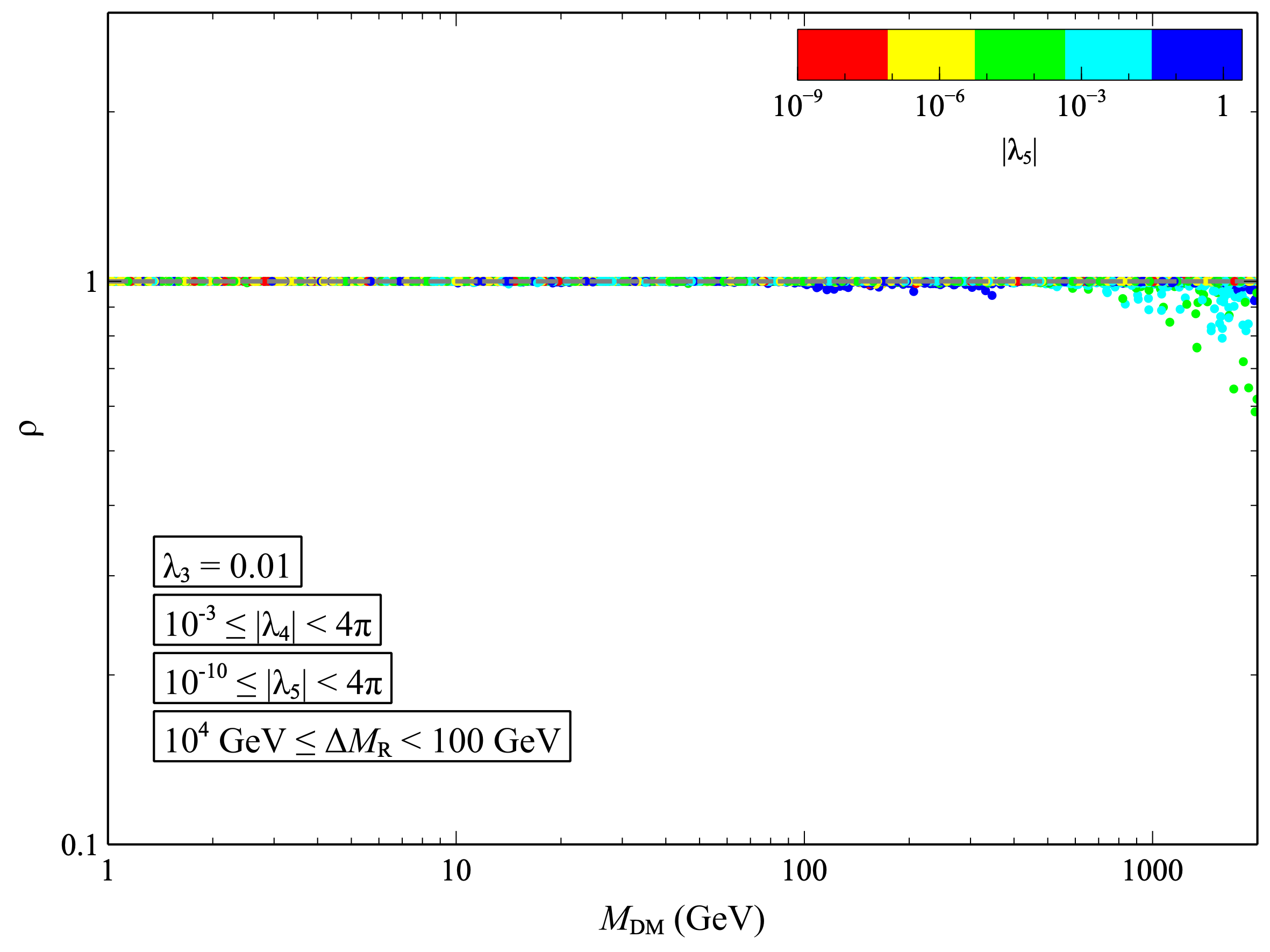}
    \caption{The parameter $\rho$ is shown as a function of the DM mass. Here, $\rho$ is defined as the ratio of the relic density obtained by including all processes to that obtained when all processes except the co-scattering contribution are taken into account.
}\label{fig:rho_param}
\end{figure}

In order to check the effect of co-scattering on DM relic density, we solve the BEs in Eqs. (\ref{eq:Y1}) and (\ref{eq:Y2}) with and without the co-scattering term denoted by the term \texttt{"2010"} in Eq.~(\ref{eq:gamma21}). Our parameters are chosen as before by fixing the $\lambda_3=0.01$ and varying $\lambda_4$ and $\lambda_5$ in the range $[10^{-3},4\pi]$ and $[10^{-10},4\pi]$, respectively. The $\Delta M_R$ is also varied in the range $[10^{-4},10^{2}]$ GeV. In order to segregate the impact of co-scattering, we define a parameter,
\begin{equation}
    \rho=\frac{\Omega_{\rm DM}h^2}{\Omega_{\rm DM}h^2~({\rm Co-scattering~off})},
\end{equation}
where $\Omega_{\rm DM}h^2$ represents the relic density by solving the BEs in Eqs. (\ref{eq:Y1}) and (\ref{eq:Y2}) while $\Omega_{\rm DM}h^2~({\rm Co-scattering~off})$ is obtained by solving the same BEs but by switching off the co-scattering term. 

The resulting values of the $\rho$ parameter are displayed in Fig.~\ref{fig:rho_param}, where we observe that $\rho$ remains approximately equal to $1$ across the entire DM mass range considered. A value of $\rho \approx 1$ indicates the absence of any significant co-scattering effects, whereas deviations from unity (i.e., $\rho < 1$) signal that co-scattering contributes non-negligibly to the final relic density. Our results therefore imply that the impact of co-scattering is subdominant compared to the decay and inverse-decay processes within the conversion-driven framework. Consequently, the conversion-driven dynamics can be effectively analyzed by focusing solely on the decay and inverse-decay contributions.

\section{Decay Width of $\eta$ particle}
Since $N_1$ is the lightest of the dark sector particles, other heavy particles can decay to DM if kinematically allowed. For instance, the $\eta$ particle, whether it be $\eta^+,\eta_R$ or $\eta_I$ can decay to $N_1$ and SM leptons. The decay width is calculated as
\begin{equation}
        \Gamma_{\eta\rightarrow N_{1}l_{\alpha}} =\frac{y_{\alpha1}^2}{8\pi~M_{\eta}} \left(M_{\eta}^2-(M_{\alpha}+M_{N_1})^2\right)\sqrt{1-\frac{(M_{N_1}-M_{\alpha})^2}{M_{\eta}^2}} \sqrt{1-\frac{(M_{N_1}+M_{\alpha})^2}{M_{\eta}^2}},
    \label{eq:decay_width_eta}
\end{equation}
where $\alpha\in\{e,\mu,\tau\}$ represents the SM lepton and $M_{\alpha}$ represents the mass of $\alpha$ lepton.

%%%%%%%%%%%%%%%%%%%%%%%%%%%%%%%%%%%%%%%%%%
%\bibliography{ref}

\begin{thebibliography}{10}
	
	\bibitem{Ma:2006km}
	E.~Ma, \emph{{Verifiable radiative seesaw mechanism of neutrino mass and dark
			matter}}, \href{https://doi.org/10.1103/PhysRevD.73.077301}{\emph{Phys. Rev.
			D} {\bfseries 73} (2006) 077301}
	[\href{https://arxiv.org/abs/hep-ph/0601225}{{\ttfamily hep-ph/0601225}}].
	
	\bibitem{Barbieri:2006dq}
	R.~Barbieri, L.J.~Hall and V.S.~Rychkov, \emph{{Improved naturalness with a
			heavy Higgs: An Alternative road to LHC physics}},
	\href{https://doi.org/10.1103/PhysRevD.74.015007}{\emph{Phys. Rev. D}
		{\bfseries 74} (2006) 015007}
	[\href{https://arxiv.org/abs/hep-ph/0603188}{{\ttfamily hep-ph/0603188}}].
	
	\bibitem{Kubo:2006yx}
	J.~Kubo, E.~Ma and D.~Suematsu, \emph{{Cold Dark Matter, Radiative Neutrino
			Mass, $\mu \to e\gamma$, and Neutrinoless Double Beta Decay}},
	\href{https://doi.org/10.1016/j.physletb.2006.08.085}{\emph{Phys. Lett. B}
		{\bfseries 642} (2006) 18}
	[\href{https://arxiv.org/abs/hep-ph/0604114}{{\ttfamily hep-ph/0604114}}].
	
	\bibitem{Vicente:2014wga}
	A.~Vicente and C.E.~Yaguna, \emph{{Probing the scotogenic model with lepton
			flavor violating processes}},
	\href{https://doi.org/10.1007/JHEP02(2015)144}{\emph{JHEP} {\bfseries 02}
		(2015) 144} [\href{https://arxiv.org/abs/1412.2545}{{\ttfamily 1412.2545}}].
	
	\bibitem{Liu:2022byu}
	J.~Liu, Z.-L.~Han, Y.~Jin and H.~Li, \emph{{Unraveling the Scotogenic model at
			muon collider}}, \href{https://doi.org/10.1007/JHEP12(2022)057}{\emph{JHEP}
		{\bfseries 12} (2022) 057}
	[\href{https://arxiv.org/abs/2207.07382}{{\ttfamily 2207.07382}}].
	
	\bibitem{Avila:2025qsc}
	I.M.~{\'A}vila, A.~Karan, S.~Mandal, S.~Sadhukhan and J.W.F.~Valle, \emph{{Dark
			matter as the source of neutrino mass: Theory overview and experimental
			prospects}}, \href{https://doi.org/10.1016/j.physrep.2026.02.003}{\emph{Phys.
			Rept.} {\bfseries 1173} (2026) 1}
	[\href{https://arxiv.org/abs/2506.24027}{{\ttfamily 2506.24027}}].
	
	\bibitem{Molinaro:2014lfa}
	E.~Molinaro, C.E.~Yaguna and O.~Zapata, \emph{{FIMP realization of the
			scotogenic model}},
	\href{https://doi.org/10.1088/1475-7516/2014/07/015}{\emph{JCAP} {\bfseries
			07} (2014) 015} [\href{https://arxiv.org/abs/1405.1259}{{\ttfamily
			1405.1259}}].
	
	\bibitem{Hessler:2016kwm}
	A.G.~Hessler, A.~Ibarra, E.~Molinaro and S.~Vogl, \emph{{Probing the scotogenic
			FIMP at the LHC}}, \href{https://doi.org/10.1007/JHEP01(2017)100}{\emph{JHEP}
		{\bfseries 01} (2017) 100}
	[\href{https://arxiv.org/abs/1611.09540}{{\ttfamily 1611.09540}}].
	
	\bibitem{Baumholzer:2019twf}
	S.~Baumholzer, V.~Brdar, P.~Schwaller and A.~Segner, \emph{{Shining Light on
			the Scotogenic Model: Interplay of Colliders and Cosmology}},
	\href{https://doi.org/10.1007/JHEP09(2020)136}{\emph{JHEP} {\bfseries 09}
		(2020) 136} [\href{https://arxiv.org/abs/1912.08215}{{\ttfamily
			1912.08215}}].
	
	\bibitem{Heeck:2022rep}
	J.~Heeck, J.~Heisig and A.~Thapa, \emph{{Dark matter and radiative neutrino
			masses in conversion-driven scotogenesis}},
	\href{https://doi.org/10.1103/PhysRevD.107.015028}{\emph{Phys. Rev. D}
		{\bfseries 107} (2023) 015028}
	[\href{https://arxiv.org/abs/2211.13013}{{\ttfamily 2211.13013}}].
	
	\bibitem{Bonilla:2019ipe}
	C.~Bonilla, L.M.G.~de~la Vega, J.M.~Lamprea, R.A.~Lineros and E.~Peinado,
	\emph{{Fermion Dark Matter and Radiative Neutrino Masses from Spontaneous
			Lepton Number Breaking}},
	\href{https://doi.org/10.1088/1367-2630/ab7254}{\emph{New J. Phys.}
		{\bfseries 22} (2020) 033009}
	[\href{https://arxiv.org/abs/1908.04276}{{\ttfamily 1908.04276}}].
	
	\bibitem{DeRomeri:2022cem}
	V.~De~Romeri, J.~Nava, M.~Puerta and A.~Vicente, \emph{{Dark matter in the
			scotogenic model with spontaneous lepton number violation}},
	\href{https://doi.org/10.1103/PhysRevD.107.095019}{\emph{Phys. Rev. D}
		{\bfseries 107} (2023) 095019}
	[\href{https://arxiv.org/abs/2210.07706}{{\ttfamily 2210.07706}}].
	
	\bibitem{Chun:2023vbh}
	E.J.~Chun, A.~Roy, S.~Mandal and M.~Mitra, \emph{{Fermionic dark matter in
			Dynamical Scotogenic Model}},
	\href{https://doi.org/10.1007/JHEP08(2023)130}{\emph{JHEP} {\bfseries 08}
		(2023) 130} [\href{https://arxiv.org/abs/2303.02681}{{\ttfamily
			2303.02681}}].
	
	\bibitem{Borah:2020wut}
	D.~Borah, A.~Dasgupta, K.~Fujikura, S.K.~Kang and D.~Mahanta, \emph{{Observable
			Gravitational Waves in Minimal Scotogenic Model}},
	\href{https://doi.org/10.1088/1475-7516/2020/08/046}{\emph{JCAP} {\bfseries
			08} (2020) 046} [\href{https://arxiv.org/abs/2003.02276}{{\ttfamily
			2003.02276}}].
	
	\bibitem{Borah:2018smz}
	D.~Borah, D.~Nanda, N.~Narendra and N.~Sahu, \emph{{Right-handed neutrino dark
			matter with radiative neutrino mass in gauged $B-L$ model}},
	\href{https://doi.org/10.1016/j.nuclphysb.2019.114841}{\emph{Nucl. Phys. B}
		{\bfseries 950} (2020) 114841}
	[\href{https://arxiv.org/abs/1810.12920}{{\ttfamily 1810.12920}}].
	
	\bibitem{Roy:2025moo}
	A.~Roy and R.~Sahu, \emph{{Scrutinizing Fermionic Dark Matter in Scotogenic
			Model with Low Reheating Temperature}},
	\href{https://arxiv.org/abs/2508.14726}{{\ttfamily 2508.14726}}.
	
	\bibitem{Borah:2021khc}
	D.~Borah, M.~Dutta, S.~Mahapatra and N.~Sahu, \emph{{Lepton anomalous magnetic
			moment with singlet-doublet fermion dark matter in a scotogenic
			U(1)L{\ensuremath{\mu}}-L{\ensuremath{\tau}} model}},
	\href{https://doi.org/10.1103/PhysRevD.105.015029}{\emph{Phys. Rev. D}
		{\bfseries 105} (2022) 015029}
	[\href{https://arxiv.org/abs/2109.02699}{{\ttfamily 2109.02699}}].
	
	\bibitem{Borah:2021pet}
	D.~Borah, M.~Dutta, S.~Mahapatra and N.~Sahu, \emph{{Self-interacting dark
			matter via right handed neutrino portal}},
	\href{https://doi.org/10.1103/PhysRevD.105.015004}{\emph{Phys. Rev. D}
		{\bfseries 105} (2022) 015004}
	[\href{https://arxiv.org/abs/2110.00021}{{\ttfamily 2110.00021}}].
	
	\bibitem{Borah:2021qmi}
	D.~Borah, A.~Dasgupta, S.~Mahapatra and N.~Sahu, \emph{{Unified origin of dark
			matter self interactions and low scale leptogenesis}},
	\href{https://doi.org/10.1103/PhysRevD.106.095028}{\emph{Phys. Rev. D}
		{\bfseries 106} (2022) 095028}
	[\href{https://arxiv.org/abs/2112.14786}{{\ttfamily 2112.14786}}].
	
	\bibitem{Borah:2023hqw}
	D.~Borah, S.~Mahapatra, P.K.~Paul and N.~Sahu, \emph{{Scotogenic
			U(1)L{\ensuremath{\mu}}-L{\ensuremath{\tau}} origin of
			(g-2){\ensuremath{\mu}}, W-mass anomaly and 95~GeV excess}},
	\href{https://doi.org/10.1103/PhysRevD.109.055021}{\emph{Phys. Rev. D}
		{\bfseries 109} (2024) 055021}
	[\href{https://arxiv.org/abs/2310.11953}{{\ttfamily 2310.11953}}].
	
	\bibitem{Ganguly:2023jml}
	J.~Ganguly, J.~Gluza, B.~Karmakar and S.~Mahapatra, \emph{{Phenomenology of the
			flavor symmetric scoto-seesaw model with dark matter and TM1 mixing}},
	\href{https://doi.org/10.1103/PhysRevD.110.035012}{\emph{Phys. Rev. D}
		{\bfseries 110} (2024) 035012}
	[\href{https://arxiv.org/abs/2311.15997}{{\ttfamily 2311.15997}}].
	
	\bibitem{Borah:2024gql}
	D.~Borah, P.~Das, B.~Karmakar and S.~Mahapatra, \emph{{Discrete dark matter
			with light Dirac neutrinos}},
	\href{https://doi.org/10.1103/PhysRevD.111.035032}{\emph{Phys. Rev. D}
		{\bfseries 111} (2025) 035032}
	[\href{https://arxiv.org/abs/2406.17861}{{\ttfamily 2406.17861}}].
	
	\bibitem{Planck:2018vyg}
	{\scshape Planck} collaboration, \emph{{Planck 2018 results. VI. Cosmological
			parameters}},
	\href{https://doi.org/10.1051/0004-6361/201833910}{\emph{Astron. Astrophys.}
		{\bfseries 641} (2020) A6}
	[\href{https://arxiv.org/abs/1807.06209}{{\ttfamily 1807.06209}}].
	
	\bibitem{MEGII:2023ltw}
	{\scshape MEG II} collaboration, \emph{{A search for $\mu ^+ \rightarrow e^+
			\gamma $ with the first dataset of the MEG~II experiment}},
	\href{https://doi.org/10.1140/epjc/s10052-024-12416-2}{\emph{Eur. Phys. J. C}
		{\bfseries 84} (2024) 216}
	[\href{https://arxiv.org/abs/2310.12614}{{\ttfamily 2310.12614}}].
	
	\bibitem{Super-Kamiokande:1998kpq}
	{\scshape Super-Kamiokande} collaboration, \emph{{Evidence for oscillation of
			atmospheric neutrinos}},
	\href{https://doi.org/10.1103/PhysRevLett.81.1562}{\emph{Phys. Rev. Lett.}
		{\bfseries 81} (1998) 1562}
	[\href{https://arxiv.org/abs/hep-ex/9807003}{{\ttfamily hep-ex/9807003}}].
	
	\bibitem{SNO:2001kpb}
	{\scshape SNO} collaboration, \emph{{Measurement of the rate of $\nu_e+d \to
			p+p+e^-$ interactions produced by $^8$B solar neutrinos at the Sudbury
			Neutrino Observatory}},
	\href{https://doi.org/10.1103/PhysRevLett.87.071301}{\emph{Phys. Rev. Lett.}
		{\bfseries 87} (2001) 071301}
	[\href{https://arxiv.org/abs/nucl-ex/0106015}{{\ttfamily nucl-ex/0106015}}].
	
	\bibitem{DoubleChooz:2011ymz}
	{\scshape Double Chooz} collaboration, \emph{{Indication of Reactor
			$\bar{\nu}_e$ Disappearance in the Double Chooz Experiment}},
	\href{https://doi.org/10.1103/PhysRevLett.108.131801}{\emph{Phys. Rev. Lett.}
		{\bfseries 108} (2012) 131801}
	[\href{https://arxiv.org/abs/1112.6353}{{\ttfamily 1112.6353}}].
	
	\bibitem{DayaBay:2012fng}
	{\scshape Daya Bay} collaboration, \emph{{Observation of electron-antineutrino
			disappearance at Daya Bay}},
	\href{https://doi.org/10.1103/PhysRevLett.108.171803}{\emph{Phys. Rev. Lett.}
		{\bfseries 108} (2012) 171803}
	[\href{https://arxiv.org/abs/1203.1669}{{\ttfamily 1203.1669}}].
	
	\bibitem{RENO:2012mkc}
	{\scshape RENO} collaboration, \emph{{Observation of Reactor Electron
			Antineutrino Disappearance in the RENO Experiment}},
	\href{https://doi.org/10.1103/PhysRevLett.108.191802}{\emph{Phys. Rev. Lett.}
		{\bfseries 108} (2012) 191802}
	[\href{https://arxiv.org/abs/1204.0626}{{\ttfamily 1204.0626}}].
	
	\bibitem{Aliberti:2025beg}
	R.~Aliberti et~al., \emph{{The anomalous magnetic moment of the muon in the
			Standard Model: an update}},
	\href{https://doi.org/10.1016/j.physrep.2025.08.002}{\emph{Phys. Rept.}
		{\bfseries 1143} (2025) 1}
	[\href{https://arxiv.org/abs/2505.21476}{{\ttfamily 2505.21476}}].
	
	\bibitem{Muong-2:2025xyk}
	{\scshape Muon g-2} collaboration, \emph{{Measurement of the Positive Muon
			Anomalous Magnetic Moment to 127~ppb}},
	\href{https://doi.org/10.1103/7clf-sm2v}{\emph{Phys. Rev. Lett.} {\bfseries
			135} (2025) 101802} [\href{https://arxiv.org/abs/2506.03069}{{\ttfamily
			2506.03069}}].
	
	\bibitem{Frank:2021pkc}
	M.~Frank, E.G.~Fuakye and M.~Toharia, \emph{{Restricting the parameter space of
			type-II two-Higgs-doublet models with CP violation}},
	\href{https://doi.org/10.1103/PhysRevD.106.035010}{\emph{Phys. Rev. D}
		{\bfseries 106} (2022) 035010}
	[\href{https://arxiv.org/abs/2112.14295}{{\ttfamily 2112.14295}}].
	
	\bibitem{Griest:1990kh}
	K.~Griest and D.~Seckel, \emph{{Three exceptions in the calculation of relic
			abundances}}, \href{https://doi.org/10.1103/PhysRevD.43.3191}{\emph{Phys.
			Rev. D} {\bfseries 43} (1991) 3191}.
	
	\bibitem{DAgnolo:2017dbv}
	R.T.~D'Agnolo, D.~Pappadopulo and J.T.~Ruderman, \emph{{Fourth Exception in the
			Calculation of Relic Abundances}},
	\href{https://doi.org/10.1103/PhysRevLett.119.061102}{\emph{Phys. Rev. Lett.}
		{\bfseries 119} (2017) 061102}
	[\href{https://arxiv.org/abs/1705.08450}{{\ttfamily 1705.08450}}].
	
	\bibitem{Paul:2024prs}
	P.K.~Paul, S.K.~Sahoo and N.~Sahu, \emph{{Anatomy of singlet-doublet dark
			matter relic: annihilation, co-annihilation, co-scattering, and freeze-in}},
	\href{https://doi.org/10.1088/1475-7516/2025/10/053}{\emph{JCAP} {\bfseries
			10} (2025) 053} [\href{https://arxiv.org/abs/2412.02607}{{\ttfamily
			2412.02607}}].
	
	\bibitem{Paul:2025spm}
	P.K.~Paul, S.K.~Sahoo and N.~Sahu, \emph{{Impact of conversion-driven processes
			on singlet-doublet Majorana dark matter relic}},
	\href{https://arxiv.org/abs/2511.14571}{{\ttfamily 2511.14571}}.
	
	\bibitem{Garny:2017rxs}
	M.~Garny, J.~Heisig, B.~L{\"u}lf and S.~Vogl, \emph{{Coannihilation without
			chemical equilibrium}},
	\href{https://doi.org/10.1103/PhysRevD.96.103521}{\emph{Phys. Rev. D}
		{\bfseries 96} (2017) 103521}
	[\href{https://arxiv.org/abs/1705.09292}{{\ttfamily 1705.09292}}].
	
	\bibitem{Maity:2019hre}
	T.N.~Maity and T.S.~Ray, \emph{{Exchange driven freeze out of dark matter}},
	\href{https://doi.org/10.1103/PhysRevD.101.103013}{\emph{Phys. Rev. D}
		{\bfseries 101} (2020) 103013}
	[\href{https://arxiv.org/abs/1908.10343}{{\ttfamily 1908.10343}}].
	
	\bibitem{Liu:2025swd}
	A.~Liu, Z.-L.~Han, F.~Huang, F.-L.~Shao and W.~Wang, \emph{{Coscattering Dark
			Matter in Scotogenic Models}},
	\href{https://arxiv.org/abs/2510.13231}{{\ttfamily 2510.13231}}.
	
	\bibitem{Kannike:2012pe}
	K.~Kannike, \emph{{Vacuum Stability Conditions From Copositivity Criteria}},
	\href{https://doi.org/10.1140/epjc/s10052-012-2093-z}{\emph{Eur. Phys. J. C}
		{\bfseries 72} (2012) 2093}
	[\href{https://arxiv.org/abs/1205.3781}{{\ttfamily 1205.3781}}].
	
	\bibitem{Belyaev:2016lok}
	A.~Belyaev, G.~Cacciapaglia, I.P.~Ivanov, F.~Rojas-Abatte and M.~Thomas,
	\emph{{Anatomy of the Inert Two Higgs Doublet Model in the light of the LHC
			and non-LHC Dark Matter Searches}},
	\href{https://doi.org/10.1103/PhysRevD.97.035011}{\emph{Phys. Rev. D}
		{\bfseries 97} (2018) 035011}
	[\href{https://arxiv.org/abs/1612.00511}{{\ttfamily 1612.00511}}].
	
	\bibitem{Peskin:1990zt}
	M.E.~Peskin and T.~Takeuchi, \emph{{A New constraint on a strongly interacting
			Higgs sector}}, \href{https://doi.org/10.1103/PhysRevLett.65.964}{\emph{Phys.
			Rev. Lett.} {\bfseries 65} (1990) 964}.
	
	\bibitem{Peskin:1991sw}
	M.E.~Peskin and T.~Takeuchi, \emph{{Estimation of oblique electroweak
			corrections}}, \href{https://doi.org/10.1103/PhysRevD.46.381}{\emph{Phys.
			Rev. D} {\bfseries 46} (1992) 381}.
	
	\bibitem{Grimus:2007if}
	W.~Grimus, L.~Lavoura, O.M.~Ogreid and P.~Osland, \emph{{A Precision constraint
			on multi-Higgs-doublet models}},
	\href{https://doi.org/10.1088/0954-3899/35/7/075001}{\emph{J. Phys. G}
		{\bfseries 35} (2008) 075001}
	[\href{https://arxiv.org/abs/0711.4022}{{\ttfamily 0711.4022}}].
	
	\bibitem{Babu:2019mfe}
	K.S.~Babu, P.S.B.~Dev, S.~Jana and A.~Thapa, \emph{{Non-Standard Interactions
			in Radiative Neutrino Mass Models}},
	\href{https://doi.org/10.1007/JHEP03(2020)006}{\emph{JHEP} {\bfseries 03}
		(2020) 006} [\href{https://arxiv.org/abs/1907.09498}{{\ttfamily
			1907.09498}}].
	
	\bibitem{ParticleDataGroup:2024cfk}
	{\scshape Particle Data Group} collaboration, \emph{{Review of particle
			physics}}, \href{https://doi.org/10.1103/PhysRevD.110.030001}{\emph{Phys.
			Rev. D} {\bfseries 110} (2024) 030001}.
	
	\bibitem{Casas:2001sr}
	J.A.~Casas and A.~Ibarra, \emph{{Oscillating neutrinos and $\mu \to e,
			\gamma$}}, \href{https://doi.org/10.1016/S0550-3213(01)00475-8}{\emph{Nucl.
			Phys. B} {\bfseries 618} (2001) 171}
	[\href{https://arxiv.org/abs/hep-ph/0103065}{{\ttfamily hep-ph/0103065}}].
	
	\bibitem{deSalas:2020pgw}
	P.F.~de~Salas, D.V.~Forero, S.~Gariazzo, P.~Mart{\'\i}nez-Mirav{\'e}, O.~Mena,
	C.A.~Ternes et~al., \emph{{2020 global reassessment of the neutrino
			oscillation picture}},
	\href{https://doi.org/10.1007/JHEP02(2021)071}{\emph{JHEP} {\bfseries 02}
		(2021) 071} [\href{https://arxiv.org/abs/2006.11237}{{\ttfamily
			2006.11237}}].
	
	\bibitem{Lindner:2016bgg}
	M.~Lindner, M.~Platscher and F.S.~Queiroz, \emph{{A Call for New Physics : The
			Muon Anomalous Magnetic Moment and Lepton Flavor Violation}},
	\href{https://doi.org/10.1016/j.physrep.2017.12.001}{\emph{Phys. Rept.}
		{\bfseries 731} (2018) 1} [\href{https://arxiv.org/abs/1610.06587}{{\ttfamily
			1610.06587}}].
	
	\bibitem{Petcov:1976ff}
	S.T.~Petcov, \emph{{The Processes $\mu \rightarrow e + \gamma, \mu \rightarrow
			e + \overline{e}, \nu' \rightarrow \nu + \gamma$ in the Weinberg-Salam Model
			with Neutrino Mixing}}, {\emph{Sov. J. Nucl. Phys.} {\bfseries 25} (1977)
		340}.
	
	\bibitem{Ma:2001mr}
	E.~Ma and M.~Raidal, \emph{{Neutrino mass, muon anomalous magnetic moment, and
			lepton flavor nonconservation}},
	\href{https://doi.org/10.1103/PhysRevLett.87.011802}{\emph{Phys. Rev. Lett.}
		{\bfseries 87} (2001) 011802}
	[\href{https://arxiv.org/abs/hep-ph/0102255}{{\ttfamily hep-ph/0102255}}].
	
	\bibitem{Gondolo:1990dk}
	P.~Gondolo and G.~Gelmini, \emph{{Cosmic abundances of stable particles:
			Improved analysis}},
	\href{https://doi.org/10.1016/0550-3213(91)90438-4}{\emph{Nucl. Phys. B}
		{\bfseries 360} (1991) 145}.
	
	\bibitem{Alguero:2022inz}
	G.~Alguero, G.~Belanger, S.~Kraml and A.~Pukhov, \emph{{Co-scattering in
			micrOMEGAs: A case study for the singlet-triplet dark matter model}},
	\href{https://doi.org/10.21468/SciPostPhys.13.6.124}{\emph{SciPost Phys.}
		{\bfseries 13} (2022) 124}
	[\href{https://arxiv.org/abs/2207.10536}{{\ttfamily 2207.10536}}].
	
	\bibitem{Alguero:2023zol}
	G.~Alguero, G.~Belanger, F.~Boudjema, S.~Chakraborti, A.~Goudelis, S.~Kraml
	et~al., \emph{{micrOMEGAs 6.0: N-component dark matter}},
	\href{https://doi.org/10.1016/j.cpc.2024.109133}{\emph{Comput. Phys. Commun.}
		{\bfseries 299} (2024) 109133}
	[\href{https://arxiv.org/abs/2312.14894}{{\ttfamily 2312.14894}}].
	
	\bibitem{McDonald:2001vt}
	J.~McDonald, \emph{{Thermally generated gauge singlet scalars as
			selfinteracting dark matter}},
	\href{https://doi.org/10.1103/PhysRevLett.88.091304}{\emph{Phys. Rev. Lett.}
		{\bfseries 88} (2002) 091304}
	[\href{https://arxiv.org/abs/hep-ph/0106249}{{\ttfamily hep-ph/0106249}}].
	
	\bibitem{McDonald:2008ua}
	J.~McDonald and N.~Sahu, \emph{{keV Warm Dark Matter via the Supersymmetric
			Higgs Portal}}, \href{https://doi.org/10.1103/PhysRevD.79.103523}{\emph{Phys.
			Rev. D} {\bfseries 79} (2009) 103523}
	[\href{https://arxiv.org/abs/0809.0247}{{\ttfamily 0809.0247}}].
	
	\bibitem{Hall:2009bx}
	L.J.~Hall, K.~Jedamzik, J.~March-Russell and S.M.~West, \emph{{Freeze-In
			Production of FIMP Dark Matter}},
	\href{https://doi.org/10.1007/JHEP03(2010)080}{\emph{JHEP} {\bfseries 03}
		(2010) 080} [\href{https://arxiv.org/abs/0911.1120}{{\ttfamily 0911.1120}}].
	
	\bibitem{Ibarra:2016dlb}
	A.~Ibarra, C.E.~Yaguna and O.~Zapata, \emph{{Direct Detection of Fermion Dark
			Matter in the Radiative Seesaw Model}},
	\href{https://doi.org/10.1103/PhysRevD.93.035012}{\emph{Phys. Rev. D}
		{\bfseries 93} (2016) 035012}
	[\href{https://arxiv.org/abs/1601.01163}{{\ttfamily 1601.01163}}].
	
	\bibitem{Babu:2025czb}
	K.S.~Babu, P.S.B.~Dev and A.~Thapa, \emph{{Large Neutrino-Dark Matter
			Interactions: From Effective Field Theory to Ultraviolet Completions}},
	\href{https://arxiv.org/abs/2512.25035}{{\ttfamily 2512.25035}}.
	
	\bibitem{Jungman:1995df}
	G.~Jungman, M.~Kamionkowski and K.~Griest, \emph{{Supersymmetric dark matter}},
	\href{https://doi.org/10.1016/0370-1573(95)00058-5}{\emph{Phys. Rept.}
		{\bfseries 267} (1996) 195}
	[\href{https://arxiv.org/abs/hep-ph/9506380}{{\ttfamily hep-ph/9506380}}].
	
	\bibitem{HERMES:2006jyl}
	{\scshape HERMES} collaboration, \emph{{Precise determination of the spin
			structure function g(1) of the proton, deuteron and neutron}},
	\href{https://doi.org/10.1103/PhysRevD.75.012007}{\emph{Phys. Rev. D}
		{\bfseries 75} (2007) 012007}
	[\href{https://arxiv.org/abs/hep-ex/0609039}{{\ttfamily hep-ex/0609039}}].
	
	\bibitem{XENON:2025vwd}
	{\scshape XENON} collaboration, \emph{{WIMP Dark Matter Search Using a 3.1
			Tonne-Year Exposure of the XENONnT Experiment}},
	\href{https://doi.org/10.1103/msw4-t342}{\emph{Phys. Rev. Lett.} {\bfseries
			135} (2025) 221003} [\href{https://arxiv.org/abs/2502.18005}{{\ttfamily
			2502.18005}}].
	
	\bibitem{LZ:2024zvo}
	{\scshape LZ} collaboration, \emph{{Dark Matter Search Results from
			4.2{\,}{\,}Tonne-Years of Exposure of the LUX-ZEPLIN (LZ) Experiment}},
	\href{https://doi.org/10.1103/4dyc-z8zf}{\emph{Phys. Rev. Lett.} {\bfseries
			135} (2025) 011802} [\href{https://arxiv.org/abs/2410.17036}{{\ttfamily
			2410.17036}}].
	
	\bibitem{PandaX:2024qfu}
	{\scshape PandaX} collaboration, \emph{{Dark Matter Search Results from
			1.54{\,}{\,}Tonne{\textperiodcentered}Year Exposure of PandaX-4T}},
	\href{https://doi.org/10.1103/PhysRevLett.134.011805}{\emph{Phys. Rev. Lett.}
		{\bfseries 134} (2025) 011805}
	[\href{https://arxiv.org/abs/2408.00664}{{\ttfamily 2408.00664}}].
	
	\bibitem{Hoferichter:2017olk}
	M.~Hoferichter, P.~Klos, J.~Men{\'e}ndez and A.~Schwenk, \emph{{Improved limits
			for Higgs-portal dark matter from LHC searches}},
	\href{https://doi.org/10.1103/PhysRevLett.119.181803}{\emph{Phys. Rev. Lett.}
		{\bfseries 119} (2017) 181803}
	[\href{https://arxiv.org/abs/1708.02245}{{\ttfamily 1708.02245}}].
	
	\bibitem{CMS:2024trg}
	{\scshape CMS} collaboration, \emph{{Search for long-lived particles using
			displaced vertices and missing transverse momentum in proton-proton
			collisions at s=13{\,}{\,}TeV}},
	\href{https://doi.org/10.1103/PhysRevD.109.112005}{\emph{Phys. Rev. D}
		{\bfseries 109} (2024) 112005}
	[\href{https://arxiv.org/abs/2402.15804}{{\ttfamily 2402.15804}}].
	
	\bibitem{ATLAS:2022gbw}
	{\scshape ATLAS} collaboration, \emph{{Search for events with a pair of
			displaced vertices from long-lived neutral particles decaying into hadronic
			jets in the ATLAS muon spectrometer in pp collisions at $\sqrt
			s$=13{\,}{\,}TeV}},
	\href{https://doi.org/10.1103/PhysRevD.106.032005}{\emph{Phys. Rev. D}
		{\bfseries 106} (2022) 032005}
	[\href{https://arxiv.org/abs/2203.00587}{{\ttfamily 2203.00587}}].
	
	\bibitem{MATHUSLA:2019qpy}
	{\scshape MATHUSLA} collaboration, \emph{{Explore the lifetime frontier with
			MATHUSLA}},
	\href{https://doi.org/10.1088/1748-0221/15/06/C06026}{\emph{JINST} {\bfseries
			15} (2020) C06026} [\href{https://arxiv.org/abs/1901.04040}{{\ttfamily
			1901.04040}}].
	
	\bibitem{Chen:2015gaa}
	C.-Y.~Chen, S.~Dawson and Y.~Zhang, \emph{{Complementarity of LHC and EDMs for
			Exploring Higgs CP Violation}},
	\href{https://doi.org/10.1007/JHEP06(2015)056}{\emph{JHEP} {\bfseries 06}
		(2015) 056} [\href{https://arxiv.org/abs/1503.01114}{{\ttfamily
			1503.01114}}].
	
	\bibitem{Batra:2022pej}
	A.~Batra, S.K.~A, S.~Mandal, H.~Prajapati and R.~Srivastava, \emph{{CDF-II
			W-boson mass anomaly in the canonical Scotogenic neutrino{\textendash}dark
			matter model}}, \href{https://doi.org/10.1142/S0217732323500906}{\emph{Mod.
			Phys. Lett. A} {\bfseries 38} (2023) 2350090}
	[\href{https://arxiv.org/abs/2204.11945}{{\ttfamily 2204.11945}}].
	
	\bibitem{Batra:2022arl}
	A.~Batra, P.~Bharadwaj, S.~Mandal, R.~Srivastava and J.W.F.~Valle,
	\emph{{W-mass anomaly in the simplest linear seesaw mechanism}},
	\href{https://doi.org/10.1016/j.physletb.2022.137408}{\emph{Phys. Lett. B}
		{\bfseries 834} (2022) 137408}
	[\href{https://arxiv.org/abs/2208.04983}{{\ttfamily 2208.04983}}].
	
	\bibitem{LHCHiggsCrossSectionWorkingGroup:2016ypw}
	{\scshape LHC Higgs Cross Section Working Group} collaboration, \emph{{Handbook
			of LHC Higgs Cross Sections: 4. Deciphering the Nature of the Higgs Sector}},
	\href{https://doi.org/10.23731/CYRM-2017-002}{\emph{CERN Yellow Rep. Monogr.}
		{\bfseries 2} (2017) 1} [\href{https://arxiv.org/abs/1610.07922}{{\ttfamily
			1610.07922}}].
	
	\bibitem{CMS:2022qva}
	{\scshape CMS} collaboration, \emph{{Search for invisible decays of the Higgs
			boson produced via vector boson fusion in proton-proton collisions at
			s=13{\,}{\,}TeV}},
	\href{https://doi.org/10.1103/PhysRevD.105.092007}{\emph{Phys. Rev. D}
		{\bfseries 105} (2022) 092007}
	[\href{https://arxiv.org/abs/2201.11585}{{\ttfamily 2201.11585}}].
	
	\bibitem{ATLAS:2022yvh}
	{\scshape ATLAS} collaboration, \emph{{Search for invisible Higgs-boson decays
			in events with vector-boson fusion signatures using 139 fb$^{-1}$ of
			proton-proton data recorded by the ATLAS experiment}},
	\href{https://doi.org/10.1007/JHEP08(2022)104}{\emph{JHEP} {\bfseries 08}
		(2022) 104} [\href{https://arxiv.org/abs/2202.07953}{{\ttfamily
			2202.07953}}].
	
\end{thebibliography}
\bibliographystyle{JHEP}

\providecommand{\href}[2]{#2}\begingroup\raggedright\endgroup

%%%%%%%%%%%%%%%%%%%%%%%%%%%%%%%%%%%%%%%%%%

\end{document}